\documentclass[a4paper,11pt]{article}

\usepackage[table]{xcolor}
\usepackage{lmodern}
\usepackage{jcappub} 
\usepackage[toc,page]{appendix}
\usepackage{subcaption}
\usepackage{mathtools}
\usepackage{orcidlink}
\usepackage{multirow}
\usepackage[T1]{fontenc} 
\usepackage{float}
\usepackage{bm}
\usepackage{soul}
\usepackage{tensor}

\newcommand{\semibold}[1]{{\fontseries{b}\selectfont{#1}}}

\newcommand{\para}[1]{\par\vspace{2mm}\noindent\semibold{{#1.}---}\ignorespaces} 

\DeclareMathOperator{\Or}{O}

\renewcommand{\leq}{\leqslant}

\definecolor{verde}{rgb}{0,0.5,0}

\definecolor{SussexCobaltBlue}{HTML}{1d4289}
\definecolor{SussexDeepAquamarine}{HTML}{007a78}
\definecolor{SussexPowderBlue}{HTML}{7da1c4}
\definecolor{SussexCornYellow}{HTML}{f2c75c}
\definecolor{SussexChinaRose}{HTML}{be84a3}
\definecolor{SussexBurntOrange}{HTML}{dc582a}
\definecolor{SussexGrape}{HTML}{59315F}
\definecolor{SussexDarkMagenta}{HTML}{AC145A}
\definecolor{SussexVividRed}{HTML}{A6192E}
\definecolor{SussexVibrantGreen}{HTML}{007A33}
\definecolor{SussexMidBlue}{HTML}{006BA6}

\hypersetup{colorlinks=true,
citecolor=SussexDeepAquamarine,
linkcolor=SussexMidBlue,
urlcolor=SussexBurntOrange}

\newcommand{\etal}{\emph{et al.}}

\renewcommand{\d}{\mathrm{d}}
\newcommand{\e}[1]{\mathrm{e}^{{#1}}}
\newcommand{\im}{\mathrm{i}}
\newcommand{\Mp}{M_{\mathrm{P}}}
\newcommand{\vect}[1]{\bm{\mathrm{{#1}}}}

\newcommand{\dimlessP}{\mathcal{P}}
\newcommand{\fNL}{f_{\mathrm{NL}}}

\newcommand{\bigBox}{L}

\newcommand{\smallBox}{\ell}

\newcommand{\kevent}{k_{\text{event}}}
\newcommand{\tevent}{t_{\text{event}}}

\newcommand{\ttransition}{t_{\text{tr}}}
\newcommand{\etatransition}{\eta_{\text{tr}}}
\newcommand{\ktransition}{k_{\text{tr}}}

\newcommand{\tinit}{t_{\circ}}
\newcommand{\tstar}{t^\ast}
\newcommand{\tend}{t_{\text{end}}}
\newcommand{\qend}{q_{\text{end}}}

\newcommand{\qmin}{q_{\text{min}}}
\newcommand{\qmax}{q_{\text{max}}}

\newcommand{\Nlookback}{N_{\text{lookback}}}

\newcommand{\smoothScale}[2]{\mathcolor{SussexBurntOrange}{\bm{[}}{{#2}}\mathcolor{SussexBurntOrange}{\bm{]}_{{#1}}}}
\newcommand{\smoothBig}[1]{\smoothScale{\bigBox}{{#1}}}
\newcommand{\smoothSmall}[1]{\smoothScale{\smallBox}{{#1}}}

\newcommand{\massmatrix}{\mathcal{M}}
\newcommand{\kpeak}{k_\text{peak}}

\newcommand{\Operator}[1]{\mathcal{{#1}}}
\newcommand{\OpO}{\Operator{O}}

\hypersetup{pdftitle={Decoupling of large-scale, adiabatic inflationary perturbations
from enhanced small-scale modes at one-loop}}

\title{Decoupling of large-scale, adiabatic inflationary perturbations
from enhanced small-scale modes at one-loop}

\author[1,2]{Laura Iacconi\,$^{\text{\orcidlink{0000-0002-1152-3056}}}$,}
\author[1]{David Mulryne,}
\author[3]{David Seery\,$^{\text{\orcidlink{0000-0003-3421-6080}}}$}

\affiliation[1]{Astronomy Unit, Queen Mary University of London, \\
Mile End Road, London, E1 4NS, UK}
\affiliation[2]{Institute of Cosmology and Gravitation, University of Portsmouth, \\
Burnaby Road, Portsmouth, PO1 3FX, UK}
\affiliation[3]{Astronomy Centre, University of Sussex, \\
Falmer, Brighton, BN1 9QH, UK}

\emailAdd{l.iacconi@qmul.ac.uk}
\emailAdd{d.mulryne@qmul.ac.uk}
\emailAdd{D.Seery@sussex.ac.uk}

\abstract{
We reconsider back-reaction
from large amplitude, short-scale perturbations
onto a long wavelength adiabatic mode.
In a loop expansion of the long-mode power spectrum,
this back-reaction appears first at 1-loop. 
Due to the separation between the long and short scales,
the separate universe method
provides a simple and efficient framework for
this computation. 
In this paper,
building on our earlier work,
we employ a $\delta N$ formula
for the long mode, which captures the effect of short scales. 
We show that back-reaction at 1-loop is due to either
(i) non-linearity of the $\delta N$ formula, 
or (ii) 1-loop corrections to the initial conditions.
We argue that contributions of type (ii) cannot themselves
be described within the separate universe framework, but
their properties can be constrained using soft theorems
and a ``multi-point propagator'' expansion.
When applied to a band of enhanced short-scale perturbations
that crossed the horizon during inflation,
our result shows that the loop correction decouples
from their detailed properties.
Furthermore, the back-reaction we obtain is scale-invariant.
Its magnitude is model-dependent, but is
degenerate with effects from
modes that were still sub-horizon at the end of inflation.
In this scenario
(but not necessarily in all scenarios),
we conclude that the effect is not observable.
}

\begin{document}
\maketitle
\newpage

\section{Introduction}
\label{sec: introduction}

Recently there has been much
interest in evaluation of
loop corrections to the correlation functions
of the inflationary
curvature perturbation $\zeta$.
These corrections are interesting
because (among other effects),
at each scale,
they
measure back-reaction
due to structure on much shorter
scales~\cite{Riotto:2023hoz,Iacconi:2023ggt}.
At small wavenumbers,
the statistics of the curvature perturbation
are tightly constrained
by CMB and galaxy-clustering observations,
and are consistent
with tree-level predictions from single-field, slow-roll inflation.
To maintain this success, the backreaction onto
these wavenumbers should not disrupt their statistical properties.
This may be a nontrivial
constraint.
In scenarios where the amplitude of short-scale
fluctuations is enhanced
(for example, to produce a population of early compact objects such
as primordial black holes),
there is a reasonable concern that
significant back-reaction may occur.

Estimates of the
back-reaction effect have been obtained at 1-loop level.
However, its amplitude continues to be debated in the literature~\cite{Cheng:2021lif, Inomata:2022yte, Kristiano:2022maq, Riotto:2023hoz, Choudhury:2023vuj, Kristiano:2023scm, Riotto:2023gpm, Motohashi:2023syh, Firouzjahi:2023ahg, Franciolini:2023agm, Tasinato:2023ukp, Cheng:2023ikq, Fumagalli:2023zzl, Tada:2023rgp, Maity:2023qzw, Firouzjahi:2023bkt, Davies:2023hhn, Iacconi:2023ggt, Inomata:2024lud,Firouzjahi:2024psd, Braglia:2024zsl, Kawaguchi:2024lsw, Ballesteros:2024zdp, Kristiano:2024vst, Kawaguchi:2024rsv, Fumagalli:2024jzz, Ruiz:2024weh, Firouzjahi:2024sce, Frolovsky:2025qre, Inomata:2025bqw, Fang:2025vhi, Firouzjahi:2025gja, Firouzjahi:2025ihn,  Inomata:2025pqa}.
Kristiano \& Yokoyama~\cite{Kristiano:2022maq} studied
the case of a transient ultra-slow-roll (USR)
phase~\cite{Kinney:2005vj, Dimopoulos:2017ged, Pattison:2018bct}
in a single-field model.
This is
sometimes regarded as a possible scenario
leading to primordial black hole (PBH) production.
They estimated that an enhancement of small-scale
power sufficient to produce
interesting abundances of PBHs could induce
a 1-loop power spectrum comparable to the tree-level one,
which we would usually interpret to mean that backreaction
effects were not under adequate control.
The amplitude of their effect was proportional to the peak
amplitude of the short-scale power spectrum.

Kristiano \& Yokoyama worked in a simplified
model designed for analytic treatment.
Their work was followed by many authors,
who attempted to obtain refined estimates
by relaxing assumptions made in Ref.~\cite{Kristiano:2022maq},
or by accounting
for extra physical effects.
Like many subsequent authors,
they
made use of the ``in--in'' formalism
of nonequilibrium quantum field theory,
and in particular its diagrammatic expansion
into Green's functions.
This is powerful, but complicated.
Some---but, as we shall see,
usually not all---of the back-reaction
we wish to capture takes place when quantum
effects are not significant.
This suggests that it may be
possible to simplify the calculation by
replacing the Green's function
approach with a classical procedure.

In  an earlier paper, Ref.~\cite{Iacconi:2023ggt},
we suggested that
it should be possible to use the separate universe framework
to build a classical back-reaction model.%
    \footnote{See also  Ref.~\cite{Firouzjahi:2023ahg}
    for an earlier application of the $\delta N$ framework,
    although the details differ from those of Ref.~\cite{Iacconi:2023ggt}.}
In particular, we studied the back-reaction onto
long-wavelength modes with
wavenumber $p$ from ultra-slow-roll scales, 
near wavenumber $q$.
In common with many similar
processes characterized by decoupling and separation of scales,
there is an EFT-like description
expressed in terms of effects entering at different powers of $p/q$.
In Ref.~\cite{Iacconi:2023ggt}
we evaluated
the different
1-loop contributions due to non-linear super-horizon evolution,
finding one which was volume-suppressed (i.e., scaling as $(p/q)^3$),
and two which were unsuppressed.
The unsuppressed effects are supported by long--short mode
couplings~\cite{Iacconi:2023ggt}.

In this paper we return to
the topic of back-reaction, building on the results of Ref.~\cite{Iacconi:2023ggt}.
Our aim is now to include effects from a broad band of enhanced fluctuations.
For realistic models this is usually represented by a localized peak in
$\dimlessP_{\zeta,\text{tree}}$,
including modes rising to and falling away from the main peak.
Further, we account for 1-loop contributions
in the initial conditions of the separate universe computation,
which were not included in Ref.~\cite{Iacconi:2023ggt}. 
After dropping volume-suppressed contributions at 1-loop,
we investigate the mathematical structure of
the remaining effects.
The result yields a compact and illuminating final expression.
We use this to assess under what conditions a broad peak can lead to
observable effects on larger scales,
and whether the importance of these effects
depends on the peak amplitude, as concluded in Ref.~\cite{Kristiano:2022maq}.

Our analysis depends on two fundamental assumptions. 
These are that:
(i) the field configuration described by the long mode is adiabatic, and
(ii) there is sufficient separation of scales between the
long mode and the short scales whose effect we want to capture.
Assumption (ii) is needed to allow the application of soft theorems
to determine
the response of short-scale correlation functions
to the long-wavelength mode.
Apart from the assumption
that these soft theorems are themselves valid,
no further assumptions
are made, e.g. regarding the nature of the transient non-attractor phase,
or the type and duration of the transitions into and out of this phase.
To make contact with the literature, we sometimes give explicit expressions
for the case of a transient USR era.
However, our discussion is not limited to this scenario.
Our analysis is applicable to any single-field model which realizes assumptions (i) and (ii).
Finally, although we focus on the case of single-field inflation,
we briefly discuss the applicability of our results to more general
scenarios in {\S}\ref{sec: discussion}.

\subsection{Roadmap}
In {\S}\ref{sec: properties of long and short modes} we briefly summarize
the properties of the long mode
for which we wish to estimate the change due to back-reaction,
and of the short-scale modes that induce the effect.
We spell out the two fundamental assumptions of adiabaticity
of the long mode,
and separation of scales.
These provide the foundation for the remainder of our analysis. 

In {\S}\ref{sec: separate universe} we review the separate universe framework.
We comment on its application to models containing a sudden deviation
from slow-roll dynamics in {\S}\ref{sec: separate universe and sudden transitions}. 
In~\S\ref{sec:separate-universe-correlations} we review the
$\delta N$ formula including back-reaction,
and explain how it can be used to compute the
long-mode power spectrum at 1-loop~\cite{Iacconi:2023ggt}.
In~\S\ref{sec: does separate universe compare with in-in}
we argue that the loops computed by applying separate universe
capture the same effects as ``in--in'' loop integrals
(at least for the specific computation of interest here),
and discuss the correct choice of the $\delta N$ initialization time.
In {\S}\ref{sec: the 1-loop contributions} we collect all 1-loop contributions,
and distinguish between those where the loop is due to
(i) non-linearity of the $\delta N$
formula,
or (ii) initial conditions at 1-loop.

We present our main computations in
{\S}\ref{sec: 1-loop from non-linear superhorizon evolution is a boundary term}
and {\S}\ref{sec: 1-loop from quantum initial conditions is a boundary term}. 
For both cases
(i) (in {\S}\ref{sec: 1-loop from non-linear superhorizon evolution is a boundary term})
and (ii) (in {\S}\ref{sec: 1-loop from quantum initial conditions is a boundary term})
we show that the non-volume-suppressed contributions can be written
in a compact form, where the integrand is the total derivative of a 
function whose explicit form we provide.
We conclude in {\S}\ref{sec: discussion} by discussing
the implications of our results. 
In Appendix~\ref{app: separate universe computation of tree level power spectrum}
and~\ref{app: squeezed bispectrum at t_k} we provide additional materials. 

\subsection{Conventions and definitions}
We work in natural units where $c = \hbar = 1$.
The reduced Planck mass is defined by $\Mp = (8\pi G)^{-1/2}$,
where $G$ is Newton's gravitational constant.

\para{Phase space}
For a generic inflation model with $M$ fields,
we collect the fields and their momenta
into a $2M$-dimensional phase-space vector 
\begin{equation}
    \label{eq:phase space coordinates}
    X^I(\vect{x}, \, t) \equiv
    \Big(
        \phi^1(\vect{x}, t),
        \ldots,
        \phi^M(\vect{x}, t),
        {\pi^1}(\vect{x}, t),
        \ldots,
        {\pi^M}(\vect{x}, t)
    \Big) ,
\end{equation}
where $\pi\equiv \phi'$, and a prime $^{\prime}$
indicates a derivative with respect to the e-fold number $N$,
defined by $\d N \equiv H \d t$. 
Latin indices $I$, $J$, \ldots run over $1$, \ldots, $2M$
and label phase-space coordinates.
Each phase-space coordinate can be decomposed into a background component
and a linear perturbation, viz.,
$X^I(\vect{x}, t) = \bar{X}^I(t) + \delta X^I(\vect{x}, t)$.

\para{Fourier transforms}
Our Fourier transform convention is
\begin{equation*}
    f(\vect{k}) = \int \d^3 x \; f(\vect{x})
        \e{-\im \vect{k}\cdot\vect{x}}
    \quad
    \text{and}
    \quad
    f(\vect{x}) = \int \frac{\d^3 k}{(2\pi)^3} \; f(\vect{k})
        \e{\im \vect{k}\cdot\vect{x}} ,
\end{equation*}
where $f(\vect{x})$ and $f(\vect{k})$ represent an arbitrary
function and its Fourier transform, respectively.
We sometimes use the shorter notation
$f_{\vect{k}} = f(\vect{k})$,
or $[ \OpO ]_{\vect{k}}$
for the Fourier mode of a composite quantity $\OpO$.

\para{Wavevectors}
We use $\vect{p}$
to label the comoving wavevector of a long mode,
and $\vect{q}$ to label the wavevectors of enhanced short modes.
The back-reaction effect we aim to compute is a measure of the aggregate
effect $\vect{q}$-modes have on the longer $\vect{p}$-mode.
If there is a band of enhanced
short-scale modes, rather than a single well-defined mode,
we use $\kpeak$ to label the most strongly enhanced mode.

In single field models with a non-attractor phase, we use
$t_s$ and $t_e$ to label the start and end times of the
non-attractor behaviour.
The corresponding scales crossing the horizon at these times
are $k_s$ and $k_e$,
where
(as usual) the horizon-crossing time $t_k$ for a scale $k$ is defined by the
condition $k = (a H)_{t = t_k}$.

\para{Correlation functions}
We define the
spectrum $P_\zeta$ and bispectrum $B_\zeta$
for the curvature perturbation $\zeta$
in terms of the equal-time 2- and 3-point correlation functions,
\begin{subequations}
\begin{align}
	\label{eq: def 2pt zeta}
	\langle
        \zeta_{\vect{k}_1}(t)
        \zeta_{\vect{k}_2}(t)
    \rangle
    & \equiv
    (2\pi)^3 \delta(\vect{k}_1 + \vect{k}_2)
    P_\zeta(k_1; t) \;, \\
	\label{eq: def 3pt zeta}
	 \langle
        \zeta_{\vect{k}_1}(t)
        \zeta_{\vect{k}_2}(t)
        \zeta_{\vect{k}_3}(t)
    \rangle
    & \equiv
    (2\pi)^3 \delta(\vect{k}_1 + \vect{k}_2 + \vect{k}_3)
    B_\zeta(k_1, \, k_2,\, k_3; t) .
\end{align}
\end{subequations}
The spectrum and bispectrum depend only on the magnitudes
of their wavevector arguments, and not their orientation,
as a consequence of statistical
translation invariance and isotropy.

We also define a dimensionless power spectrum
and a reduced (dimensionless) bispectrum,
\begin{subequations}
\begin{align}
	\label{eq: dimless Pz def}
	\dimlessP_\zeta(k; t)
    & \equiv
    \frac{k^3}{2\pi^2} P_\zeta(k; t) , \\
	\label{eq: dimless Bz def}
    \fNL(k_1,k_2,k_3;\, t)
    & \equiv
    \frac{5}{6}
    \frac{B_\zeta(k_1, k_2, k_3; t)}
    {
        P_\zeta(k_1; t) P_\zeta(k_2; t)
        + P_\zeta(k_1; t) P_\zeta(k_3; t)
        + P_\zeta(k_2; t) P_\zeta(k_3; t)
    } . 
\end{align} 
\end{subequations}
Finally,
we introduce
equal-time 2- and 3-point correlation
functions in phase space 
\begin{subequations}
\begin{align}
	\label{eq: def 2pt phase space}
	\langle
        \delta X^I_{\vect{k}_1}(t)
        \delta X^J_{\vect{k}_2}(t)
    \rangle
    & \equiv
    (2\pi)^3 \delta(\vect{k}_1 + \vect{k}_2)
    P^{IJ}(k_1; t)  \;, \\
	\label{eq: def 3pt phase space}
	 \langle
        \delta X^I_{\vect{k}_1}(t)
        \delta X^J_{\vect{k}_2}(t)
        \delta X^K_{\vect{k}_3}(t)
    \rangle
    & \equiv
    (2\pi)^3 \delta(\vect{k}_1 + \vect{k}_2 + \vect{k}_3)
    \tensor{\alpha}{^I^J^K}(k_1, \, k_2,\, k_3; t) . 
\end{align}
\end{subequations}

\section{Properties of long and short modes}
\label{sec: properties of long and short modes}

Our objective is to investigate back-reaction
onto a large-scale mode $\vect{p}$,
due to small-scale modes
with enhanced amplitude
and
typical
wavenumbers of order $q$.
We usually have in mind a scenario
where $\vect{p}$ contributes to
the CMB anisotropy or the statistics of galaxy clustering,
and therefore is
constrained by observation.
However, our results
are more general
and require only a
separation of scales, so that $p \ll q$.
We write the horizon exit time for $\vect{p}$
as $t_p$,
and the horizon exit time for a typical
enhanced mode
as $t_q$.
Eventually we are interested in the scenario
where a broad band of modes near $q$
are enhanced,
in which case $t_q$ may not be well defined.
However, here and below,
we use this notation where it simplifies the presentation,
but our analysis does not depend on it.

We focus on single-field models.
In such scenarios, enhanced small-scale fluctuations can be
produced when slow-roll evolution is temporarily interrupted
by an ultra-slow-roll phase%
\footnote{Notice that, despite the name,
    ultra-slow-roll is \emph{not} a sub-case of slow-roll.}
(where $\epsilon_1 \ll 1$ and
$\epsilon_2 \sim -6$), or by other non-attractor dynamics.
Here, $\epsilon_1 \equiv -H'/H$,
$\epsilon_2 \equiv \epsilon_1'/\epsilon_1$,
and a prime $^{\prime}$ denotes a derivative with respect to
$\d N = H \, \d t$.
The non-attractor phase is itself assumed
to be followed by
another era, which may be a second slow-roll period or something else.
The mode $\vect{p}$ crosses the horizon during the
first slow-roll period.
In~\S\ref{sec: discussion} we comment briefly on the applicability
of our results to more general scenarios, including multiple-field
models.

In this section we review the physical properties of
the small- and large-scale modes.
We assume these are coupled by long--short correlations
that control how small scales respond to changes in
the long-wavelength fields.
As explained in~\S\ref{sec: introduction},
we have two main requirements:
(i) the long-wavelength configuration
is \emph{adiabatic};
and
(ii) there is an appreciable
separation of scales, so that
the short-scale fields experience the long-wavelength
disturbance as a nearly-constant shift.
We now consider each of these conditions in more detail.

\para{Long-wavelength disturbance is adiabatic}
The long-wavelength mode $\vect{p}$
will typically
pass outside the horizon some time before
any enhanced short-scale perturbations.
In certain models the
corresponding
field configuration
may be adiabatic,
in the sense
that the perturbation in each relevant field
(or their momenta)
at wavenumber $\vect{p}$
is derived from
a shift along the background trajectory.
Here, ``relevant'' means that the field carries
a non-negligible fraction of the cosmological energy budget.

In a single-field
(or effectively single-field)
model the condition of adiabaticity
entails further simplifications.
The relevant phase space~\eqref{eq:phase space coordinates}
will be spanned by some field-space direction $\phi$
and its momentum $\pi$.
For the models we will consider,
the adiabatic condition relates $\delta \pi$ to $\delta \phi$
via
\begin{equation}
    \label{eq: delta pi and delta phi long mode}
    \delta \pi_{\vect{p}}(t)
    =
    \frac{\epsilon_2}{2} \delta \phi_{\vect{p}}(t)
    +
    \text{decaying}, 
    \quad
    (t > t_p)
\end{equation}
where the perturbations $\delta \phi$, $\delta \pi$
are defined on
spatially flat hypersurfaces.
Eq.~\eqref{eq: delta pi and delta phi long mode} is valid
after horizon crossing, once
decaying contributions can be neglected.
Note that this includes
smooth transitions between the different dynamical phases.

The condition~\eqref{eq: delta pi and delta phi long mode}
guarantees that $\delta\phi$ and $\delta\pi$
combine to displace the field configuration
along the \emph{original} phase space trajectory.
In any region where the evolution of the field value
$\phi$ is monotonic,
we can
choose to measure position along this trajectory
using $\phi$.
The rate of change of any phase-space function
$F(\phi, \pi)$
on the trajectory can then be obtained using the usual
advective derivative,
\begin{equation}
\label{eq: total derivative along unpert trajectory}
    \frac{\d}{\d\phi} \equiv \frac{\partial}{\partial \phi} + \frac{\epsilon_2}{2} \frac{\partial}{\partial \pi}  \;.
\end{equation}
Eq.~\eqref{eq: total derivative along unpert trajectory}
will play an important role in the emergence
of a total derivative in the 1-loop computation.
(See {\S}\ref{sec: delta N 12 and 13 loops}.)

\para{Reaction of short-scale perturbations}
Some time after horizon exit of $\vect{p}$
(perhaps a very long time afterwards),
the enhanced short-scale perturbations exit the horizon.
We take this to happen at time $t \sim t_q$.
The short modes exit into a background that is
locally disturbed by the long-wavelength $\vect{p}$-mode.
Because of the hierarchy of scales,
modes
with wavenumbers in the vicinity of $q$
experience this disturbance as a nearly-constant
shift in the background fields and their momenta.

We now
collect $\phi$ and $\pi$ into a single vector $X^L$.
For a pure shift $\delta X^L$
in the background
fields
we could evaluate the response of the 2-point function
via a Taylor
expansion~\cite{Maldacena:2002vr, Kenton:2015lxa, Kenton:2016abp}
\begin{equation}
    \label{eq: short scale P with reaction}
	\langle
        \delta X^I_{\vect{q}}
        \delta X^J_{-\vect{q}}
    \rangle^{\prime}_{t}
    = 
    \langle
        \delta X^I_{\vect{q}}
        \delta X^J_{-\vect{q}}
    \rangle^{\prime}_{t} \Big|_{0}
    +
    \left.
    \frac{\partial
        \langle
            \delta X^I_{\vect{q}}
            \delta X^J_{-\vect{q}}
        \rangle^{\prime}_{t}}
        {\partial X^L(t)}
    \right|_{0} \delta X^L(t)
    + \cdots ,
\end{equation}
where ``$\cdots$'' denotes corrections of higher orders in
$\delta X^L$.
The notation $|_0$ indicates that the attached quantity
is to be
evaluated in the absence of the background disturbance,
and $\langle \cdots \rangle_t$ denotes an equal-time correlation function
evaluated at time $t$.
Finally, a prime $'$ attached to a correlation function indicates
that factors of $(2\pi)^3$
and the momentum conservation $\delta$-function
should be removed.
Similar expressions can be written for
each higher correlation function.

\para{Long-short mode coupling}
We can regard Eq.~\eqref{eq: short scale P with reaction}
as a simple example of an operator product expansion.
This determines the behaviour of the
composite operator $\delta X^I \delta X^J$
when correlated with
perturbations
at wavenumbers much smaller than $q$.
Promoting~\eqref{eq: short scale P with reaction} in this way,
we obtain the operator
relation~\cite{Kenton:2015lxa,Kenton:2016abp,Byrnes:2015dub}
\begin{equation}
    \label{eq:delta-X-OPE}
    \delta X^I_{\vect{q}}
    \delta X^J_{\vect{p}-\vect{q}}
    \Big|_{t}
    \approx
    (2\pi)^3 \delta(\vect{p})
    \langle
        \delta X^I_{\vect{q}}
        \delta X^J_{-\vect{q}}
    \rangle^{\prime}_{t} \Big|_{0}
    +
    \left.
    \frac{\partial
        \langle
            \delta X^I_{\vect{q}}
            \delta X^J_{-\vect{q}}
        \rangle^{\prime}_{t}}
        {\partial X^L(t)}
    \right|_{0} \delta X^L_{\vect{p}}(t)
    +
    \cdots .
\end{equation}
Here, ``$\cdots$'' takes the place of the higher-order
corrections in~\eqref{eq: short scale P with reaction},
and now indicates all omitted
higher-dimensional operators.
We assume these to be defined so that
they have zero expectation value.
In addition to higher powers of the
perturbations $\delta X^L$,
which would
already appear in the Taylor
expansion~\eqref{eq: short scale P with reaction},
the omitted operators will include gradients such
as $\partial^2 \delta X^L_{\vect{p}}$
that introduce corrections of order $p^2$.
We will shortly consider the significance of these.
The symbol $\approx$
denotes a weak equality,
in the sense that Eq.~\eqref{eq:delta-X-OPE}
should be regarded
as valid when used inside a correlation function.
However, Eq.~\eqref{eq:delta-X-OPE}
is expected to hold as an \emph{operator} statement.
Therefore, the same expansion
can be used in any
correlation function.

The operator product expansion controls
all long--short correlations~\cite{Kehagias:2012pd,Assassi:2012et}.
One important and familiar
example controls the ``squeezed''
limit of the three-point function
$\langle \delta X_{\vect{p}} \delta X_{\vect{q}} \delta X_{-\vect{p}-\vect{q}} \rangle$.
Applying~\eqref{eq:delta-X-OPE}
to the bilinear
$\delta X_{\vect{q}} \delta X_{-\vect{p}-\vect{q}}$
yields
\begin{equation}
\label{eq: 3-point function squeezed}
    \langle
        \delta X^M_{\vect{p}}
        (\delta X^I_{\vect{q}}
        \delta X^J_{-\vect{p}-\vect{q}})
    \rangle_{t}
    =
    \langle
        \delta X^M_{\vect{p}}
        \delta X^L_{-\vect{p}}
    \rangle_{t}
    \frac{\partial
        \langle
            \delta X^I_{\vect{q}}
            \delta X^J_{-\vect{q}}
        \rangle^{\prime}_{t}}
        {\partial X^L(t)}
    \Big|_{0}
    +
    \cdots .
\end{equation}
Eq.~\eqref{eq: 3-point function squeezed}
was given by Kenton \& Mulryne~\cite{Kenton:2015lxa}.
The corrections indicated by ``$\cdots$''
include contributions with higher gradients $\sim p^2$
\emph{and} higher powers of the fluctuation amplitude.
Which of these is the leading correction
may be model-dependent. In this paper,
we do not retain either.

Corrections to
Eq.~\eqref{eq: 3-point function squeezed}
from operators containing gradients
must involve
dimensionless ratios of $p$
with some comparison scale.
This ratio determines their relevance.
If the background is smooth, the only
available scale is the Hubble scale $H$.
The gradient corrections will therefore scale as
$p/aH$.
This is typically assumed in the simplest presentations
of the ``separate universe framework''
(to be discussed in more detail in~\S\ref{sec: loops in the separate universe framework}),
which can also be regarded as an example of
an operator product expansion.

Alternatively, if the background is not smooth but contains
transitions or other events associated with
a particular time $\tevent$,
and hence a scale $\kevent = (aH)_{t = \tevent}$,
then
these scales are also available to form dimensionless ratios.
In this case, there may be corrections that scale as
powers of
$p/\kevent$,
if the initial time is chosen too early~\cite{Jackson:2023obv}.
These effects are described in~\S\ref{sec: separate universe and sudden transitions}.
When interpreted within the separate universe framework,
demanding that these gradient-dependent
terms do not spoil
the background evolution
is equivalent to the well-known
requirement that we should wait for all
relevant scales to pass outside the horizon.%
    \footnote{For example, this condition was already
    recognized by Lyth
    in the early literature on loop corrections;
    see, Ref.~\cite{Lyth:2006gd},
    especially the discussion in~{\S}III.E.}
We will see further examples of this requirement
in~\S\ref{sec: loops in the separate universe framework}.
In some cases,
the discussion in~\S\ref{sec: loops in the separate universe framework}
will show that this constitutes the primary limitation
on the utility of separate universe methods
to evaluate loop corrections.

\section{Loops in the separate universe framework}
\label{sec: loops in the separate universe framework}

In addition to the key properties
described in {\S}\ref{sec: properties of long and short modes},
we must supply
a means
to model
the time evolution
of the perturbations and their correlations.
By default, if we wish to accurately model
all effects, we should apply a method
from non-equilibrium field theory,
such as the diagrammatic expansion
of the in--in path integral into Green's
functions.
Such methods are powerful but technically
complex, and it is not always easy to
interpret the expressions they produce.
As we now explain, under some
circumstances the separate universe framework
provides an alternative approximate
approach.
As with any approximation scheme,
it has advantages and disadvantages.

\subsection{The separate universe framework}
\label{sec: separate universe}
The separate universe
framework~\cite{Starobinsky:1985ibc,Sasaki:1995aw,
Wands:2000dp,Lyth:2003im,Lyth:2005fi}
is a tool to describe the non-linear
evolution of large-scale
cosmological perturbations.
To use it, we
write the number of e-folds accumulated
between an initial time $t_i$ and
some later time $t$
as $N^{(t_i,t)} \equiv \int_{t_i}^{t} H(t') \, \d t'$.
Now let $\zeta(t)$ denote the curvature perturbation at time
$t$, evaluated in some smoothed patch.
In the separate universe approach,
the subsequent evolution of this patch
is assumed to
follow
one of the trajectories of the background phase
space.
In the simplest scenarios,
this trajectory is
fixed uniquely
by initial conditions set at
$t_i$.
The evolution of any perturbed quantity may
then be
determined by
studying how one such patch
evolves relative to another.
This entails
solving its background equations of motion with
displaced initial conditions.
The leading corrections to this procedure
will involve spatial gradients,
which (as discussed
in~\S\ref{sec: separate universe and sudden transitions}
below)
are sometimes, but not always, suppressed on
superhorizon scales.

We identify $\zeta(t)$
with the variation $\delta N$
in the duration of inflation in each patch
relative to the mean in a larger volume,
now taking $t_i$ to label a spatially-flat initial hypersurface
and $t$ to label a uniform-density final
hypersurface~\cite{Starobinsky:1985ibc, Sasaki:1995aw, Sasaki:1998ug, Lyth:2004gb}.
We will often regard $t$ as the end of inflation, although
this is not required; in principle,
$t$ could be
any time up to the present,
provided it is later than $t_i$
and gradients can be neglected.
The initial conditions at $t_i$ are the values of the
fields and their momenta.

In principle,
this procedure
is nonperturbative in the amplitude of each perturbation.
In practice, as a matter of convenience,
we often invoke an expansion in powers of the initial
fluctuation.
This yields the well-established
``$\delta N$'' Taylor
expansion~\cite{Starobinsky:1985ibc,Sasaki:1995aw,Lyth:2005fi}.
It provides a
description
equivalent to cosmological perturbation theory,
provided that the
time dependence of perturbations with nonzero wavevector
$\vect{k}$ can be described by the growing and decaying solutions
of the homogeneous background.
Notice that
this does not require the perturbations to
be
$\vect{k}$-independent, but
rather that
\emph{any $\vect{k}$-dependence
must appear only in the initial amplitude
of the growing and decaying modes}.

It does not always happen that each patch follows
a single background trajectory.
Under certain circumstances,
patches may jump stochastically between different 
available trajectories
due to emerging short-scale
perturbations.
Vennin \& Starobinsky
described this as the ``stochastic $\delta N$''
formalism~\cite{Fujita:2013cna,Fujita:2014tja,Vennin:2015hra,Ezquiaga:2019ftu}.
In this picture,
the back-reaction effect we are considering
could be interpreted roughly as a
single stochastic ``kick''
near time $t = t_q$,
which displaces each patch to a (possibly)
different
trajectory.
However, away from $t = t_q$, the evolution
is deterministic.

When it can be
applied,
the separate universe framework
is very appealing. It provides a clear
and simple description
that accounts for time dependence
and the influence of isocurvature modes.
The key criterion
for applicability
is that all $\vect{k}$-dependence is captured in
the initial data.
There is no guarantee that this is always possible.
At a minimum, it requires that the initial time $t_i$
is chosen to be sufficiently late.
As we explain below,
when
back-reaction is neglected,
$t_i$ can usually be chosen a little after horizon
exit for the $\vect{k}$ mode being studied,
or after the last transition event of the type to be discussed
in~{\S}\ref{sec: separate universe and sudden transitions}.
When back-reaction is included
$t_i$ must be chosen later,
at least
after all modes that back-react have left
the horizon.
However, we will see that it cannot be chosen too late,
otherwise we risk large corrections
to the correlation functions at $t_i$,
which
would have
to be computed using some other method.
Although this is possible
in principle,
much of the simplicity of the separate universe
approach would be lost.
To avoid this it is necessary to balance
a number of effects, and it may
not be possible to find an initial hypersurface that satisfies them all;
see~\S\ref{sec:initial-time}.
As we discuss below,
in this situation
it might be necessary to abandon the separate universe
framework in favour of an alternative.

\subsection{The separate universe framework for models with a non-attractor phase}
\label{sec: separate universe and sudden transitions}
In~\S\ref{sec:separate-universe-correlations}
we use the
separate universe method to write down
explicit formulae for correlation functions.
However, before doing so,
we illustrate the necessity of choosing
$t_i$ sufficiently late
by studying models featuring a transition.
Models in which slow-roll evolution is interrupted by an
ultra-slow-roll phase (or other non-attractor epoch)
belong to this category.
A related, complementary discussion has recently been
given by Briaud~{\etal}~\cite{Briaud:2025ayt}.

It was explained above
that to apply the separate universe framework
we must wait until all relevant scales have left the
horizon.
There is always a \emph{minimum} requirement
for all wavenumbers to be in the superhorizon regime,
so that gradient corrections scaling as $p/aH$ are small.
However, if there are other
distinguished events associated with a scale $\kevent$, then
gradient corrections scaling as $p/\kevent$ may also enter
if $t_i$ is chosen too early.%
    \footnote{In presentations of the separate universe
    idea, it is sometimes stated or assumed
    that gradient corrections always occur in the combination
    $p/aH$.
    As explained here,
    failure of this property does not itself invalidate the
    separate universe
    principle,
    although it may restrict how
    it can be applied.}
This
issue has recently been emphasized by Jackson {\etal}~\cite{Jackson:2023obv}.
In this section we
briefly
review their analysis.
Our main aim is to highlight how gradient data can
pass through the calculation
to produce $\Or(p/\kevent)$ contributions,
even on superhorizon scales,
when the background evolution is not smooth.

\para{Superhorizon evolution in slow-roll}
The discussion of~\S\ref{sec: separate universe}
shows that the (deterministic) separate universe framework
requires the future evolution of a
smoothed, superhorizon-scale spacetime
volume
to be uniquely predictable from its initial data.

In this framework we would normally
determine perturbations in a field $\phi$
by solving for the non-perturbative
evolution of $\phi$ with displaced boundary
conditions.
However, one can also study perturbation
equations directly.
Here, we follow the second approach.
The appropriate equation of motion
for a ``homogeneous'' perturbation
(that is, covering an entire smoothed patch) is
\begin{equation}
\label{eq: delta phi eom conformal time}
    \frac{\d^2\delta {\phi}_k(\eta) }{\d\eta^2}+ 2 a H \, \frac{\d\delta {\phi}_k(\eta) }{\d\eta} + \left({k^2} + a^2 \massmatrix\right)\delta \phi_k(\eta)=0 
\end{equation}
in the limit $k\to 0$,
where $\eta$ is the conformal time, satisfying
$\d\eta \equiv \d t/a(t)$.
The mass $\massmatrix$
satisfies
\begin{equation}
\label{eq: mass of delta phi def}
    \massmatrix
    =
    V_{\phi \phi}
    - \frac{1}{a^3}\frac{\d}{\d t}
    \bigg(
        a^3 \frac{\dot{\phi}^2}{H}
    \bigg)
    = H^2 \times \Or(\epsilon) \;,
\end{equation}
where $\Or(\epsilon)$ stands for the magnitude
of a generic slow-roll parameter.
(For definitions,
see the beginning of {\S}\ref{sec: properties of long and short modes}.)
We caution that the slow-roll parameters cannot be assumed
to be always small;
e.g., $\epsilon_2\sim-6$ in ultra-slow-roll inflation.

During a slow-roll epoch,
all slow-roll parameters \emph{are} suppressed
and the mass $\massmatrix$ can be neglected.
The equation of motion is then
\begin{equation}
    \frac{\d^2\delta {\phi}_{k\to0}(\eta)}{\d\eta^2}
    - \frac{2}{\eta} \frac{\d\delta {\phi}_{k\to0}(\eta)}{\d\eta}
    \approx 0
    ,
\end{equation}
whose general solution can be written
\begin{equation}
\label{eq: delta phi background}
    \delta {\phi}_{k\to0}(\eta) = A + B (-\eta)^3 . 
\end{equation}
The $A$-mode is the ``growing''
mode (here actually a constant), which scales like $(-\eta)^0$.
The $B$-mode is the ``decaying'' mode,
which scales like $(-\eta)^3$.
For the separate universe principle to be valid,
the time dependence of a perturbation at finite
wavevector $\vect{k}$
must be a linear combination of these
two modes, not necessarily just the ``growing'' mode.
Further, the amplitudes
$A$ and $B$ must be predictable from
data
at the initialization time.
Any dependence on $\vect{k}$
can be inherited only
from the assigned values of $A$ and $B$.

Let us see how these properties
are usually satisfied outside the horizon.
Still assuming slow-roll,
the $\vect{k}$-dependent solution to Eq.~\eqref{eq: delta phi eom conformal time} is 
\begin{equation}
    \label{eq: delta phi k dependent}
    \delta \phi_k(\eta)
    =
    \frac{1}{a(\eta)}
    \left[
        \alpha
        \left(
            1-\frac{\im}{k\eta}
        \right)
        \e{-ik\eta}
        +
        \beta
        \left(
            1+\frac{\im}{k\eta}
        \right)
        \e{+ik\eta}  \right] ,
\end{equation}
where $\alpha$ and $\beta$ are constants encoding the initial data.
We typically match to the Bunch--Davies vacuum at past infinity,
where $k\eta \rightarrow -\infty$.
This gives
$\alpha = 1/\sqrt{2k}$ and $\beta=0$. 

Outside the horizon $|k\eta| \sim |k/aH| \ll 1$.
Assuming Bunch--Davies initial conditions,
and making a Taylor expansion of~\eqref{eq: delta phi k dependent}
in this limit, yields
\begin{equation}
    \label{eq: delta phi series}
    \delta \phi_k(\eta)
    \approx
    \frac{\im H}{\sqrt{2}k^{3/2}}
    \left\{
        \mathcolor{SussexVividRed}{1}
        \left[
            \mathcolor{SussexVividRed}{1}
            + \mathcolor{SussexVibrantGreen}{\frac{1}{2} (-k\eta)^2}
            + \Or(-k\eta)^4
        \right]
        +
        \mathcolor{SussexVividRed}{\frac{\im}{3} (-k\eta)^3}
        \left[
            \mathcolor{SussexVividRed}{1}
            - \mathcolor{SussexVibrantGreen}{\frac{1}{10} (-k\eta)^2}
            + \Or(-k\eta)^4
        \right]
    \right\} ,
\end{equation}
where we are temporarily taking $H$ to be constant.
We recognize the background ``growing''
and ``decaying'' modes in \textcolor{SussexVividRed}{\semibold{red}},
with time-dependence matching Eq.~\eqref{eq: delta phi background}.
The leading gradient corrections to both
``growing'' and ``decaying'' modes
are shown in \textcolor{SussexVibrantGreen}{\semibold{green}}.

Only the \textcolor{SussexVividRed}{\semibold{red}}
part of Eq.~\eqref{eq: delta phi series}
is captured by the
separate universe method.
The corresponding $A$- and $B$-coefficients,
as defined
in~\eqref{eq: delta phi background},
are
\begin{subequations}
\begin{align}
    \label{eq:slow-roll-A-soln}
    A & = \frac{\im H}{\sqrt{2}} k^{-3/2} , \\
    \label{eq:slow-roll-B-soln}
    B & = \frac{H}{3 \sqrt{2}} k^{3/2} .
\end{align}
\end{subequations}
The dominant behaviour as $|k\eta| \rightarrow 0$
comes from the constant ``growing'' mode.
If slow-roll were to continue indefinitely,
this solution would give the required
evolution outside the horizon.

The leading correction
to the separate universe result
is of order $(-k\eta)^2$.
It
comes from the gradient
correction to the ``growing'' mode,
not the ``decaying'' mode.
(See also the discussion
in Ref.~\cite{Dias:2012qy}.)

\para{Transient non-attractor phase}
Let us now allow a
transition to a
short non-attractor phase.
To be concrete, but without significant
loss of generality, we use
the example of an ultra-slow-roll phase.
We take
the transition into ultra-slow-roll to be instantaneous,
occurring at cosmic time $\ttransition$,
or equivalently conformal time $\etatransition$.
This event defines a distinguished scale,
corresponding to $\kevent$ above,
via $\ktransition = (aH)_{\ttransition}$.
Consider a wavevector $\vect{k}$ that crosses
the horizon just prior to the transition. 
In this example,
the wavenumber $k$ would be located in the region
where the power spectrum is growing sharply, 
like $\dimlessP_\zeta(k)\propto k^4$.
The solution up to $\etatransition$
is given by~\eqref{eq: delta phi series}.
Up to
and including the first contribution from the ``decaying'' mode,
it is
\begin{equation}
    \label{eq: sol before transition}
    \delta \phi^{\text{SR}}_{\vect{k}}(\eta)
    =
    \frac{\im H \alpha }{k}
    \bigg(
        1
        +
        \frac{1}{2}(-k \eta)^2
        +
        \frac{\im}{3} (-k \eta)^3
    \bigg)
    + \Or(-k\eta)^4
    \quad
    \text{if $\eta < \etatransition$}
    . 
\end{equation}

As is now well-known,
during ultra-slow-roll,
the mass term~\eqref{eq: mass of delta phi def}
turns out to have the same value as in slow-roll,
even though $\epsilon_2 \sim -6$.
(This is an example of \emph{Wands duality}~\cite{Wands:1998yp}.)
Therefore the solution to the $\delta\phi$
mode equation will be the same as~\eqref{eq: delta phi k dependent},
but with differing coefficients $\alpha$, $\beta$.
Labelling these with a tilde,
viz. $\tilde{\alpha}$, $\tilde{\beta}$,
we have
\begin{equation}
    \label{eq: sol after transition}
    \delta \phi^{\text{USR}}_{\vect{k}}(\eta)
    =
    \frac{\im H}{k}
    \big( \tilde{\alpha} - \tilde{\beta} \big)
    \bigg(
        1
        +
        \frac{1}{2} (-k \eta)^2
    \bigg)
    -
    \frac{H}{3k}
    \big( \tilde{\alpha} + \tilde{\beta} \big)
    (-k \eta)^3
    + \Or(-k\eta)^4
    \quad
    \text{if $\eta > \etatransition$}
    .
\end{equation}
To determine $\tilde{\alpha}$, $\tilde{\beta}$,
we require matching conditions at $\etatransition$.
The required conditions are%
    \footnote{Notice that while $\delta \phi_{\vect{k}}(\eta)$
    is continuous at the transition, its derivative is not.
    The matching condition can be obtained by modelling the slow-roll to ultra-slow-roll
    transition using Starobinsky's piece-wise linear model~\cite{Starobinsky:1992ts}.
    This yields 
    \begin{equation}
        \label{eq:starobinsky-matching}
        \frac{\d\delta \phi^\text{USR}_{\vect{k}}}{\d\eta}\Big|_{\eta=\etatransition}
        =
        \bigg(
            \frac{\d\delta \phi^\text{SR}_{\vect{k}}}{\d\eta}
            + 3 \frac{\Delta A}{A_+} \ktransition
            \delta \phi^{\text{SR}}_{\vect{k}}
        \bigg)_{\eta=\etatransition}
        ,
    \end{equation} 
    where $A_+>0$ $(A_->0)$ is the slope of the potential before
    (after) the transition, and $\Delta A \equiv A_- - A_+$.
    Ultra-slow-roll is realized when $A_-\ll A_+$. In this limit,
    Eq.~\eqref{eq:starobinsky-matching}
    matches the second condition given in
    Eq.~\eqref{eq: matching condition derivative at transition}.}
\begin{subequations}
\begin{align}
    \label{eq: matching condition field at transition}
    \delta \phi^{\text{USR}}_{\vect{k}}\Big|_{\eta=\etatransition}
    & =
    \delta \phi^{\text{SR}}_{\vect{k}}\Big|_{\eta=\etatransition}
    \\
    \label{eq: matching condition derivative at transition}
    \frac{\d\delta \phi^{\text{USR}}_{\vect{k}}}{\d\eta}\Big|_{\eta=\etatransition}
    & =
    \bigg(
        \frac{\d\delta \phi^{\text{SR}}_{\vect{k}}}{\d\eta}
        -
        3 \ktransition \delta \phi^{\text{SR}}_{\vect{k}}
    \bigg)\Big|_{\eta=\etatransition}
    .
\end{align}
\end{subequations}
The jump in the derivative,
enforced by Eq.~\eqref{eq: matching condition derivative at transition},
is responsible for the difference between
$\alpha$, $\beta$
and
$\tilde{\alpha}$, $\tilde{\beta}$.
Solving for $\tilde{\alpha}$, $\tilde{\beta}$
and inserting these
into~\eqref{eq: sol after transition},
we find
\begin{multline}
    \label{eq: sol after transition determined}
    \delta \phi^{\text{USR}}_{\vect{k}}(\eta)
    =
    \frac{\im H}{\sqrt{2}k^{3/2}}
    \Bigg\{
        \bigg[
            {-\frac{2}{5}} \left(\frac{k}{\ktransition} \right)^2
            +
            \Or\left(\frac{k}{\ktransition}\right)^3
        \bigg]
        \bigg(
            \mathcolor{SussexVividRed}{1}
            +
            \mathcolor{SussexVibrantGreen}{\frac{1}{2}(-k\eta)^2}
            +
            \Or(-k\eta)^4
        \bigg) \\
        +
        \bigg[
            \left(\frac{k}{\ktransition}\right)^{-3}
            +
            \Or\left(\frac{k}{\ktransition}\right)^{-1}
        \bigg]
        \bigg(
            \mathcolor{SussexVividRed}{(-k\eta)^3}
            +
            \Or(-k\eta)^5
        \bigg)
    \Bigg\} ,
\end{multline}
where $\ktransition$
labels the horizon scale at the transition time,
i.e. $\ktransition/aH \approx 1$ at $\eta = \etatransition$.
As before, we have
highlighted the lowest-order ``growing'' and ``decaying'' modes
in \textcolor{SussexVividRed}{\semibold{red}},
and the leading gradient correction in
\textcolor{SussexVibrantGreen}{\semibold{green}}.
Note that Eq.~\eqref{eq: sol after transition determined}
is well-behaved as $k \rightarrow 0$,
despite the appearance of inverse powers of $k/\ktransition$
in $\tilde{\alpha} + \tilde{\beta}$.

The corresponding
separate universe $A$- and $B$-coefficients are
\begin{subequations}
\begin{align}
    \label{eq:post-transition-A-coeff}
    A & \approx
    -
    \frac{2\im H}{5 \sqrt{2}}
    \bigg(
        \frac{k}{\ktransition}
    \bigg)^2 k^{-3/2} ,
    \\
    \label{eq:post-transition-B-coeff}
    B & \approx
    -
    \frac{\im H}{\sqrt{2}}
    \bigg(
        \frac{k}{\ktransition}
    \bigg)^{-3}
    k^{3/2} .
\end{align}
\end{subequations}
For a long-wavelength mode
that did not cross much before the transition,
$k/\ktransition$ will not be too much
smaller than unity.%
    \footnote{Our aim here is only to illustrate the effect of
    gradient terms in models with a transition,
    and discuss the correct choice of initialization time.
    For this reason, we restrict our attention to
    modes that crossed the horizon just before the transition.
    In order to model the behaviour of substantially
    larger scales,
    such as CMB scales,
    or the scale corresponding to the dip in $\dimlessP_\zeta$,
    one would need to keep track of all factors
    (e.g. $\Delta A \neq -A_+$ in Eq.~\eqref{eq:starobinsky-matching}),
    as done in Ref.~\cite{Jackson:2023obv}.}
At least a few e-folds after the transition,
we also have $|\ktransition \eta| \ll 1$,
so both the ``decaying'' mode
and the gradient correction
to the ``growing'' mode
can be dropped.
In the power spectrum, it follows that
the $A$-coefficient
for the ``growing'' mode will yield
the expected growing behaviour $\dimlessP_\zeta \propto k^4$.

Eq.~\eqref{eq: sol after transition determined}
satisfies the separate universe principle
at times later than the transition.
It yields $A$ and $B$ coefficients
that do not agree with the
pre-transition
solution~\eqref{eq: sol before transition},
which clearly is to be expected.
The critical feature
of~\eqref{eq: sol after transition determined}--\eqref{eq:post-transition-A-coeff}
is not the difference as such,
but rather that
the $(k/\ktransition)^2$
scaling of the $A$-coefficient
in~\eqref{eq:post-transition-A-coeff}
cannot be reproduced by the $k \rightarrow 0$
limit of
\eqref{eq: delta phi eom conformal time}
with initial condition~\eqref{eq:slow-roll-A-soln}.
A similar statement applies to the $B$ coefficient.
The appearance of $k/\ktransition$ is attributable
to imposition
of the jump condition~\eqref{eq: matching condition derivative at transition}
at a localized
time $\etatransition$. In other words,
it follows from the presence of a sudden
transition in the background.

The
conclusion is that the separate universe framework
correctly describes the time dependence of the perturbation
for each
$\vect{k}$-mode, but only
if we use initial data
determined \emph{after} the transition.
In the language of~\S\ref{sec: separate universe},
this ensures that the subsequent evolution of
each superhorizon volume
is uniquely predictable
from its initial conditions.
This conclusion was emphasized by Ref.~\cite{Jackson:2023obv}.
Although not shown here,
the choice $t_i > \ttransition$
actually allows a correct description of all large
scales, not just those that exit not too long before the transition.
In Appendix~\ref{app: separate universe computation of tree level power spectrum}
we show that, for two independent choices of initial time
$t_i > \ttransition$,
the separate universe principle successfully reproduces the numerical tree-level
power spectrum for the toy model considered in Ref.~\citep{Iacconi:2023ggt}.

\subsection{Correlation functions}
With these considerations in mind,
we now explain how to
compute correlation functions
in the scenario of~\S\ref{sec: properties of long and short modes},
where there is a back-reaction event
at some time $t_q$.

\label{sec:separate-universe-correlations}
\para{Tree-level $\delta N$ formula in phase space}
To begin, we work at tree level.
We choose an initial time $t_i$
just after some large-scale mode $\vect{p}$
has left the horizon, and smooth on the scale $\bigBox \sim p^{-1}$.
As explained above,
each resulting patch
evolves like a locally unperturbed universe with displaced
initial conditions.
Making a $\delta N$
Taylor expansion in these displacements,
and dropping gradients,
we find
that the change
$\delta X^I$
from patch to patch
in some field $X^I$ is
\begin{multline}
    \label{eq:deltaN-loop-step1}
    \smoothBig{\delta X^I(t, \vect{x})}
    =
    \frac{\partial X^I(t)}{\partial X^M(t_i)}
    \smoothBig{\delta X^M(t_i, \vect{x})}
    +
    \frac{1}{2!}
    \frac{\partial^2 X^I(t)}{\partial X^M(t_i) \partial X^N(t_i)}
    \smoothBig{\delta X^M(t_i, \vect{x})}
    \smoothBig{\delta X^N(t_i, \vect{x})}
    \\
    +
    \frac{1}{3!}
    \frac{\partial^3 X^I(t)}{\partial X^M(t_i) \partial X^N(t_i) \partial X^R(t_i)}
    \smoothBig{\delta X^M(t_i, \vect{x})}
    \smoothBig{\delta X^N(t_i, \vect{x})}
    \smoothBig{\delta X^R(t_i, \vect{x})}
    + \cdots ,
\end{multline}
where
$\smoothBig{\cdots}$ denotes smoothing on the scale $\bigBox$.
The smoothing implies that
the spatial coordinate $\vect{x}$
should be regarded as labelling each patch.
The notation
$\partial X^I(t) / \partial X^M(t_i)$
(and its higher-order generalizations)
denotes the
variation
of a late-time field $X^I(t)$
with respect to a
displacement
in the early-time initial condition
$X^M(t_i)$.
Eq.~\eqref{eq:deltaN-loop-step1}
enables us to determine
correlation functions of the
perturbation $\smoothBig{\delta X(t)}$ at
wavevector $\vect{p}$,
at any time up to the point where
enhanced small-scale modes emerge.
This is analogous to the transition time
$\ttransition$ from~\S\ref{sec: separate universe and sudden transitions}.
For convenience, we continue to label this time
$t_q$, with the understanding that all
back-reacting modes should have exited the horizon by
that time.
Eq.~\eqref{eq:deltaN-loop-step1}
is invalidated
once back-reaction becomes important,
because the fields
are no longer determined only by initial data at $t_i$.
This parallels the difference between the pre-
and post-transition
solutions~\eqref{eq:slow-roll-A-soln}
and~\eqref{eq:post-transition-A-coeff}.

To build correlation functions
from~\eqref{eq:deltaN-loop-step1},
we use it to
define
Fourier modes
and 
evaluate suitable expectation values.
When we do so, we encounter
many possible terms.
These can be classified according to the number
of unconstrained momentum integrals they contain.
`Tree-level' contributions
contain no unconstrained momentum integrals.
We may also encounter
`loop-level' terms.
By analogy with quantum field theory,
a term is said to be at $n^{\text{th}}$ order in the
loop expansion when it contains $n$ unconstrained integrals.
Smoothing on the scale $\bigBox \sim p^{-1}$
implies that
these integrals will
effectively be cut off for comoving momenta
$\gtrsim p$.
The tree-level expression,
Eq.~\eqref{eq:deltaN-loop-step1},
has no knowledge of the short-scale modes.
It therefore neglects any back-reaction.

At this stage, the notion
of `loop-level'
used here
is simply a formal definition
and does not imply any relationship
with loops generated by the
expansion of ``in--in'' quantum field theory into
Green's functions.

\para{$\delta N$ formula with back-reaction}
To capture back-reaction
we make two changes.
First,
Eq.~\eqref{eq: sol after transition determined}
shows that when back-reaction
modifies the background
we must choose new initial data,
to capture effects involving powers of $p/q$.
This ratio
is analogous to $k/\ktransition$.
We are therefore forced to move the initial time
$t_i$ to be later than $t_q$.
Second, we must calculate the relevant
$(p/q)$-corrected effects.
To do so, we
introduce averages over
the short-scale modes
by
working beyond tree-level in the loop expansion.
This procedure was described in
Ref.~\cite{Iacconi:2023ggt},
and we briefly summarize the steps below.
Up to limitations set by our approximations,
it is
valid provided $p/q \ll 1$,
so that there is an appreciable separation of scales.

We continue to label the new initial hypersurface
with time label $t_i$.
It must be chosen late enough to allow horizon exit
for all modes that contribute to the back-reaction.
We populate this hypersurface
with smoothed horizon volumes of
size $\smallBox$.
Separation of scales means that $\smallBox$
is much smaller than the
initial smoothing scale $\bigBox$.
The distribution of Fourier modes
in these volumes includes contributions from 
all scales that exited the horizon between (at least)
$t_p$ and $t_q$.
Applying Eq.~\eqref{eq:deltaN-loop-step1},
now specialized to the curvature perturbation $\zeta$,
and extracting a long-wavelength Fourier wavevector $\vect{p}$,
we obtain
\begin{equation}
    \label{eq: zeta with backreaction Fourier}
    \begin{multlined}
        \zeta_{\vect{p}}(t)
        =
        N_I^{(t_i, t)}
        \Big[ \smoothSmall{\delta X^I(t_i, \vect{x})} \Big]_{\vect{p}}
        +
        \frac{1}{2!}
        N_{IJ}^{(t_i, t)}
        \int \frac{\d^3 r}{(2\pi)^3}
        \Big[ \smoothSmall{\delta X^I(t_i, \vect{x})} \Big]_{\vect{p}-\vect{r}}
        \Big[ \smoothSmall{\delta X^J(t_i, \vect{x})} \Big]_{\vect{r}}
        \\
        +
        \frac{1}{3!}
        N_{IJK}^{(t_i, t)}
        \int \frac{\d^3 r}{(2\pi)^3} \frac{\d^3 s}{(2\pi)^3}
        \Big[ \smoothSmall{\delta X^I(t_i, \vect{x})} \Big]_{\vect{p}-\vect{r}-\vect{s}}
        \Big[ \smoothSmall{\delta X^J(t_i, \vect{x})} \Big]_{\vect{r}}
        \Big[ \smoothSmall{\delta X^K(t_i, \vect{x})} \Big]_{\vect{s}}
        \\
        +
        \Or(\delta X)^4 .
    \end{multlined}
\end{equation}
Each integral should now be cut off
for wavenumbers $\gtrsim \smallBox^{-1}$.
The notation
$[ \smoothSmall{\delta X^I} ]_{\vect{p}}$
indicates that we \emph{first} smooth on the scale
$\smallBox$, and \emph{then} extract the Fourier mode $\vect{p}$.
This smoothing
procedure may already disturb
the
$\vect{p}$-mode of $\delta X^I$,
when compared to the value that would be
predicted by~\eqref{eq:deltaN-loop-step1}.
This corresponds to the possibility of a 
non-negligible loop correction
to $\delta X^I$ at time $t_i$.
We discuss this in more detail
in~\S\ref{sec: 1-loop from quantum initial conditions is a boundary term}.

\para{1-loop contributions}
Eq.~\eqref{eq: zeta with backreaction Fourier}
yields four contributions to the 2-point function
$\langle \zeta_{\vect{p}}(t) \zeta_{-\vect{p}}(t) \rangle'$.
These are
\begin{equation}
    \label{eq: total 1-loop correction}
    \langle \zeta _{\vect{p}}(t) \zeta_{-\vect{p}}(t)\rangle'_{\text{1-loop}}
    =
    \underbrace{\langle \zeta_{\vect{p}} \zeta_{-\vect{p}}\rangle'_{11}}_
    {\substack{\text{1-loop in} \\ \text{initial conditions}}}
    +
    \quad
    \underbrace{\langle \zeta_{\vect{p}} \zeta_{-\vect{p}}\rangle'_{12}
                + \langle \zeta_{\vect{p}} \zeta_{-\vect{p}}\rangle'_{22}
                + \langle \zeta_{\vect{p}} \zeta_{-\vect{p}}\rangle'_{13}
    }_{\text{1-loop due to non-linearity of $\delta N$}} , 
\end{equation}
where we have labelled each piece based on
the contributions
from Eq.~\eqref{eq: zeta with backreaction Fourier};
``1'' labels the linear term,
``2'' the quadratic term,
and ``3'' the cubic term.
Explicitly, these
can be written
\begin{subequations}
\begin{equation}
\label{eq: 11 type loop initial expression}
    \langle
        \zeta_{\vect{p}}
        \zeta_{-\vect{p}}
    \rangle'_{11}
    = 
    N_I^{(t_i,t)} N_J^{(t_i,t)}
    \Big\langle
        \Big[\smoothSmall{\delta X^I(t_i, \vect{x})}\Big]_{\vect{p}}
        \Big[\smoothSmall{\delta X^J(t_i, \vect{x})}\Big]_{-\vect{p}}
    \Big\rangle'_\text{1-loop} , 
\end{equation}
\begin{multline}
\label{eq: 12 type base expression}
    \langle
        \zeta_{\vect{p}}
        \zeta_{-\vect{p}}
    \rangle'_{12}
    =
    \frac{1}{2}
    N_I^{(t_i,t)} N_{JK}^{(t_i,t)}
    \int \frac{\d^3 r}{(2\pi)^3} \;
    \bigg(
        \Big\langle
            \delta X^I _{\vect{p}}(t_i)
            \delta X^J_{\vect{r}}(t_i)
            \delta X^K_{-\vect{p}-\vect{r}}(t_i)
        \Big\rangle'_\text{tree}
        \\
        +
        \Big\langle
            \delta X^J _{\vect{r}}(t_i)
            \delta X^K_{\vect{p}-\vect{r}}(t_i)
            \delta X^I_{-\vect{p}}(t_i)
        \Big\rangle'_\text{tree}
    \bigg) , 
\end{multline}
\begin{equation}
\label{eq: 22 base expression}
    \langle
        \zeta_{\vect{p}}
        \zeta_{-\vect{p}}
    \rangle'_{22}
    =
    \frac{1}{4}
    N_{IJ}^{(t_i,t)} N_{KM}^{(t_i,t)}
    \int \frac{\d^3 r}{(2\pi)^3} \frac{\d^3 s}{(2\pi)^3} \;
    \Big\langle
        \delta X^I _{\vect{r}}(t_i)
        \delta X^J_{\vect{p}-\vect{r}}(t_i)
        \delta X^K_{\vect{s}}(t_i)
        \delta X^M_{-\vect{p}-\vect{s}}(t_i)
    \Big\rangle'_\text{tree} , 
\end{equation}
\begin{multline}
\label{eq: 13 base expression}
    \langle
        \zeta_{\vect{p}}
        \zeta_{-\vect{p}}
    \rangle'_{13}
    =
    \frac{1}{6} N_{I}^{(t_i,t)} N_{JKM}^{(t_i,t)}
    \int \frac{\d^3 r}{(2\pi)^3} \frac{\d^3 s}{(2\pi)^3}
    \bigg(
        \Big\langle
            \delta X^I _{\vect{p}}(t_i)
            \delta X^J_{\vect{r}}(t_i)
            \delta X^K_{\vect{s}}(t_i)
            \delta X^M_{-\vect{p}-\vect{r}-\vect{s}}(t_i)
        \Big\rangle'_\text{tree}
        \\
        +
        \Big\langle
            \delta X^J_{\vect{r}}(t_i)
            \delta X^K_{\vect{s}}(t_i)
            \delta X^M_{\vect{p}-\vect{r}-\vect{s}}(t_i)
            \delta X^I_{-\vect{p}}(t_i)
        \Big\rangle'_\text{tree}
    \bigg)
    . 
\end{multline}
\end{subequations}
At this
order in the loop expansion,
only the ${11}$-type contribution includes
a loop correction to the initial $\delta X$
correlation functions.
Such corrections can be ignored
in the remaining contributions.
We have therefore
approximated
$[ \smoothSmall{\delta X^I} ]_{\vect{k}} \approx \delta X^I_{\vect{k}}$.
In what follows,
we distinguish between the two types of contribution
in Eq.~\eqref{eq: total 1-loop correction}
by describing the $(12)$-, $(22)$- and $(13)$-type loops as
``$\delta N$ loops''~\cite{Lyth:2006qz}.

\subsection{Does the separate universe framework capture ``in--in'' loop effects?}
\label{sec: does separate universe compare with in-in}

In the language
of~\S\ref{sec: separate universe and sudden transitions},
the time-dependent separate-universe
factors
$N_I^{(t_i, t)}$,
$N_{IJ}^{(t_i, t)}$
and $N_{IJK}^{(t_i, t)}$
in Eqs.~\eqref{eq: 11 type loop initial expression}--\eqref{eq: 13 base expression}
contain the
\textcolor{SussexVividRed}{\semibold{red}}-highlighted
time-dependent growing and decaying
modes
in~\eqref{eq: sol after transition determined}.
Meanwhile,
the $\delta X^I$
correlation functions
determine the initial data.
They
correspond to
(expectation values of)
the $\vect{k}$-dependent
$A$- and $B$-coefficients
of Eqs.~\eqref{eq:post-transition-A-coeff}--\eqref{eq:post-transition-B-coeff}.
In the full theory $A$ and $B$ are stochastic variables,
which inherit their stochasticity from the initial data
at time $t_p$.
The loop integrals average over the short-scale modes,
and
measure their influence
on the evolution
of the large-scale mode $\vect{p}$.
Dependence on $p/q$
arises from these integrals
and from
the
$\delta X^I$
correlation functions.
It follows that
loop-level formulae
such as~\eqref{eq: 11 type loop initial expression}--\eqref{eq: 13 base expression}
can be regarded as
an approximation that captures
some features of
back-reaction.
As usual, whether this approximation
is adequate
for any particular purpose
depends on the model being studied and the observable
under discussion.

We have already emphasized that,
as presented here, separate-universe loop-level
expressions such as Eqs.~\eqref{eq: 11 type loop initial expression}--\eqref{eq: 13 base expression}
do not yet have a clear relationship with
the loop expansion of
non-equilibrium quantum field theory,
represented by (for example)
the loop expansion of
the ``in--in'' path integral.
In Ref.~\cite{Iacconi:2023ggt} it was assumed,
but not demonstrated,
that the ``separate-universe'' loops
of Eq.~\eqref{eq: 11 type loop initial expression}--\eqref{eq: 13 base expression}
represented a subset of the ``in--in'' loops,
so that it was meaningful
to compare calculations performed in these different frameworks.
For example,
if the enhanced modes contributing
to these integrals
are on superhorizon scales,
and behaving nearly
classically,
one would expect that their contribution could be captured
in either formalism.

This expectation can be made more precise.
A comprehensive explanation of the relation
between the separate universe
framework and ``in-in'' perturbation theory
requires more elaboration
than can be given here, and we will return
to it in a separate publication~\cite{in_preparation}.
Here, we simply summarize the key ideas.
Starting from either
the operator Heisenberg equation of motion,
or an explicit 1-loop formula for the correlation function,
it is possible to derive a 1-loop transport equation
for the two-point function.
Working in a compact de Witt notation, where
index summation implies integration over $\vect{k}$ labels
(for details, see, e.g., Ref.~\cite{Dias:2016rjq}),
this equation can be written,
for generic fields $X^I$,
\begin{multline}
    \label{eq:1loop-2f-transport}
    \frac{\d}{\d N}
    \langle
        \tensor{X}{^I}
        \tensor{X}{^J}
    \rangle
    =
    \tensor{u}{^I_M}
    \langle
        \tensor{X}{^M}
        \tensor{X}{^J}
    \rangle
    +
    \tensor{u}{^J_M}
    \langle
        \tensor{X}{^I}
        \tensor{X}{^M}
    \rangle
    \\
    \mbox{}
    +
    \frac{1}{2}
    \tensor{u}{^I_M_N}
    \langle
        \tensor{X}{^M}
        \tensor{X}{^N}
        \tensor{X}{^J}
    \rangle
    +
    \frac{1}{2}
    \tensor{u}{^J_M_N}
    \langle
        \tensor{X}{^I}
        \tensor{X}{^M}
        \tensor{X}{^N}
    \rangle
    \\
    \mbox{}
    +
    \frac{1}{2}
    \tensor{u}{^I_M_N_R}
    \langle
        \tensor{X}{^M}
        \tensor{X}{^J}
    \rangle
    \langle
        \tensor{X}{^N}
        \tensor{X}{^R}
    \rangle
    +
    \frac{1}{2}
    \tensor{u}{^J_M_N_R}
    \langle
        \tensor{X}{^I}
        \tensor{X}{^M}
    \rangle
    \langle
        \tensor{X}{^N}
        \tensor{X}{^R}
    \rangle
    .
\end{multline}
The tensors
$\tensor{u}{^I_M}$,
$\tensor{u}{^I_M_N}$,
$\tensor{u}{^I_M_N_R}$
depend on
the background trajectory,
and the wavenumbers carried by their indices.
Explicit expressions for
$\tensor{u}{^I_M}$
and $\tensor{u}{^I_M_N}$
were given in Dias~{\etal}~\cite{Dias:2016rjq},
valid
for the case that the $X^I$ are
perturbations in the fields and momenta
on a spatially flat slicing.
The zero-wavenumber limit of
$\tensor{u}{^I_M_N_R}$
was obtained by Andersen {\etal}~\cite{Anderson:2012em}.
An exact wavenumber-dependent expression for
it
is not yet known,
but is not needed at the level of the present discussion.
Eq.~\eqref{eq:1loop-2f-transport} should be supplemented
by a tree-level transport equation for the
3-point function, described in Ref.~\cite{Dias:2016rjq}.
This system of coupled transport equations
for the 2- and 3-point functions
can be integrated
by writing
\begin{equation}
    \label{eq:1loop-MPP-ansatz}
    \Big( X^I \Big)_t
    =
    \tensor{\Gamma}{^I_M} \Big( X^M \Big)_{\tstar}
    +
    \frac{1}{2!}
    \tensor{\Gamma}{^I_M_N} \Big(X^M X^N \Big)_{\tstar}
    +
    \frac{1}{3!}
    \tensor{\Gamma}{^I_M_N_R} \Big(X^M X^N X^R \Big)_{\tstar}
    +
    \cdots
    ,
\end{equation}
where $\tstar < t$,
and
$( \cdots )_t$ denotes evaluation of the enclosed
operator at time $t$.
Eq.~\eqref{eq:1loop-MPP-ansatz}
can be regarded as a form of operator product expansion,
in which the
operator $(X^I)_t$ is interpreted as
composite with respect to a basis of operators defined
at $\tstar$.
The $\Gamma$-tensors play the role of OPE coefficients.
For convenience we assume
the operators
$( X^M X^N )_{\tstar}$,
$( X^M X^N X^R )_{\tstar}$
to be symmetrized.

The momentum-dependent tensors
(technically bitensors)
$\tensor{\Gamma}{^I_M}$,
$\tensor{\Gamma}{^I_M_N}$,
$\tensor{\Gamma}{^I_M_N_R}$
satisfy
evolution equations
that symbolically
match those already known at tree-level,
although now with momentum-dependent $u$-tensors.
For the $\tensor{\Gamma}{^I_M}$ tensor the required equation
is~\cite{Seery:2012vj,Mulryne:2013uka}
\begin{equation}
    \frac{\d}{\d N}
    \tensor{\Gamma}{^I_M}
    =
    \tensor{u}{^I_J}
    \tensor{\Gamma}{^J_M}
    .
\end{equation}
Corresponding equations for the
$\tensor{\Gamma}{^I_M_N}$
and
$\tensor{\Gamma}{^I_M_N_R}$
tensors
can be found in
Dias~{\etal}~\cite{Dias:2016rjq}
and Anderson~{\etal}~\cite{Anderson:2012em}.
The initial conditions require
$\tensor{\Gamma}{^I_M} = \tensor{\delta}{^I_M}$
at $t = \tstar$, with all other $\Gamma$-tensors
equal to zero.
We discuss the $\Gamma$-tensors
in more detail
in~\S\ref{sec: 1-loop from quantum initial conditions is a boundary term}.

We can now connect this picture with the separate universe
framework.
The initial time $\tstar$
appearing in Eq.~\eqref{eq:1loop-MPP-ansatz}
is arbitrary, because it
merely corresponds to breaking the integration
of~\eqref{eq:1loop-2f-transport}
into two steps:
an integral up to $\tstar$,
followed by an integral from $\tstar$ to $t$.
As a result,
Eq.~\eqref{eq:1loop-MPP-ansatz}
solves the transport equation
in the sense that the 2-point function computed using it
matches direct solution of Eq.~\eqref{eq:1loop-2f-transport},
provided correlations of the operators
on the right-hand side of~\eqref{eq:1loop-MPP-ansatz}
are computed using the same system of transport equations,
or equivalently, to 1-loop for 2-point functions
and tree-level for 3-point functions.
However we choose $\tstar$,
the result matches the 1-loop ``in--in'' expression
(before imposing a cutoff) provided
all momentum-dependent effects are retained in the
$\Gamma$-tensors and the $X^I$ correlation functions at $\tstar$.

We now choose $\tstar$ to match the initial time $t_i$
used in
Eqs.~\eqref{eq: 11 type loop initial expression}--\eqref{eq: 13 base expression}.
The $u$-tensors
$\tensor{u}{^I_M}$,
$\tensor{u}{^I_M_N}$
and
$\tensor{u}{^I_M_N_R}$
have local expressions
and may depend on $p/(aH)$ and $q/(aH)$, but not on the ratio
$q/p$.%
    \footnote{This can be demonstrated explicitly for
    $\tensor{u}{^I_M}$ and
    $\tensor{u}{^I_M_N}$, for which complete momentum-dependent expressions
    are known; see Ref.~\cite{Dias:2016rjq}.
    The formulae quoted there for $\tensor{u}{^I_M_N}$ appear non-local,
    but this can be removed using the Hamiltonian constraint,
    as discussed in Dias {\etal}~\cite{Dias:2014msa}.
    One expects this property to apply at all orders.
    It must be true for $\tensor{u}{^I_M_N_R}$ in order 
    to smoothly connect with the zero-momentum expressions
    quoted by Anderson {\etal}~\cite{Anderson:2012em}.}
The same dependences will be inherited by the
$\Gamma$-tensors.
With initial conditions set after all back-reacting modes
leave the horizon,
their values will therefore be close to those predicted by the separate
universe framework because integration of the
$\Gamma$-tensors in the zero-momentum limit is known
to reproduce the Taylor expansion in initial
conditions~\cite{Seery:2012vj}.
At 1-loop order,
this includes a slight shift in
$\tensor{u}{^I_M}$
corresponding to a tadpole.
After making a gauge transformation to $\zeta$,
the result is that Eq.~\eqref{eq:1loop-MPP-ansatz}
will lead to a set of 1-loop expressions
equivalent to
Eqs.~\eqref{eq: 11 type loop initial expression}--\eqref{eq: 13 base expression},
up to corrections of order $p/(aH)$ and $q/(aH)$.
As we have already emphasized,
the initial correlation functions appearing
in Eqs.~\eqref{eq: 11 type loop initial expression}--\eqref{eq: 13 base expression}
must include loop corrections to the 11 initial condition,
evaluated up to time $t_i$.

These loop integrals clearly
will \semibold{not} capture contributions to the full
``in--in'' loops
from regions of the momentum integration where $q/aH \gtrsim 1$.
This is not only
expected but
guaranteed, because the separate universe method
does not accurately capture effects occurring on subhorizon scales.
However, for the problem at hand we wish only to capture
effects from the band of enhanced short-scale modes,
which at this stage have been inflated to superhorizon scales.
In this region, as we have seen,
the $\Gamma$-tensors are nearly independent of wavenumber
and closely approximate the values predicted by the separate universe
framework.
Contributions to the complete ``in--in'' loop integrals from back-reacting modes
on superhorizon scales should therefore be captured equally well
using either method.
Provided the final renormalization scale is
reasonably larger than the ultraviolet end of
the enhanced band,
we expect this conclusion will not be
significantly
altered by ultraviolet regularization and renormalization.

Use of the separate universe
formulae~\eqref{eq: 11 type loop initial expression}--\eqref{eq: 13 base expression}
has a number of advantages
compared to the full ``in--in'' calculation.
These expressions are substantially
less complex than
those produced by an expansion of ``in--in''
diagrams into Green's functions.
As a result,
the physical mechanism of back-reaction is more explicit.
Further, it is much easier to include the effect of
multiple scalar fields,
and to track time dependence accurately outside the horizon.

\para{Choice of initial time}
\label{sec:initial-time}
The main drawback
with the programme outlined above
is the necessity of including loop
corrections to the 11 initial condition.
Its 1-loop correction
represents disturbance of the long $\vect{p}$
mode by short-scale structure even before
the initial time $t_i$, as already pointed out
below Eq.~\eqref{eq: 13 base expression}.
This implies that we must balance competing requirements
when choosing the initial time $t_i$.

We have seen that the $\delta N$ loops of
Eqs.~\eqref{eq: 12 type base expression}--\eqref{eq: 13 base expression}
provide an accurate estimate for loop momenta
in the enhanced band only
if $t_i$ is chosen
after all back-reacting scales exit the horizon.
In fact,
if the peak is produced by modes enhanced
during an ultra-slow-roll phase,
modes exiting after
a later return to slow-roll
may also contribute to 
the broad peak in the power
spectrum.
To capture the effect of all these modes it may be
necessary to choose $t_i$ \emph{substantially} later than the
time of the transition.
We will comment on the inclusion of the whole peak
in {\S}\ref{sec: interpretation and discussion}.

On the other hand,
the later we choose $t_i$,
the more significant loop corrections to the
initial correlation function in~\eqref{eq: 11 type loop initial expression}
are likely to become.
Such corrections are not expected to be large during slow-roll
evolution~\cite{Senatore:2009cf},
but it is plausible that there could be a significant effect 
in the ultra-slow-roll epoch.
To make our application of the separate
universe method as simple
as possible, the initial time should be set sufficiently
\emph{early} that the initial correlations are
affected minimally by such effects.%
    \footnote{We emphasize that this is not
    an issue of principle, but only convenience.
    If it is not possible to choose $t_i$ early enough that
    the $\delta X^I$ are not contaminated by loop corrections,
    one could always choose to evaluate them
    using the in--in formalism or an equivalent.
    However, in this case, the simplicity of the separate universe
    method is largely lost.
    Whatever method we choose to evaluate the loop corrections
    to the $\delta X^I$,
    we might as well calculate the entire loop correction
    to $\zeta$ in the same way.}
Clearly, these desiderata
are in some tension,
and it may not be possible to find a $t_i$
that satisfies all of these requirements.

\subsection{The 1-loop contributions}
\label{sec: the 1-loop contributions}
For future convenience, we collect reduced
expressions for the contribution to the power spectrum
from the
$(11)$-, $(12)$-, $(22)$-, and $(13)$-type
loops,
given by Eqs.~\eqref{eq: 11 type loop initial expression}--%
\eqref{eq: 13 base expression}.

Note
that we do not need
to account for further loop contributions.
In particular,
the initial correlation functions absorb all loop corrections
up to the initial time $t_i$.
The separate universe contributions
(such as
Eqs.~\eqref{eq: 12 type base expression}--\eqref{eq: 13 base expression} at 1-loop)
absorb corrections
generated between $t_i$ and the time of observation, $t$.
This includes all loop corrections
generated by nonlinear evolution of the underlying fields and
momenta $X^I$, \emph{and}
nonlinearities associated with the gauge transformation into $\zeta$.
The $\delta N$ formula accounts for all these sources of nonlinearity,
not merely those associated with the gauge transformation to $\zeta$.

\para{$(11)$-type loop}
The $(11)$-type loop is given by Eq.~\eqref{eq: 11 type loop initial expression}.
We can write it in the more economical form
\begin{equation}
    \label{eq: 11 loop master}
    \langle
        \zeta_{\vect{p}}(t)
        \zeta_{-\vect{p}}(t)
    \rangle'_{11}
    =
    N_I^{(t_i, t)}
    N_J^{(t_i, t)}
    \langle
        \delta X^I_{\vect{p}}(t_i)
        \delta X^J_{-\vect{p}}(t_i)
    \rangle'_\text{1-loop} ,
\end{equation} 
in which the explicit smoothing
has been dropped.
As explained above,
this is replaced
by the 1-loop correction.

\para{$(12)$-, $(22)$-, and $(13)$-type loops}
The $(12)$-, $(22)$- and $(13)$-type diagrams are
associated with
non-linear terms in the $\delta N$
formula~\eqref{eq: zeta with backreaction Fourier},
which
induce correlations between the long- and short-scale modes.
The primary expressions
are Eqs.~\eqref{eq: 12 type base expression}--\eqref{eq: 13 base expression}.
Below,
the loop momentum variable
($\vect{r}$ or $\vect{s}$ in
Eqs.~\eqref{eq: 12 type base expression}--\eqref{eq: 13 base expression})
has been relabelled $\vect{q}$.

The expressions
for these terms
can be simplified
in our assumed momentum configuration,
where there is a large hierarchy $p \ll q$.
In this configuration,
the operators
$\delta \hat X^I_{\vect{p}}$ and $\delta \hat X^J_{\vect{q}}$
commute.
Also,
the 3-point correlation function~\eqref{eq: def 3pt phase space}
depends only on the magnitude of the momenta.
Dropping corrections of relative order $(p/q)^2$,
so that we can write $|\vect{p}+\vect{q}| \approx q$,
it follows that
the $(12)$-type loop~\eqref{eq: 12 type base expression} can be written
\begin{equation}
\label{eq: 12 simple}
    P_\zeta(p;t)_{12}
    =
    N_I^{(t_i,t)}
    N_{JK}^{(t_i,t)}
    \int \d^3 q \;
    \tensor{\alpha}{^I^J^K}(p,q,q;t_i) . 
\end{equation}
Making the same approximations,
the $(22)$-type diagram~\eqref{eq: 22 base expression} reduces to 
\begin{equation}
    P_\zeta(p;t)_\text{22}
    =
    \frac{1}{2}
    N_{IJ}^{(t_i,t)}
    N_{KL}^{(t_i,t)}
    \int \d^3 q \; P^{IK}(q;t_i)\, P^{JL}(q;t_i) . 
\end{equation}
Finally,
after
performing Wick contractions in Eq.~\eqref{eq: 13 base expression}
we find that the $(13)$-contribution can be rewritten
\begin{equation}
    P_\zeta(p;t)_\text{13}
    =
    \frac{1}{2}
    N_{I}^{(t_i,t)}
    N_{JKL}^{(t_i,t)}
    \Big[
        P^{IJ}(p;t_i)
        + P^{JI}(p;t_i)
    \Big]
    \int \d^3 q \; P^{KL}(q;t_i) .
\end{equation}
During inflation, the commutator
of the field operator and its momentum
decays rapidly after horizon crossing,
\begin{equation}
    \Big[
        \delta \hat \phi_{\vect{k}}(t),
        \delta \hat \pi_{\vect{k}'}(t')
    \Big]
    =
    \im \, (2\pi)^3
    \delta(\vect{k} + \vect{k}')
    \delta (t-t')
    a(t)^{-2}
    . 
\end{equation}
Since the initial time $t_i$
will typically be much later than the horizon exit
time $t_p$ for $\vect{p}$,
we can take operators associated with the long mode to commute
at time $t_i$.
It follows that the power spectrum $P^{IJ}(p; t_i)$ will be symmetric.
Hence, we find
\begin{equation}
\label{eq: 13 simple}
    P_\zeta(p;t)_\text{13}
    =
    N_{I}^{(t_i,t)}
    N_{JKL}^{(t_i,t)}
    P^{IJ}(p;t_i)
    \int \d^3 q \; P^{KL}(q;t_i)
    .
\end{equation}
The $(22)$-type loop can be regarded as
an average over uncorrelated
noise on the scale $q^{-1}$~\cite{Iacconi:2023ggt}.
When averaged over a spacetime region of size $p^{-1}$
we obtain $\mathcal{N} \sim (q/p)^3$ independent samples
of this noise.
If they are uncorrelated,
the central limit theorem
implies that
the variance of their average is suppressed by
$1/\mathcal{N} \sim (p/q)^3$.
In the literature, this is described as
\emph{volume suppression}.
Meanwhile,
the $(12)$- and $(13)$-type loops~\eqref{eq: 12 type base expression}
and~\eqref{eq: 13 base expression}
incorporate
long--short couplings,
in the form of initial conditions on squeezed configurations.
As explained in Ref.~\cite{Iacconi:2023ggt},
these correlations invalidate
a na\"{\i}ve application of the
central limit theorem.
The result is that the $(12)$- and $(13)$-type loops need not
exhibit the same volume suppression.

In the remainder of this paper,
we discard the $(22)$-type diagram and focus on the
$(12)$- and $(13)$-type contributions.

\section{1-loop from non-linear superhorizon evolution is a boundary term}
\label{sec: 1-loop from non-linear superhorizon evolution is a boundary term}

In~\S\ref{sec: delta N 12 and 13 loops} we will derive
one of our main results.
Starting from
the $\delta N$ formula~\eqref{eq: zeta with backreaction Fourier}
for the long-wavelength
mode,
we show that the $(12)$- and $(13)$-type loops
can be unified into a single loop integral.
The integrand is a total derivative of a function
we are able to identify explicitly. 
We elaborate on the interpretation of our result
in {\S}\ref{sec: interpretation and discussion}.

\subsection{Combining the $(12)$- and $(13)$-type loops}
\label{sec: delta N 12 and 13 loops}
From Eqs.~\eqref{eq: 12 simple} and~\eqref{eq: 13 simple},
the sum of the $(12)$- and $(13)$-type loops is
\begin{multline}
    \label{eq:12 13 step 1}
    P_\zeta(p;t)_\text{12+13} = N_I^{(t_i,t)} N_{JK}^{(t_i,t)} \int \d^3 q \;
    \tensor{\alpha}{^I^J^K}(p,q,q;t_i) \\
    +  N_{I}^{(t_i,t)} N_{JKL}^{(t_i,t)} \int \d^3 q\;   P^{IJ}(p;t_i) P^{KL}(q;t_i) \;.
\end{multline}
The squeezed
bispectrum $\tensor{\alpha}{^I^J^K}(p,q,q;t_i)$ can be estimated by
soft-limit arguments~\cite{Maldacena:2002vr,Kenton:2015lxa,Kenton:2016abp};
see {\S}\ref{sec: properties of long and short modes}.  
At leading order in $p \ll q$,
Eq.~\eqref{eq: 3-point function squeezed}
and the definition~\eqref{eq: def 3pt phase space}
yield%
    \footnote{The rationale for
    Eq.~\eqref{eq:soft limit alpha at t_i}
    is fairly clear at horizon exit of $q$.
    One might be concerned that evolution between this
    time and $t_i$ could change the relationship, but in
    Appendix~\ref{app: squeezed bispectrum at t_k}
    we show this is not the case.}
\begin{equation}
    \label{eq:soft limit alpha at t_i}
    \tensor{\alpha}{^I^J^K}(p,q,q;t_i)
    \supseteq
    P^{IL}(p;t_i)
    \frac{\partial P^{JK}(q;t_i)}{\partial X^L(t_i)}
    .
\end{equation}
Substitution
of Eq.~\eqref{eq:soft limit alpha at t_i}
in Eq.~\eqref{eq:12 13 step 1} yields
\begin{equation}
    \label{eq:12 13 step 2}
    P_\zeta(p;t)_\text{12+13}
    =
    N_I^{(t_i,t)} P^{IL}(p;t_i)
    \int \d^3 q \;
    \Big[
        N_{JK}^{(t_i,t)}
        \frac{\partial P^{JK}(q;t_i)}{\partial X^L(t_i)}
        + 
        N_{LJK}^{(t_i,t)}
        P^{JK}(q;t_i)
    \Big]
    \;,
\end{equation}
where we have relabelled
summation
indices in the second term.
Note that the $\delta N$ coefficients
$\tensor{N}{_J}$,
$\tensor{N}{_J_K}$,
and
$\tensor{N}{_L_J_K}$
have no momentum dependence;
they carry information only about the background evolution,
and therefore commute with the integral. 
By rearranging the derivatives we obtain
\begin{equation}
\label{eq:12 13 step 3}
    P_\zeta(p;t)_\text{12+13}
    =
    N_I^{(t_i,t)}
    \int \d^3 q \;
    P^{IL}(p;t_i)
    \frac{\partial}{\partial X^L(t_i)}
    \Big[
        N_{JK}^{(t_i,t)}  P^{JK}(q;t_i)
    \Big]
    \; .
\end{equation}
On the other hand, the derivative
$\partial/\partial X^L(t_i)$ with respect to the initial
conditions at $t_i$
does \emph{not} commute with the integral,
even though the notation suggests that it is momentum
independent.
This is because, under a shift in its initial data,
the change in the quantity enclosed by square
brackets $[ \cdots ]$
can be $q$-dependent even if the shift in $X^L(t_i)$ is not.
The fact the 1-loop correction can be organized in
this form
shows clearly
that it measures
the correlation
between
the response of the short scale modes
to the long mode,
and the original long mode.
Moreover,
the response of the short modes
is expressed in terms of a shift in their effective background.
This is a consequence of our assumed separation of
scales, so that modes in the enhanced band have $q \gg p$.

The shift of $X^L(t_i)$
does not change the numerical value of $q$,
so we may freely move a factor $q^{-3}$
past the derivative.
This exchanges the short-scale power spectrum
$P^{JK}(q)$ 
for its dimensionless counterpart
$\dimlessP^{JK}(q)$,
\begin{equation}
    \label{eq:12 13 step 3b}
    P_\zeta(p;t)_\text{12+13}
    =
    N_I^{(t_i,t)}
    \int \d \ln q \;
    P^{IL}(p;t_i)
    \frac{\partial}{\partial X^L(t_i)}
    \Big[
        N_{JK}^{(t_i,t)}
        \dimlessP^{JK}(q;t_i)
    \Big]
    \;. 
\end{equation}
In what follows, we give a concrete calculation by
specializing to a single-field scenario in which the
phase-space indices $I$, $J$, \ldots, run over
the coordinates $\{ \phi, \pi \}$.
We expect effectively
the same analysis to apply whenever the
field configuration for the mode $\vect{p}$
is adiabatic,
but we leave an explicit demonstration for future work.

Writing out the sum over $L$ explicitly yields
\begin{equation}
\label{eq:12 13 step 4}
    P_\zeta(p;t)_\text{12+13}= N_I^{(t_i,t)}
    \int \d \ln q \; 
    \bigg(
        P^{I\phi}(p;t_i)
        \frac{\partial}{\partial \phi (t_i)}
        +
        P^{I\pi}(p;t_i)
        \frac{\partial}{\partial \pi (t_i)}
    \bigg)
    \Big[
        N_{JK}^{(t_i,t)}
        \dimlessP^{JK}(q;t_i)
    \Big]
    \;.
\end{equation}
The assumption of adiabaticity means that
fluctuations in the field and its momentum
are not independent.
Using 
Eqs.~\eqref{eq: delta pi and delta phi long mode}
and~\eqref{eq:12 13 step 4} allows us to write
\begin{equation}
\label{eq:12 13 step 5}
    P_\zeta(p;t)_\text{12+13}
    =
    N_I^{(t_i,t)}
    \int \d \ln q \;
    P^{I\phi}(p;t_i)
    \bigg(
        \frac{\partial}{\partial \phi (t_i)}
        +
        \frac{\epsilon_2}{2}\frac{\partial}{\partial \pi (t_i)}
    \bigg)
    \Big[
        N_{JK}^{(t_i,t)}
        \dimlessP^{JK}(q;t_i)
    \Big]
    . 
\end{equation}
This combination of partial derivatives is
merely the total derivative
with respect to $\phi$, evaluated along
the unperturbed trajectory;
see
Eq.~\eqref{eq: total derivative along unpert trajectory}.
Therefore,
\begin{equation}
    \label{eq:12 13 step 6.1}
    \dimlessP_\zeta(p;t)_\text{12+13}
    =
    \Big(
        N_\phi^{(t_i,t)}
        +
        \frac{\epsilon_2}{2}
        N_\pi^{(t_i,t)}
    \Big)
    \dimlessP^{\phi\phi}(p;t_i)
    \int \d \ln q \;
    \frac{\d}{\d\phi(t_i)}
    \Big[
        N_{JK}^{(t_i,t)}
        \dimlessP^{JK}(q;t_i)
    \Big]
    ,
\end{equation}
in which we have also expanded the sum over $I$.
To repeat,
the emergence of a single derivative
along the unperturbed trajectory
is
a consequence of adiabaticity
in the long-wavelength
field configuration.
The field and velocity variations combine to
produce a shift
along the unperturbed trajectory at $t_i$.
The (apparently) independent variations $\delta \phi$
and $\delta \pi$
are really
parametrized by a single variable, $\phi(t_i)$.

The prefactor of the integral
is proportional to the
linear power spectrum of the long mode $\vect{p}$
at the end of inflation, $\dimlessP_\zeta(p;t)_\text{tree}$.
This can be obtained using the separate universe approach,
with initial surface set
either soon after horizon crossing
for $\vect{p}$,
or at some later time,
e.g., during the non-attractor phase,
provided the initial conditions are chosen appropriately. 
In Ref.~\cite{Iacconi:2023ggt} we
computed $\dimlessP_\zeta(p;t)_\text{tree}$
in the case of ultra-slow-roll,
and explicitly showed that these two different choices
of initialization time led to equivalent results.
In particular, for an adiabatic configuration at $\vect{p}$
we have
\begin{equation}
\label{eq: Pzeta p tree}
    \dimlessP_\zeta(p;t)_\text{tree}
    =
    \Big(
        N_\phi^{(\tinit,t)}
        +
        \frac{\epsilon_2}{2}
        N_\pi^{(\tinit,t)}
    \Big)^2
    \dimlessP^{\phi \phi}(p;\tinit)_\text{tree}
    =
    \bigg(
        \frac{\d N^{(\tinit,t)}}{\d\phi(\tinit)}
    \bigg)^2
    \dimlessP^{\phi \phi}(p;\tinit)_\text{tree}
    ,
\end{equation}
where $\tinit \gtrsim t_p $
is an arbitrary
initialization time.
Substituting Eq.~\eqref{eq: Pzeta p tree}
in Eq.~\eqref{eq:12 13 step 6.1} with
$\tinit = t_i$,
and defining the relative
correction
$\Delta \dimlessP \equiv \dimlessP_{\text{12}+\text{13}} / \dimlessP_{\text{tree}}$,
we obtain
\begin{equation}
    \label{eq:Delta 12 13 step 1}
    \Delta \dimlessP_\zeta(p;t)_\text{12+13}
    \equiv
    \frac{\dimlessP_\zeta(p;t)_\text{12+13}}{\dimlessP_\zeta(p;t)_\text{tree}}
    =
    \bigg(
        \frac{\d N^{(t_i,t)}}{\d\phi(t_i)}
    \bigg)^{-1}
    \int \d \ln q \;
    \frac{\d}{\d\phi(t_i)}
    \Big[
        N_{JK}^{(t_i,t)}
        \dimlessP^{JK}(q;t_i)
    \Big]
    .
\end{equation}

We must now evaluate the change in the bracket $[ \cdots ]$
induced by a shift in its initial data.
In order to assign wavenumbers to inflationary
perturbations,
and therefore define the spectrum
$\dimlessP^{JK}(q)$,
we must wait until inflation has ended,
taken to occur on a uniform energy hypersurface at time $\tend$.
The horizon scale at this time defines some physical scale,
associated with comoving wavenumber $\qend = (a H)_{\tend}$.
Looking backwards from this hypersurface,
we assign wavenumbers to the perturbations generated
during inflation
by the rule
\begin{equation}
    \label{eq:lookback-equation}
    \ln \frac{q}{\qend}
    =
    - \Nlookback
    + \Or(\epsilon) ,
\end{equation}
where $\Nlookback > 0$ is the lookback time
to the horizon exit point of the corresponding
fluctuation.

At time $t_i$, when the background configuration is
shifted in Eq.~\eqref{eq:Delta 12 13 step 1},
the band of enhanced modes
contributing to the peak has already been generated.
The change induced by
$\d/\d \phi(t_i)$ therefore corresponds to insertion or
removal
of a short segment of the trajectory, 
which changes the lookback time.
It follows that the change in the bracket
$[ \ldots ]$,
for the modes of interest,
merely amounts to a relabelling of wavenumbers.

A change $\delta \phi(t_i)$ in the field configuration
produces a change $-\delta\phi(t_i) / \phi'(t_i)$
in the duration of inflation.
If this is positive (corresponding to insertion
of a small trajectory segment),
then the required relabelling must shift the assignment
of perturbations
to smaller wavenumbers.
Therefore
the necessary relabelling is
\begin{equation}
    \delta \ln q
    = 
    \frac{\delta q}{q}
    =
    \frac{\delta \phi(t_i)}{\phi'(t_i)}
    +
    \Or(\delta \phi)^2 ,
\end{equation}
where $\phi ' = \d \phi / \d N$
represents the field derivative with respect to the number
of e-folds.
As explained above, this relabelling is $q$-dependent,
even if the shift $\delta \phi(t_i)$ is $q$-independent.
It follows that
$\d/\d \phi(t_i)$ in Eq.~\eqref{eq:Delta 12 13 step 1}
can be recast into a derivative with respect to $\ln q$,
\begin{equation}
    \label{eq: operational expression for derivative wrt phi(ti) step3}
    \frac{\d}{\d\phi(t_i)}
    \Big[
        N_{JK}^{(t_i,t)}
        \dimlessP^{JK}(q;t_i)
    \Big]
    =
    \frac{1}{\phi'(t_i)}
    \frac{\d}{\d\ln q}
    \Big[
        N_{JK}^{(t_i,t)}
        \dimlessP^{JK}\left(q;t_i\right)
    \Big]
    . 
\end{equation}
The result is that we can rewrite Eq.~\eqref{eq:Delta 12 13 step 1} as 
\begin{equation}
    \label{eq:Delta 12 13 step 3}
    \Delta \dimlessP_\zeta(p;t)_\text{12+13}
    =
    \int_{\qmin}^{\qmax} \d \ln q \;
    \frac{\d }{\d\ln q}
    \Big[
        N_{JK}^{(t_i,t)}
        \dimlessP^{JK}(q;t_i)_\text{tree}
    \Big]
    , 
\end{equation}
where we have marked explicitly the limits of
integration and reintroduced the ``tree'' label for the power spectrum.

Eq.~\eqref{eq:Delta 12 13 step 3} represents
one of our main results. 
Before discussing its consequences,
let us comment on what we believe is
interesting about Eq.~\eqref{eq:Delta 12 13 step 3}. 
The fundamental theorem of calculus relates
integration to differentiation,
in the sense that if $\int f(x) \, \d x = F(x)$,
then $\d F(x) / \d x = f(x)$.
Therefore it is hardly remarkable that the integral in
Eq.~\eqref{eq:Delta 12 13 step 3}
can be expressed as the antiderivative of \emph{some}
function $F(x)$.
The useful feature
is not the emergence of a total derivative \emph{in itself},
but rather that we are able to identify
this antiderivative explicitly.
In particular, it is simply the
combination $N_{JK} \dimlessP^{JK}(q)$.

\subsection{Interpretation and discussion}
\label{sec: interpretation and discussion}

To interpret
Eq.~\eqref{eq:Delta 12 13 step 3} it will be
helpful to refer to explicit examples of single-field models
leading to enhanced short-scale fluctuations.
As we have explained,
this is typically achieved by inclusion of a transient
non-attractor (``NA'') phase,
during which the decaying mode of $\zeta$ is promoted to a
rapidly growing mode on superhorizon scales.

Dynamics of this kind
constitutes a sub-class of constant-roll (``CR'')
inflation~\cite{Motohashi:2025qgd}.
During constant roll, the Klein--Gordon equation is
$\ddot \phi /(H \dot \phi) \approx \beta$,
with constant $\beta$,
and $\epsilon_2 \approx 2\beta$.  
When $\beta < -3/2$ $(\epsilon_2< -3)$
there is no attractor in phase space. 
In this regime, the curvature perturbation evolves
according to
\begin{equation}
    \zeta(k, N)_{k\ll a H}
    =
    c_1
    +
    c_2
    \int^N \d N' \,
    \exp
    \Big(
        {-\int^{N'}} \d N'' \; (3-\epsilon_1+\epsilon_2)
    \Big)
    .
\end{equation}
For $\epsilon_1\ll 1$ and $\epsilon_2< -3$, $\zeta$
displays exponential growth.
For most potentials
that have already been studied,
the transient non-attractor constant-roll
dynamics are characterized by
$\epsilon_2\leq -6$, with $\epsilon_2=-6$
(equivalently $\beta=-3$) being ultra-slow-roll.

Note that for $\beta > -3/2$ $(\epsilon_2>-3)$, the constant-roll model
has attractor behaviour,
and $\zeta$ is frozen on superhorizon scales.
For example, ordinary slow-roll corresponds to $\beta = 0$. 
To avoid confusion,
in the following we will always specify the value of
$\epsilon_2$ when we label a phase of dynamics as constant-roll.

After the non-attractor constant-roll phase,
we assume the background evolves smoothly into
a subsequent period, which we label ``A''.
Although other options are possible,
we take this to be the Wands dual of the
non-attractor
phase~\cite{Wands:1998yp, Motohashi:2014ppa, Atal:2018neu}.
This
is again a constant-roll phase,
now characterized by a dynamical attractor in phase-space,
which quenches the growth of perturbations.
Eventually, inflation ends.
The second slow-roll parameter in the
``A''
phase is $\epsilon_{2,\text{A}} = -6 -\epsilon_{2,\text{NA}}$,
where $\epsilon_{2,\text{NA}}$ is the value
of $\epsilon_2$ in the non-attractor phase. 
 
We select two examples of typical single-field models
leading to enhanced fluctuations: a ultra-slow-roll
model with $\epsilon_{2,\text{NA}}=-6$,
and a constant-roll model with $\epsilon_{2,\text{NA}}=-7$.
These are representative of generic non-attractor constant-roll models. 
The critical difference between these models
is the value of $\epsilon_{2}$ in the subsequent
dual phase.
For ultra-slow-roll,
$\epsilon_{2,\text{A}}\approx 0$, while for constant-roll with $\epsilon_{2,\text{NA}}=-7$ one has
$\epsilon_{2,\text{A}}\approx 1$.

\begin{figure}
    \centering
    \captionsetup[subfigure]{justification=centering}
        \begin{subfigure}{.48\textwidth}
            \includegraphics[width=\textwidth]{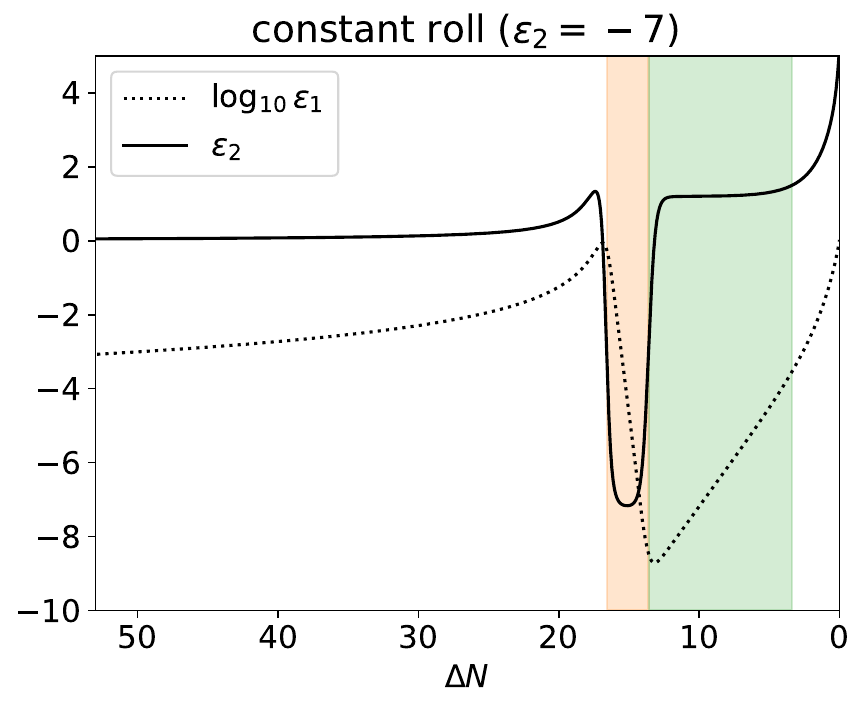}
        \end{subfigure}
        \begin{subfigure}{.48\textwidth}
            \includegraphics[width=\textwidth]{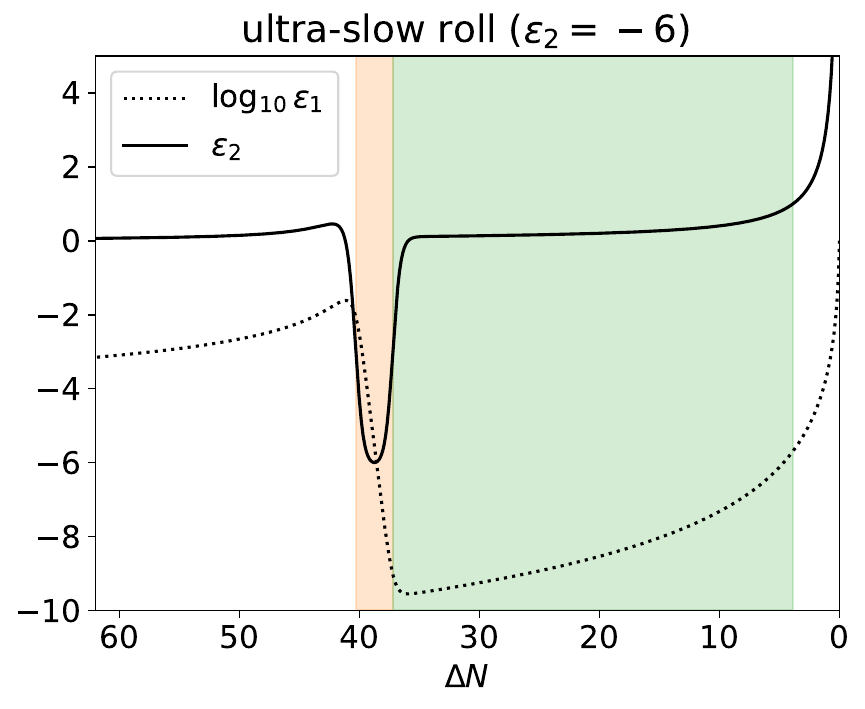}
        \end{subfigure}
    \caption{
        Time-evolution of $\epsilon_1$ and $\epsilon_2$
        for two models leading to amplified fluctuations
        on small scales; see main text for details.  
        In both panels the horizontal axis is
        $\Delta N \equiv N-N_\text{end}$,
        where $N_\text{end}$ labels the end of inflation. 
        We highlight times during the non-attractor phase
        $(\epsilon_2<-3)$ in orange,
        and the following dual attractor phase in green. 
    }
    \label{fig: background examples} 
\end{figure}

\begin{figure}
    \centering
    \captionsetup[subfigure]{justification=centering}
        \begin{subfigure}{.47\textwidth}
            \includegraphics[width=\textwidth]{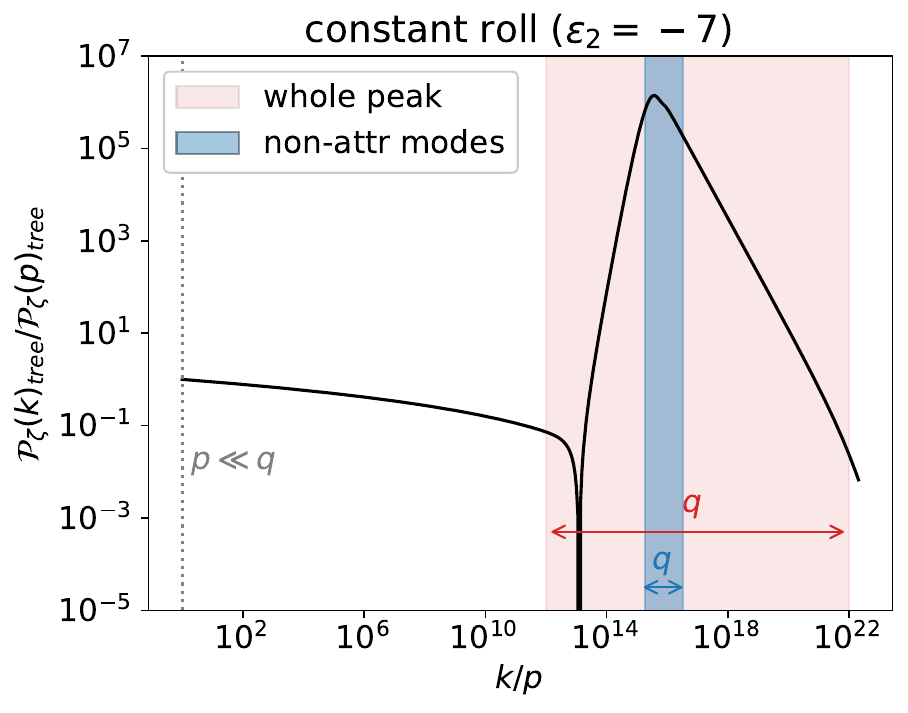}
        \end{subfigure}
        \begin{subfigure}{.49\textwidth}
            \includegraphics[width=\textwidth]{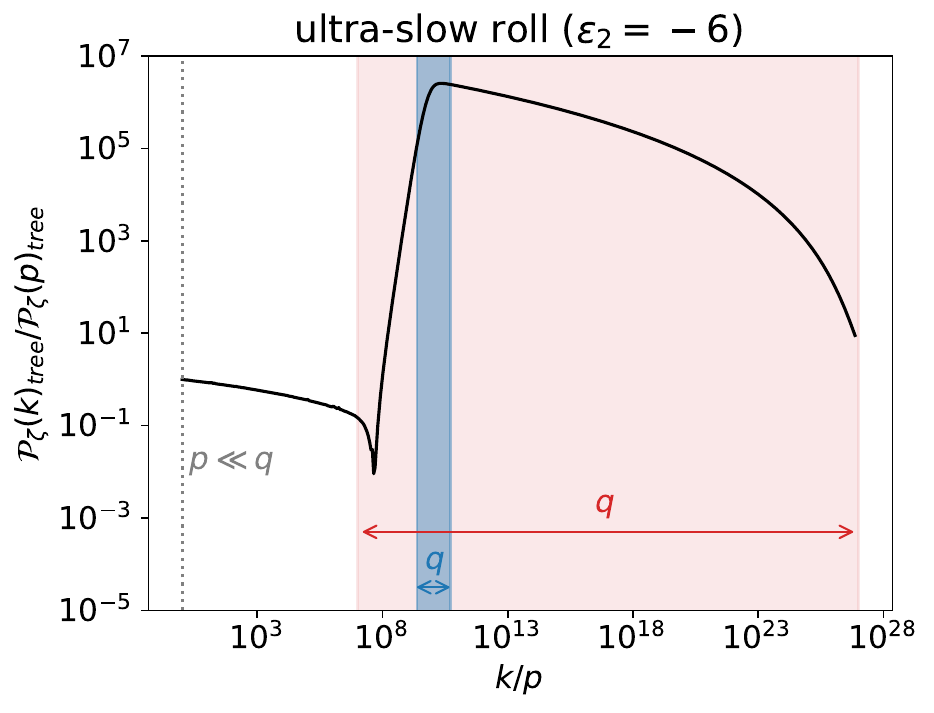}
        \end{subfigure}
    \caption{
        Tree-level scalar power spectrum
        ${\dimlessP_\zeta(k)}_\text{tree}$,
        for the models of Fig.~\ref{fig: background examples};
        see main text for details. 
        In both panels, the vertical dotted line identifies the long
        adiabatic mode $p$, for which we wish to estimate
        the back-reaction. 
        Modes highlighted in blue cross the horizon during the non-attractor
        CR phase, defined by the condition $\epsilon_2<-3$,
        as in Fig.~\ref{fig: background examples}. 
        We label the first and last scale of the blue band as
        $k_s$ and $k_e$, respectively. 
        We highlight the broad peak in red.
        }
    \label{fig:integral boundaries} 
    \end{figure}

To obtain numerical results
we have
used \texttt{PyTransport}~\cite{Costantini:2025tek}
to
implement
the ultra-slow-roll model
presented by Germani \& Prokopec~\cite{Germani:2017bcs},
which was
adapted from Garc\'{\i}a-Bellido \&
Ruiz Morales~\cite{Garcia-Bellido:2017mdw}.
We have also implemented the constant-roll model
introduced by Cicoli {\etal}~\cite{Cicoli:2018asa}. 
For both models we make the same parameter choices
used by Cole~{\etal}~\cite{Cole:2023wyx}.
In Fig.~\ref{fig: background examples}
we show
the time-evolution of the first two slow-roll parameters
for each model,
and
in Fig.~\ref{fig:integral boundaries} we show
the corresponding tree-level scalar power spectrum
computed at the end of inflation.

\para{Only the integral boundaries contribute}
Formally, one can use Eq.~\eqref{eq:Delta 12 13 step 3}
to estimate the contribution to the 1-loop
correction from the band of enhanced modes.
This yields
\begin{equation}
\label{eq: evaluation at boundaries}
    \Delta \dimlessP_\zeta(p;t)_\text{12+13}
    =
    N_{JK}^{(t_i,t)}
    \dimlessP^{JK}(q;t_i)_\text{tree}
    \Big|_{q=\qmax}
    -
    N_{JK}^{(t_i,t)}
    \dimlessP^{JK}(q;t_i)_\text{tree}
    \Big|_{q=\qmin}
    . 
\end{equation}
We conclude that the
contribution from this band
can be expressed in terms
of the phase-space power spectrum evaluated at its
boundaries,
$\qmin$ and $\qmax$.
The loop correction receives contributions
from all modes with $\qmin \leq q \leq \qmax$,
but it decouples from the precise details of their behaviour.
In a full calculation,
Eq.~\eqref{eq: evaluation at boundaries}
would be accompanied by
integrals over the
infrared and ultraviolet
regions that would subtract
any dependence on $\qmin$
and $\qmax$.
In this sense these limits
are arbitrary,
but we can exploit this arbitrariness
for the purpose of making an estimate.

As explained in~\S\ref{sec: separate universe},
because our calculation is performed within the
separate universe framework,
we must choose the
scale $\qmax$ 
to be smaller than the inverse
of the separate universe smoothing scale. 
Meanwhile,
the scale $\qmin$ 
can be any wavenumber sufficiently
larger than $p$,
i.e., $\qmin \gg p$.
One possible choice is the first scale that crossed the horizon
during the non-attractor phase,
which we label $k_s$
(see Fig.~\ref{fig:integral boundaries}).
To obtain the best estimate,
in the sense of minimizing large subtractions
from adjoining regions,
we should choose
$\qmin$ and $\qmax$ to bracket the whole enhanced band,
represented by the red-shaded
region in Fig.~\ref{fig:integral boundaries}.
In particular, we must choose the initial time $t_i$
of~\S\ref{sec: separate universe}
substantially
later than the
transition into the final slow-roll era,
represented by the right-hand edge of the
orange-shaded region in Fig.~\ref{fig: background examples}.
We have already seen that
this raises the possibility
of a substantial loop correction to the (11) initial condition at $t_i$.

\para{The case of a transient ultra-slow-roll phase}
In some models, further simplifications are possible.
For the choices of initial time discussed above,
the $q$-modes included in the loop
integral are on
superhorizon scales.
Further, because $t_i$ lies during an attractor phase,
$\zeta_{\vect{q}}$ is constant, and therefore
\begin{equation}
\label{eq: delta phi and delta pi at t_i}
    \delta \phi_{\vect{q}}
    \propto
    \sqrt{\epsilon_1}
    \quad \text{and} \quad
    \delta \pi_{\vect{q}}
    =
    \frac{\epsilon_2}{2}
    \delta \phi_{\vect{q}}
    +
    \text{decaying}
    . 
\end{equation}
The relative importance of field and velocity fluctuations
at $t_i$ is determined by the second slow-roll parameter
$\epsilon_2(t_i)$.
Reference to the panels of
Fig.~\ref{fig: background examples}
shows that,
if the non-attractor
phase is of ultra-slow-roll type,
there is a period
where $\epsilon_2(t_i) \approx 0$.
(In the right-hand panel
of Fig.~\ref{fig: background examples}
this matches most of the
green region, exluding its right-hand edge
where $\epsilon_2$ is growing.)
On the other hand, the left-hand
panel of Fig.~\ref{fig: background examples}
shows that if the non-attractor
phase is of constant-roll type,
there is nowhere to locate
$t_i$ within the green region where
$\epsilon_2(t_i)$ is negligible.

If it is possible to find
a $t_i$ for which
$\epsilon_2(t_i)$ is sufficiently
small,
but where
$\qmax$ can still be chosen
to bracket most of the
enhanced band,
then
it is possible to simplify
Eq.~\eqref{eq: evaluation at boundaries}
by including only field fluctuations.
In the right-hand panel of
Fig.~\ref{fig: background examples}
this would correspond to choosing $t_i$
somewhere near the middle of the green region.
One could then not choose $\qmax$
to cover the whole band of enhanced
modes
(see the corresponding panel of
Fig.~\ref{fig: background examples}).
However, $\dimlessP_\zeta(\qmax, t_i)$
would still be significantly
smaller than its value at the peak.
One would then have to accept the possibility
of some cancellation with the ultraviolet
region, but presumably this not
an $\Or(1)$ effect.
Accepting these limitations and
setting $J = K = \phi$, we obtain
\begin{equation}
    \label{eq:Delta 12 13 USR step 1}
    \Delta \dimlessP_\zeta(p;t)_\text{12+13}
    =
    \int_{\qmin}^{\qmax} \d \ln q \;
    \frac{\d }{\d\ln q}
    \Big[
        N_{\phi \phi }^{(t_i,t)}
        \dimlessP^{\phi \phi}(q;t_i)_\text{tree}
    \Big]
    +
    \Or[\epsilon_2(t_i)]
    .
\end{equation}
Where the $\epsilon_2(t_i)$ term can be neglected,
it is possible to give a simple
interpretation of the remaining integral.
Since this seems mostly possible in
ultra-slow-roll type scenarios,
we will indicate this simplified version of
Eq.~\eqref{eq:Delta 12 13 step 3} with a ``USR'' label. 

After multiplying and dividing the function inside
square brackets $[ \cdots ]$
by
$[ N_{\phi}^{(t_i,t)} ]^2$,
it is possible to reconstruct the amplitude of the local
bispectrum generated from non-linear evolution
(which we label $\fNL^{N_{\phi\phi}}(k_i; t)$),
and the short-scale scalar power spectrum,
$\dimlessP_\zeta(q;t)_\text{tree}$. 
In particular,  
\begin{equation}
\begin{split}
    \label{eq:Delta 12 13 USR step 2}
    \Delta \dimlessP_\zeta(p;t)_\text{12+13, USR}
    & \approx
    \int_{\qmin}^{\qmax} \d \ln q \;
    \frac{\d}{\d\ln q}
    \Bigg[ 
        \underbrace{
            \frac{N_{\phi\phi}^{(t_i,t)}}{{N_{\phi}^{(t_i,t)}}^2}
        }_{6 \fNL^{N_{\phi\phi}}(k_i; t) / 5} 
        \underbrace{
            {N_{\phi}^{(t_i,t)}}^2
            \dimlessP^{\phi\phi}(q;t_i)_\text{tree}
        }_{\dimlessP_\zeta(q;t)_\text{tree}}
    \Bigg]
    \\[1ex]
    & =
    \frac{6}{5}
    \int_{\qmin}^{\qmax} \d \ln q \;
    \frac{\d}{\d\ln q}
    \Bigg[ 
        \fNL^{N_{\phi\phi}}(k_i; t)
        \dimlessP_\zeta(q;t)_\text{tree}
        \Bigg]
    ,
\end{split}
\end{equation}
where the $\Or(\epsilon_2)$ contribution has been dropped.
Here, $k_i$ is the scale that crossed the horizon at time $t_i$.
This yields a simple estimate
in terms of $\fNL^{N_{\phi\phi}}$
and $\dimlessP_\zeta$,
\begin{equation}
    \label{eq:Delta 12 13 USR step 4}
    \Delta \dimlessP_\zeta(p;t)_\text{12+13,USR}
    \approx
    \frac{6}{5} \fNL^{N_{\phi\phi}}(k_i; t)
    \Big[
        \dimlessP_\zeta(\qmax;t)_\text{tree}
        -
        \dimlessP_\zeta(\qmin;t)_\text{tree}
    \Big]
    .
\end{equation}
If $\qmin$ and $\qmax$ adequately bracket
the enhanced band, so that
$\dimlessP_\zeta(\qmax; t)_{\text{tree}}$
and
$\dimlessP_\zeta(\qmin; t)_{\text{tree}}$
are both substantially smaller than the peak
amplitude
(and we assume there is no large coupling with the
ultraviolet portion of the integral)
this represents a significant limitation
on the importance of the 1-loop backreaction.

We close this discussion of the
ultra-slow-roll case with a
comment on the $(13)$-type loop.  
Applying the
analysis of {\S}\ref{sec: delta N 12 and 13 loops} to
the $(12)$-type loop only, one obtains
\begin{equation}
    \label{eq: 12 loop step 1}
    \Delta \dimlessP_\zeta(p;t)_\text{12,USR}
    =
    N_{\phi\phi}^{(t_i,t)}
    \int \d \ln q \;
    \frac{\d}{\d\ln q}
    \Big[
        \dimlessP^{\phi \phi}(q;t_i)
    \Big]
    +
    \Or[\epsilon_2(t_i)]
    ,
\end{equation}
where again we have used~\eqref{eq: delta phi and delta pi at t_i}
to neglect the contribution of momentum
fluctuations.
Now
multiplying
and dividing
by a factor
${N_\phi^{(t_i,t)}}^2$, we obtain 
\begin{equation}
    \label{eq: 12 loop step 2}
    \Delta \dimlessP_\zeta(p;t)_\text{12,USR}
    = 
    \frac{6}{5}
    \fNL^{N_{\phi\phi}}(k_i; t)
    \int_{\qmin}^{\qmax} \d \ln q \;
    \frac{\d\dimlessP_\zeta(q;t)_\text{tree}}{\d\ln q}
    +
    \Or[\epsilon_2(t_i)]
    . 
\end{equation}
Comparing Eqs.~\eqref{eq:Delta 12 13 USR step 4}
and~\eqref{eq: 12 loop step 2} shows that,
where the $\epsilon_2(t_i)$
contribution can be neglected,
the $(13)$-type loop itself
can be at most of
$\Or[\epsilon_2(t_i)]$.
Notice
that this does not at all imply the
contribution of every
1-loop diagram with 13 topology
is always of this order;
in the language of non-equilibrium
quantum field theory,
these correspond to 1-loop diagrams mediated by one quartic interaction. 
Indeed, as we shall see in
{\S}\ref{sec: 1-loop from quantum initial conditions is a boundary term},
the (11)-type loop in Eq.~\eqref{eq: total 1-loop correction}
includes a diagram with this topology, which does not need to be small.

\para{Summary}
What should be concluded from
Eqs.~\eqref{eq: evaluation at boundaries},
\eqref{eq:Delta 12 13 USR step 4}
and~\eqref{eq: 12 loop step 2}?

First, as explained above,
the full 1-loop correction is
independent of the limits $\qmin$ and $\qmax$,
which must be compensated by infrared and ultraviolet contributions that
we have not written explicitly.
To obtain the best estimate from~\eqref{eq: evaluation at boundaries},
\eqref{eq:Delta 12 13 USR step 4}
and~\eqref{eq: 12 loop step 2}
we should choose $\qmin$ and $\qmax$ to minimize cancellations
between these different contributions.
As we have explained,
this means that $\qmin$ and $\qmax$ should bracket the
whole enhanced band.
With these choices,
and assuming the renormalization scale is
at least modestly
larger than $\qmax$,
we should not expect large cancellations
between the ultraviolet
contribution
and the estimate~\eqref{eq: evaluation at boundaries}.

\begin{figure}
    \centering
    \captionsetup[subfigure]{justification=centering}
        \begin{subfigure}{.48\textwidth}
            \includegraphics[width=\textwidth]{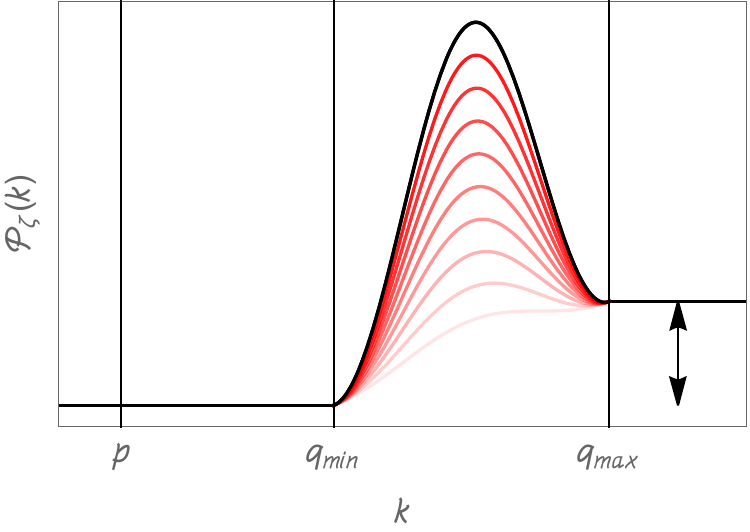}
        \end{subfigure}
        \begin{subfigure}{.48\textwidth}
            \includegraphics[width=\textwidth]{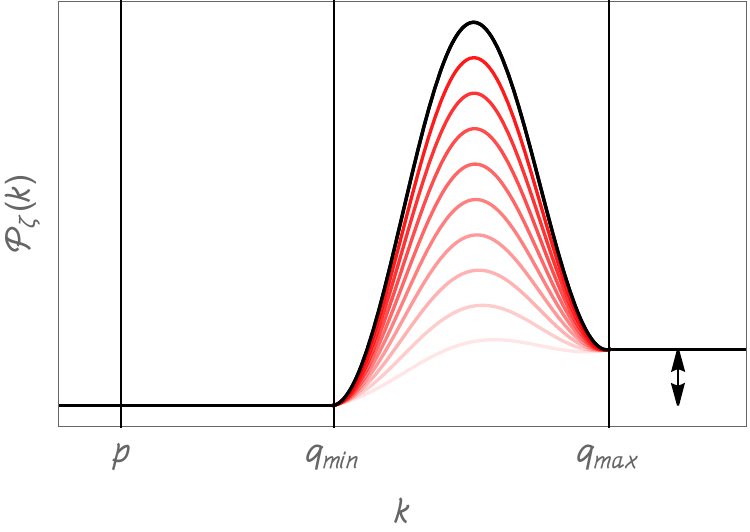}
        \end{subfigure}
    \caption{
    Schematic illustration of tree-level power spectra
    displaying a peak on short scales.  
    These are not computed from realistic inflationary models
    (such as those in Fig.~\ref{fig:integral boundaries}),
    but rather are toy examples.
    The vertical, black arrow indicates the (model-dependent) quantity $\dimlessP_\zeta(\qmax;t)_\text{tree} - \dimlessP_\zeta(\qmin;t)_\text{tree}$.
    In each panel, a reference spectrum is plotted in black. 
    The red lines represent examples of other spectra with the same
    infrared
    and
    ultraviolet
    plateaus as the reference one, but with different amplitude of the peak.
    }
    \label{fig:illustration of results}
\end{figure}

Second, given these choices for $\qmin$
and $\qmax$, the major conclusion is that
the amplitude of the 1-loop back-reaction decouples from
all detailed properties of the central peak,
including its maximum amplitude and width.
See Fig.~\ref{fig:illustration of results}.
It is not yet clear what is the spacetime picture of this
decoupling, analogous to the picture
of independent samples of $q^{-1}$-scale noise in 
a $p^{-1}$-scale region that leads to volume suppression of the (22)-type
loop from the central limit theorem.
For the case of
(12)- and (13)-type diagrams, subtle
correlations between the response of the short-scale fluctuations for
different $q$ combine to suppress the net backreaction.
It would be interesting to understand this effect in terms
of explicit realizations of the density field
produced by such a power spectrum.
Loosely speaking however, the long-wavelength description
of the averaged short-scale mode would appear as an effective isocurvature
mode.
If the wavenumber-$p$ field configuration is controlled by a theory
that does not contain such a mode, it would not seem possible for the
short-scale structure to excite one.

We note that there are longstanding results due to
Senatore \& Zaldarriaga~\cite{Senatore:2012ya}
and
Assassi~{\etal}~\cite{Assassi:2012et}
showing that the correlation functions
of $\zeta$ are constant to all orders in the loop expansion
when inflation is of the single-clock variety.
(See Senatore \& Zaldarriaga~\cite{Senatore:2009cf,Senatore:2012nq},
and Pimentel~\etal~\cite{Pimentel:2012tw} for earlier related work.)
Their results essentially amount to a demonstration that
$\d \hat{\zeta}_{\vect{k}} / \d t \rightarrow 0$ on superhorizon
scales as an operator statement.
Where these results apply, they prevent evolution
of correlation functions when later substructure emerges
from the horizon, and therefore prohibit back-reaction.

These results do not immediately apply to the scenario considered in
this paper, for which the single-clock property does not have to
hold during the non-attractor stage:
we only have this property during the early and late attractor epochs.
However, the details of our argument do not depend on the middle
epoch being of non-attractor type,
and they would continue to work if the enhanced
band were produced by a single-clock scenario.
(However, if this is the case,
it is difficult to produce an enhancement of the kind
required by PBH formation scenarios.)
If the enhanced band is produced by a single clock phase,
one can interpret our analysis as showing how the separate
universe framework is compatible
with the results of Senatore \& Zaldarriaga and Assassi~{\etal}
If they are produced by a more general phase,
our analysis is a generalization showing that adiabaticity
(effectively the same as the single clock property)
of the long-wavelength mode is sufficient, at least to 1-loop,
irrespective of the short-scale substructure.
In an effective description of the long-wavelength $\zeta$ field,
it is attractive to interpret this result as a consequence
of having no isocurvature modes that can be excited by the averaging
procedure.

Third, the integral over the enhanced band
does depend on the model, see the black arrow in the left and right panels of 
Fig.~\ref{fig:illustration of results}.
Nevertheless, 
the entire effect is
independent of the long-mode
wavenumber $p$,
as already noticed in Ref.~\cite{Iacconi:2023ggt}.
Eq.~\eqref{eq: evaluation at boundaries}
therefore represents a scale-invariant
enhancement of the power spectrum amplitude,
and it is not clear that the effect is observable. 
This model-dependence therefore does not
translate into a constraint on individual scenarios.

\section{The (11)-type loop is a boundary term}
\label{sec: 1-loop from quantum initial conditions is a boundary term}

We now return to the possibility of a significant loop correction to the $(11)$
initial condition.
In order to compute it, we must be able to express the fields at $t_i$ in terms of initial fields which have no back-reaction. 
This can be achieved by using an expansion method anchored before the transition into the non-attractor phase has taken place. 
The separate universe framework
does not provide an appropriate method,
because it would not allow the inclusion of all enhanced modes in the loop integrals.
To discuss the (11)-type loop, we
therefore consider the 1-loop computation in the context of a more general
framework that does not rely on validity of the separate universe picture.

We continue to assume the properties
of long and short modes
described in~{\S}\ref{sec: properties of long and short modes}.
The discussion in~\S\ref{sec: does separate universe compare with in-in}
shows that,
in the full framework of
non-equilibrium field thoery,
we can model time
evolution of the perturbations
using the OPE-like expansion~\eqref{eq:1loop-MPP-ansatz}.%
    \footnote{The discussion
    in~\S\ref{sec: does separate universe compare with in-in}
    shows that this is true up to 1-loop,
    which is all that is required for our present
    purposes. However, we expect that a similar
    statement can be made to all orders in the loop
    expansion.}
In Constantini {\etal}~\cite{Costantini:2025tek},
the $\Gamma$-tensors
appearing as Wilson coefficients
in this expansion
were described as ``multi-point propagators'',
following the terminology of Bernardeau {\etal}~\cite{Bernardeau:2008fa}.
As explained in~\S\ref{sec: does separate universe compare with in-in},
Eq.~\eqref{eq:1loop-MPP-ansatz}
is neither an approximation nor a model;
it is equivalent to the full ``in--in'' formalism at 1-loop,
provided all momentum-dependent effects are retained
in the $\Gamma$ coefficients,
and the initial correlations are computed appropriately.
It is therefore substantially
more flexible than the separate universe framework,
and in particular allows an arbitrary
initial time,
labelled $\tstar$ in Eq.~\eqref{eq:1loop-MPP-ansatz}.
The price to be paid for this flexibility is
a substantially more complex form of the
Wilson coefficients, represented by the multi-point propagators.
These propagators
are scale-dependent objects and their interpretation
is less straightforward than that of $\delta N$ coefficients.%
    \footnote{A point of terminology:
    in a conventional OPE, the Wilson coefficients
    capture short-distance effects,
    whereas the operators in the expansion
    capture long-distance effects.
    In Eq.~\eqref{eq:1loop-MPP-ansatz}
    we are allowing the $\Gamma$ coefficients
    to stand in for operators with higher-derivatives.
    To obtain an OPE in the usual form we should
    expand each $\Gamma$ as a Taylor series in its
    long-wavelength momenta.
    The coefficients of these series are the usual
    Wilson coefficients.}

In {\S}\ref{sec: 11-type loop} we show how
multi-point
propagators
allow us to complete the $\delta N$ 1-loop
computation of
{\S}\ref{sec: loops in the separate universe framework},
at least formally.
Specifically, we use them
to determine some properties
of
the initial condition for the (11)-type loop.
This is sourced by contributions accumulated by the 2-point
correlation functions of the long-mode
from the onset of the non-attractor phase up to the
initialization time $t_i$; see Eq.~\eqref{eq: 11 loop master}. 
We find that the integrand of the (11)-type loop
can be expressed as a total derivative of the tree-level
short-scale phase-space power spectrum contracted with a 3-index
multi-point propagator. 
In {\S}\ref{sec: lesson from completing the delta N loop computation} and {\S}\ref{sec:renormalized-deltaN}
we discuss the implications of this result
for 1-loop back-reaction.

\subsection{The (11)-type loop}
\label{sec: 11-type loop}

The initial condition
required for the (11)-type loop
in Eq.~\eqref{eq: 11 loop master} is the 2-point
correlation function at 1-loop, evaluated at $t_i$.
Below, we explain how this can be computed using the
multi-point propagator method. 

\para{Multi-point propagators}
Multi-point propagators were defined
at the 2-point level
by Crocce \& Scoccimarro
(although not named as such)
in the context of renormalized
cosmological perturbation theory~\cite{Crocce:2005xy}.
The generalization to all $n$,
and the term \emph{multi-point propagator},
were introduced by
Bernardeau, Crocce \& Scoccimarro~\cite{Bernardeau:2008fa}.
Applied to a classical field
$X^I$
they can be regarded as
momentum-dependent generalizations of the
separate universe
coefficients,
and measure the correlation
between some non-linear, evolved field
and its values at early times.
Taking $\tstar < t$, the first three
multi-point propagators would be defined by
\begin{equation}
\begin{split}
\label{eq: Gamma def}
    \frac{
        \partial X^I_{\vect{k}_1}(t)
    }{
        \partial X^J_{\vect{k}_2}(\tstar)
    }
    & \equiv
    \tensor{\Gamma}{^I_J}(k_1)^{(\tstar,t)} \;
    \delta^{(3)}( \vect{k}_1-\vect{k}_2 ) \;, \\
    \frac{
        \partial^2 X^I_{\vect{k}_1}(t)
    }{
        \partial X^J_{\vect{k}_2}(\tstar)
        \partial X^K_{\vect{k}_3}(\tstar)
    }
    & \equiv
    \tensor{\Gamma}{^I_J_K}(k_1, k_2, k_3)^{(\tstar,t)}\;
    \delta^{(3)} ( \vect{k}_1 - \vect{k}_2 - \vect{k}_3 ) \;, \\
    \frac{
        \partial^3 X^I_{\vect{k}_1}(t)
    }{
        \partial X^J_{\vect{k}_2}(\tstar)
        \partial X^K_{\vect{k}_3}(\tstar)
        \partial X^L_{\vect{k}_4}(\tstar)
    }
    & \equiv
    \tensor{\Gamma}{^I_J_K_L}(\vect{k}_1, \vect{k}_2,
        \vect{k}_3, \vect{k}_4)^{(\tstar,t)}\;
    \delta^{(3)} ( \vect{k}_1 - \vect{k}_2 - \vect{k}_3-\vect{k}_4 )
    ,
\end{split}
\end{equation}
where $\tstar$ is an arbitrary initial time,
as in Eq.~\eqref{eq:1loop-MPP-ansatz}.
When applied to a quantum field
$\hat{X}^I$,
the notion of multi-point propagators
as computable from derivatives
would be lost,
as for ordinary Wilson coefficients,
and they would have to be obtained
in some other way.
We will see below that they obey
evolution equations obtained from
the Heisenberg equation of motion for
$\hat{X}^I$, as explained
in~\S\ref{sec: does separate universe compare with in-in}.
In Ref.~\cite{Seery:2012vj}
the multi-point propagator expansion was introduced as a formal
solution for the transport equations
of inflationary
correlations~\cite{Mulryne:2009kh,Mulryne:2010rp,Mulryne:2013uka,
Dias:2016rjq,Anderson:2012em,Costantini:2025tek}.

Whether applied to a classical or quantum
field,
homogeneity and isotropy
implies that
$\tensor{\Gamma}{^I_J}$
and
$\tensor{\Gamma}{^I_J_K}$
depend only on the magnitude of their momentum labels.
Each
multi-point propagator
is a bitensor,
carrying a single momentum label (the first label)
for the Fourier mode
of the late-time field,
followed by those corresponding
to early-time phase-space variables. 
For the 2-index
multi-point propagator
$\tensor{\Gamma}{^I_J}$,
the $\delta$-function forces these momenta to have the same magnitude.
(Therefore the 2-index propagator
$\tensor{\Gamma}{^I_J}(k_1)$ carries just a single
momentum label.)
The argument
of~\S\ref{sec: does separate universe compare with in-in}
determines the multi-point propagator expansion
up to 1-loop in the 2-point function.
Restoring explicit momentum labels, it is
\begin{multline}
    \label{eq:Gamma expansion def}
    X^I_{\vect{k}}(t)
    =
    \tensor{\Gamma}{^I_J}(k)^{(\tstar,t)}
    X^J_{\vect{k}}(\tstar)
    +
    \frac{1}{2!}
    \int \d^3 k_1 \;
    \tensor{\Gamma}{^I_J_K}(
        k,
        k_1,
        |\vect{k}-\vect{k_1}|
    )^{(\tstar,t)} \,
    X^J_{\vect{k_1}}(\tstar)
    X^K_{\vect{k}-\vect{k_1}}(\tstar) \\
    +
    \frac{1}{3!}
    \int \d^3 k_1
    \int \d^3 k_2 \;
    \tensor{\Gamma}{^I_J_K_L}(
        \vect{k},
        \vect{k_1},
        \vect{k_2},
        \vect{k}-\vect{k_1}-\vect{k_2}
    )^{(\tstar,t)} \,
    X^J_{\vect{k_1}}(\tstar)
    X^K_{\vect{k_2}}(\tstar)
    X^L_{\vect{k}-\vect{k_1}-\vect{k_2}}(\tstar)
    \\
    +
    \Or({X}^4)
    .
\end{multline}

In Ref.~\cite{Costantini:2025tek} the
multi-point propagator method was implemented numerically
as an extension of the
\texttt{PyTransport} package~\cite{Ronayne:2017qzn}
for computation of 2- and 3-point correlators of $\zeta$ at tree-level.
The implementation has advantages over the previous implementation,
which calculated the evolution of each
correlation function
directly
and
was described in Ref.~\cite{Dias:2016rjq}.
In particular, for topical ultra-slow-roll models,
Ref.~\cite{Costantini:2025tek} found that,
using the standard Runge--Kutta solver implemented in
\texttt{PyTransport},
the multi-point propagator implementation
can track the decay of correlations accurately for a longer period even when traditional \texttt{PyTransport} produces erroneous results.
Also, it can obtain the bispectra in soft configurations
for values of the squeezing parameter at least one decade beyond
those attainable in the previous implementation.
Moreover, as we shall demonstrate, the
multi-point propagator
expansion allows to construct correlators
including loops.
This is equivalent to use of transport
equations directly for the phase-space correlators
beyond tree-level,
as discussed in {\S}\ref{sec: does separate universe compare with in-in}.

\para{Multi-point propagators in action: initial conditions for the (11)-type loop}
To proceed,
we apply
Eq.~\eqref{eq:Gamma expansion def}
to the perturbations
$\delta X^I_{\vect{k}}$,
and set initial conditions
at time $\tstar$
just before the transition.
This implies $\tstar \lesssim \ttransition \ll t_i$. 
In this configuration
it seems safe to assume that
the long-mode fields
at time $\tstar$ are still unaffected by short scales,
because these have not been displaced from their
vacuum state. 
With this choice of initial conditions,
we can represent a
perturbation $\delta X^I_{\vect{p}}(t_i)$
using Eq.~\eqref{eq:Gamma expansion def}
with $\vect{k} = \vect{p}$,
and $\vect{k}_1$, $\vect{k}_2$
replaced by two loop momentum labels
$\vect{q}$, $\vect{q}'$.

The most important feature of
Eq.~\eqref{eq:Gamma expansion def}
is that we can choose the initial time $\tstar$
to be \emph{before}
the onset of the non-attractor phase,
whereas $t_i$ continues to be set
some time after the last enhanced mode
has crossed the horizon.
We can therefore compute the (11)-loop
in terms of the existing correlations
at $\tstar$ and
the $\Gamma$-coefficients.
At time $\tstar$, all of the modes that will later
be enhanced by non-attractor dynamics are still in the
subhorizon vacuum state.
Notice that,
in Eq.~\eqref{eq:Gamma expansion def},
no smoothing procedure is required or implied.
Therefore, in principle, loop integrals
built using 
it can include the contribution of modes up
to an arbitrary ultraviolet scale.

Using Eq.~\eqref{eq:Gamma expansion def}
to evaluate the 2-point
correlation function
at $t_i$, we find three contributions:
the $(12)$-, $(22)$-, $(13)$-type loops.  
We do not include a $(11)$-type loop,
because we are assuming the fields at
$\tstar$ have experienced no back-reaction
and are still accurately Gaussian.
All these contributions are topologically equivalent
to the $\delta N$ 1-loop diagrams listed
in the right-hand-side of
Eq.~\eqref{eq: total 1-loop correction},
and therefore we label them using
the same convention.
In the limit $p \ll q$, we find 
\begin{equation}
    \begin{aligned}
    P^{IJ}(p;t_i)_\text{12} & =
    \frac{1}{2}
    \int \d^3q\;
    \tensor{\alpha}{^M^N^O}(p,q,q; \, \tstar)
    \left[
        \tensor{\Gamma}{^I_M}(p)^{(\tstar,t_i)} \;
        \tensor{\Gamma}{^J_N_O}(p,q,q)^{(\tstar,t_i)}
        + (I\leftrightarrow J)
    \right] \;, \\
    P^{IJ}(p;t_i)_\text{22} & =
    \frac{1}{2}
    \int \d^3q \;
    \tensor{\Gamma}{^I_M_N}(p,q,q)^{(\tstar,t_i)} \,
    \tensor{\Gamma}{^J_O_P}(p,q,q)^{(\tstar,t_i)} \,
    P^{MO}(q;\tstar) \, P^{NP}(q;\tstar) \;, \\
    \label{eq: 13+31 MPPs 1}
    P^{IJ}(p;t_i)_\text{13} & =
    \begin{multlined}[t]
        \frac{1}{2}
        \int \d^3q\; P^{ML}(p;\tstar) \, P^{NO}(q;\tstar) \,
        \Big[
            \tensor{\Gamma}{^I_M} (p)^{(\tstar,t_i)} \;
            \tensor{\Gamma}{^J_L_N_O}(-\vect{p},-\vect{p}, \vect{q},
            -\vect{q})^{(\tstar,t_i)}   \\
            +
            \tensor{\Gamma}{^I_M_N_O} (\vect{p},\vect{p}, \vect{q}, -\vect{q})^{(\tstar,t_i)} \;
            \tensor{\Gamma}{^J_L}(p)^{(\tstar,t_i)}
        \Big]
        .  
    \end{multlined}
    \end{aligned}
\end{equation}
We now assume that the $\Gamma$-coefficients
satisfy a soft theorem.
Because the long mode provides a shifted background
for the short scales, we can assume that the 4-index
$\Gamma$
needed for the $(13)$-type contribution
is related to a shift in the
3-index $\Gamma$,
\begin{equation}
\label{eq: Gamma 4 ind is der of Gamma 3 ind}
    \tensor{\Gamma}{^J_L_N_O}(-\vect{p}, -\vect{p}, \vect{q}, -\vect{q})^{(\tstar,t_i)}
    \approx
    \frac{\partial}{\partial \delta X^L_{-\vect{p}}(\tstar)}
    \tensor{\Gamma}{^J_N_O}(-\vect{p}, \vect{q}, -\vect{q})^{(\tstar,t_i)}
    \approx
    \frac{\partial}{\partial X^L(\tstar)}
    \tensor{\Gamma}{^J_N_O}(p,q,q)^{(\tstar,t_i)} \;.
\end{equation}
Under these assumptions,
Eq.~\eqref{eq: Gamma 4 ind is der of Gamma 3 ind}
follows from the properties
of the long and short modes,
described in~\S\ref{sec: properties of long and short modes}.
As a consequence,
the 4-index $\Gamma$ \emph{in this specific case}
depends only on the magnitude of its momenta.
This property, together with symmetry of the long-mode
power spectrum when it is evaluated much after horizon
crossing (since we assume $\tstar \gg t_p$),
allows to rewrite the $(13)$-type
loop in Eq.~\eqref{eq: 13+31 MPPs 1}
as 
\begin{equation}
\label{eq: 13+31 MPPs 2}
    P^{IJ}(p;t_i)_\text{13}
    =
    \frac{1}{2} \int \d^3q\;
    P^{ML}(p;\tstar) \, P^{NO}(q;\tstar) \,
    \left[
        \tensor{\Gamma}{^I_M} (p)^{(\tstar,t_i)} \;
        \tensor{\Gamma}{^J_L_N_O}(p,p,q,q)^{(\tstar,t_i)}
        + (I\leftrightarrow J)
    \right]  \;. 
\end{equation}
Now combining all contributions, we obtain  
\begin{equation}
\begin{multlined}
    \label{eq: 1loop to phase-space 2pt fct master}
    P^{IJ}(p;t_i)_\text{1-loop}
    =
    P^{IJ}(p;t_i)_\text{12}
    +
    P^{IJ}(p;t_i)_\text{22}
    +
    P^{IJ}(p;t_i)_\text{13}
    \\
    =
    \begin{aligned}[t]
        &
        \frac{1}{2} \,
        \tensor{\Gamma}{^I_M}(p)^{(\tstar,t_i)}
        \int \d^3q \;
        \tensor{\Gamma}{^J_N_O}(p,q,q)^{(\tstar,t_i)} \,
        \tensor{\alpha}{^M^N^O}(p,q,q; \, \tstar)
        + (I\leftrightarrow J)
        \\
        &
        +
        \frac{1}{2} \,
        \int \d^3q \;
        \tensor{\Gamma}{^I_M_N}(p,q,q)^{(\tstar,t_i)} \,
        \tensor{\Gamma}{^J_O_P}(p,q,q)^{(\tstar,t_i)} \,
        P^{MP}(q;\tstar) \, P^{NO}(q;\tstar)
        \\
        &
        +
        \frac{1}{2} \,
        \tensor{\Gamma}{^I_M}(p)^{(\tstar,t_i)}
        P^{ML}(p;\tstar)
        \int \d^3q\;
        \tensor{\Gamma}{^J_L_N_O}(p,p,q,q)^{(\tstar,t_i)} \,
        P^{NO}(q;\tstar)
        +
        (I\leftrightarrow J)
        .
    \end{aligned}
\end{multlined}   
\end{equation}
Each $\Gamma$-coefficient
must be regular in the limit
$p \rightarrow 0$,
otherwise the zero mode would diverge
and represent an instability of the background.
It follows that, in terms of the dimensionless
power spectrum, the
$(22)$-type diagram is suppressed for $p \rightarrow 0$.

Our discussion now parallels that of~\S\ref{sec: delta N 12 and 13 loops}.
We consider the sum of the remaining
$(12)$- and $(13)$-type diagrams.
Using the soft theorems~\eqref{eq: 3-point function squeezed}
(for the squeezed bispectrum)
and~\eqref{eq: Gamma 4 ind is der of Gamma 3 ind}
(for the 4-index $\Gamma$),
we obtain
\begin{multline}
    \label{eq: 12 13 in 11 step 1}
    \dimlessP^{IJ}(p;t_i)_{12+13}
    =
    \tensor{\Gamma}{^I_M}(p)^{(\tstar,t_i)}
    \dimlessP^{ML}(p;\tstar)
    \int \d^3q \;
    \frac{\partial}{\partial X^L(\tstar)}
    \left[
        \tensor{\Gamma}{^J_N_O}(p, q, q)^{(\tstar,t_i)} P^{NO}(q;\tstar)
    \right]
    \\
    +
    (I \leftrightarrow J)
    . 
\end{multline}
Adiabaticity of the field configuration
for the long mode
means that the combination of partial derivatives
appearing here
can be replaced by a total derivative along the
unperturbed trajectory,
in the same way as
Eqs.~\eqref{eq:12 13 step 5}--\eqref{eq:12 13 step 6.1},
\begin{multline}
    \label{eq: 12 13 in 11 step 2}
    \dimlessP^{IJ}(p;t_i)_{12+13}
    =
    \tensor{\Gamma}{^I_M}(p)^{(\tstar,t_i)} \dimlessP^{M\phi}(p;\tstar)
    \int \d \ln q \;
    \frac{\d}{\d \phi(\tstar)}
    \left[
        \tensor{\Gamma}{^J_N_O}(p, q, q)^{(\tstar,t_i)}
        \dimlessP^{NO}(q;\tstar)
    \right]
    \\
    +
    (I \leftrightarrow J)
    .
\end{multline}
In the same way, the discussion
of Eqs.~\eqref{eq:lookback-equation}--\eqref{eq:Delta 12 13 step 3}
applies,
enabling us to calculate the effect of a shift
along the trajectory in terms of a relabelling of the
short modes.
Hence we can express
the derivative $\d/\d \phi(\tstar)$
in terms of $\d/\d \ln q$.
These are all ``soft limit'' statements,
which here we assume to be properties of correlation
functions (evaluated at $\tstar$)
in the full non-equilibrium field theory.

The main result is that we conclude
\begin{multline}
\label{eq: 12 13 in 11 step 3}
    \dimlessP^{IJ}(p;t_i)_{12+13}
    = 
    \tensor{\Gamma}{^I_M}(p)^{(\tstar,t_i)}
    \dimlessP^{M\phi}(p;\tstar)
    \frac{1}{\phi'(\tstar)}
    \\
    \times
    \int \d \ln q \;
    \frac{\d}{\d \ln q}
    \left[
        \tensor{\Gamma}{^J_N_O}(p, q, q)^{(\tstar,t_i)}
        \dimlessP^{NO}(q;\tstar)
    \right]
    +
    (I \leftrightarrow J)
    .
\end{multline}
It follows that the $(11)$-type contribution
given in Eq.~\eqref{eq: 11 loop master}
can be written
\begin{multline}
    \label{eq: 12 13 in 11 step 4}
    \dimlessP_\zeta(p;t)_{11} 
    =  
    N_I^{(t_i,t)}\, N_J^{(t_i,t)}\, 
    \Bigg\{
        \tensor{\Gamma}{^I_M}(p)^{(\tstar,t_i)}
        \dimlessP^{M\phi}(p;\tstar)
        \frac{1}{\phi'(\tstar)}
        \\
        \times
        \int \d \ln q \;
        \frac{\d}{\d \ln q}
        \left[
            \tensor{\Gamma}{^J_N_O}(p, q, q)^{(\tstar,t_i)}
            \dimlessP^{NO}(q;\tstar)
        \right]
        +
        (I \leftrightarrow J)
    \Bigg\} 
    .
\end{multline}
The conclusion is that the $(11)$-type loop correction
must \emph{also} formally have the structure of
a total derivative,
as a consequence of the
multi-point propagator ``OPE'' Eq.~\eqref{eq:Gamma expansion def}
and the assumed soft-limit properties
of the $\Gamma$ coefficients with respect to the superhorizon
mode $\vect{p}$.
These are sufficient to show that the loop correction
must organize itself into the form~\eqref{eq: 12 13 in 11 step 4},
regardless of the precise
value of
$\tensor{\Gamma}{^J_N_O}$.
In this sense, Eq.~\eqref{eq: 12 13 in 11 step 4},
and the analogous separate universe
expression~\eqref{eq:Delta 12 13 step 3}---which can
be regarded as a special case of the
more general computation given here---%
can themselves be interpreted as soft limit theorems
for the 1-loop
contribution to the 2-point function in an adiabatic
scenario.

\subsection{Lessons from the loop computation}
\label{sec: lesson from completing the delta N loop computation}
Neglecting loops that are suppressed in the $p\to 0$ limit,
the 2-point function~\eqref{eq: total 1-loop correction}
for $\zeta_{\vect{p}}$ at 1-loop is 
\begin{equation}
    \label{eq: summary of total derivatives}
    \dimlessP_\zeta(p;t)_\text{1-loop}
    =
    \dimlessP_\zeta(p;t)_{11}
    +
    \dimlessP_\zeta(p;t)_{12}
    +
    \dimlessP_\zeta(p;t)_{13}
    . 
\end{equation}
We computed
$\dimlessP_\zeta(p;t)_{12} + \dimlessP_\zeta(p;t)_{13}$
in~\S\ref{sec: delta N 12 and 13 loops},
and have just finished the computation of
$\dimlessP_\zeta(p;t)_{11}$ in~\S\ref{sec: 11-type loop}.
Each contribution can be written in two ways,
either in terms of a derivative with respect to the field
configuration at time $\tstar$ or $t_i$,
or in terms of the wavenumber $q$.
For convenience we collect the final expressions below:
\begin{itemize}
    \item total derivative in terms of the background field value:
    \begin{subequations}
    \begin{multline}
        \label{eq: summary 11 field value}
        \dimlessP_\zeta(p;t)_{11} 
        =
        N_I^{(t_i,t)} N_J^{(t_i,t)}
        \,
        \Bigg\{
            \tensor{\Gamma}{^I_M}(p)^{(\tstar,t_i)}
            \dimlessP^{M\phi}(p;\tstar)
            \\
            \times
            \int_{\qmin}^{\qmax} \d \ln q \;
            \frac{\d}{\d \phi(\tstar)}
            \left[
                \tensor{\Gamma}{^J_N_O}(p, q, q)^{(\tstar,t_i)}
                \dimlessP^{NO}(q;\tstar)
            \right]
            +
            (I \leftrightarrow J)
        \Bigg\}
        , 
    \end{multline}
    \begin{equation}
        \label{eq: summary 12+13 field value}
        \dimlessP_\zeta(p;t)_{12}
        +
        \dimlessP_\zeta(p;t)_{13}
        =
        N_I^{(t_i,t)}
        \dimlessP^{I\phi}(p;t_i) 
        \int_{\qmin}^{\qmax} \d \ln q \;
        \frac{\d}{\d\phi(t_i)}
        \left[
            N_{JK}^{(t_i,t)}
            \dimlessP^{JK}(q;t_i)
        \right]
        ;
    \end{equation}
    \end{subequations}
    \item total derivative in terms of the loop comoving momentum $q$
    \begin{subequations}
    \begin{multline}
        \label{eq: summary 11 momentum}
        \dimlessP_\zeta(p;t)_{11} 
        =
        N_I^{(t_i,t)} N_J^{(t_i,t)}
        \,
        \Bigg\{
            \tensor{\Gamma}{^I_M}(p)^{(\tstar,t_i)}
            \dimlessP^{M\phi}(p;\tstar)
            \frac{1}{\phi'(\tstar)}
            \\
            \times
            \int_{\qmin}^{\qmax} \d \ln q \;
            \frac{\d}{\d \ln q}
            \left[
                \tensor{\Gamma}{^J_N_O}(p, q, q)^{(\tstar,t_i)}
                \dimlessP^{NO}(q;\tstar)
            \right]
            +
            (I \leftrightarrow J)
            \Bigg\}
            , 
    \end{multline}
    \begin{equation}
        \label{eq: summary 12+13 momentum}
        \dimlessP_\zeta(p;t)_{12}
        +
        \dimlessP_\zeta(p;t)_{13}
        =
        N_I^{(t_i,t)}
        \dimlessP^{I\phi}(p;t_i)
        \frac{1}{\phi'(t_i)}
        \int_{\qmin}^{\qmax} \d \ln q \;
        \frac{\d}{\d \ln q}
        \left[
            N_{JK}^{(t_i,t)}
            \dimlessP^{JK}(q;t_i)
        \right]
        . 
    \end{equation}
    \end{subequations}
\end{itemize}
The $(11)$-type loop
exhibits the same structure as the
$(12)$- and $(13)$-type loops,
and can be given the same interpretation
discussed in~\S\ref{sec: interpretation and discussion}.
We therefore reach the same conclusion, that the
$(11)$-type contribution \emph{also} decouples from
the precise details of the enhanced band, and in particular
from its maximum amplitude.

Clearly, there are examples in Nature where 1-loop contributions
do not have the structure required by Eq.~\eqref{eq: summary 11 momentum}.
We therefore cannot expect
this result
to apply in general:
there must be conditions under which it fails.
In the argument given here, this occurs when
the momentum configuration
does not allow us to apply soft theorems
to relate the $3$- and $4$-index
$\Gamma$ coefficients.
Therefore, Eq.~\eqref{eq: summary 11 momentum}
will not apply to
(for example)
loop corrections measured in terrestrial experiments,
allowing us to recover the usual properties of
quantum field theory
in this regime.

We can now summarize our main messages.
\begin{itemize}
    \item Given a suitable choice of initialization time for the
    separate universe framework, it is possible to choose
    $\qmin$, $\qmax$ so that the loop correction adequately
    brackets a band of enhanced short-scale modes.

    \item Assuming suitable soft-limit properties
    for squeezed correlators or $\Gamma$-coefficients,
    the loop correction over the enhanced band
    organizes itself into a total derivative.
    It therefore decouples from the detailed properties of
    these modes.
    It follows that the amplitude of the loop correction
    will not be proportional to the amplitude at the peak.

    \item As already noted in Ref.~\cite{Iacconi:2023ggt},
    the loop integral over the enhanced band is scale invariant,
    and therefore not observable.
\end{itemize}

\subsection{Renormalized $\delta N$ coefficients}
\label{sec:renormalized-deltaN}
Before closing, we note that
there is another way to understand a
1-loop formula such as Eq.~\eqref{eq:12 13 step 3b}
or Eq.~\eqref{eq: 12 13 in 11 step 1}.
In this section, we show that
the part of the 1-loop correction that is not
volume-suppressed
has a very simple interpretation
as
a \emph{tree}-level correlation
for a long-wavelength field
obtained from a separate-universe
expansion in a shifted background.
This enables us to connect our analysis
to a discussion of
loops in the $\delta N$ formalism given by Byrnes {\etal}~\cite{Byrnes:2007tm}.

In this section we return to the separate universe approach,
and
the analysis below is intended to apply only in this framework.
The argument can be given for any perturbation.
We give the discussion for
$\zeta_{\vect{p}}$,
but
exactly the same procedure can be applied to
any generic perturbation
$\delta X_{\vect{p}}$.
In the discussion of~\S\ref{sec: delta N 12 and 13 loops}
we began with a large region of spacetime
characterized by
a background field configuration
$X_0^I$
and perturbations $\delta X^I$, as in
Eq.~\eqref{eq:deltaN-loop-step1}.
In Ref.~\cite{Iacconi:2023ggt}
and~\S\ref{sec: delta N 12 and 13 loops}
we evaluated
corrections to
$\langle \zeta_{\vect{p}} \zeta_{-\vect{p}} \rangle$
by building loops immediately from~\eqref{eq: zeta with backreaction Fourier},
and then identifying contributions from different regions
of the loop momentum.

Byrnes~{\etal} noticed that the
expressions yielded by this procedure can be
interpreted in terms of ``renormalized''
$\delta N$ coefficients.
To understand their result in our language,
write the
perturbations in the large region in the form
$\delta X^I_{\text{total}} = \delta X^I_{<} + \delta X^I_{>}$,
where the subscripts $<$, $>$ indicate that the field
contains Fourier modes with wavenumbers $< p$ and $> p$,
respectively.
In the analysis of~\S\ref{sec: delta N 12 and 13 loops}
we constructed separate universe expressions
based on a Taylor expansion of $N(t)$ around
the background $X_0^I$.
This approach leads to
expressions involving correlation functions
of the total perturbation
$\delta X^I_{\text{total}}$.
We then 
identified the impact of $\delta X^I_>$ on correlations
of $\delta X^I_<$ using the loop expansion as the primary tool.

An alternative
strategy is to
break the Taylor expansion
of $\delta N$ into two steps: first
we make
an expansion
in $\delta X_<$ alone,
\begin{equation}
    \label{eq:renormalized-deltaN-preexpansion}
    N
    =
    \tilde{N}_0
    +
    \tilde{N}_I \delta X^I_<
    +
    \frac{1}{2!} \tilde{N}_{IJ} \delta X^I_< \delta X^J_<
    +
    \frac{1}{3!} \tilde{N}_{IJK} \delta X^I_< \delta X^J_< \delta X^K_<
    + \cdots ,
\end{equation}
where the notation $\tilde{N}_I$, $\tilde{N}_{JK}$ (and so on)
indicates that each derivative is to be evaluated on the
\emph{shifted}
field configuration $X_0^I + \delta X^I_>$.
In a second step, we expand each $\tilde{N}$ coefficient
in terms of the short-scale
fields $\delta X^I_>$. We then spatially average over
these modes.
This is simply a reorganization of terms in
the perturbation expansion,
and so should yield an outcome
equivalent to~\S\ref{sec: delta N 12 and 13 loops}.
We confirm this expectation explicitly below.
Hence, for example,
\begin{align}
    \langle \tilde{N}_0 \rangle_{>}
    & =
    N_0
    +
    N_{A} \langle \delta X^A_> \rangle_{>}
    +
    \frac{1}{2!} N_{AB} \langle \delta X^A_> \delta X^B_> \rangle_>
    +
    \cdots
    \\
    \langle \tilde{N}_I \rangle_{>}
    & =
    N_I
    + N_{IA} \langle \delta X_>^A \rangle_>
    + \frac{1}{2!} N_{IAB} \langle \delta X^A_> \delta X^B_> \rangle_>
    + \cdots .
\end{align}
On the right hand side we have written $N_I$, $N_{IA}$, $N_{IAB}$,
because these derivatives are to be evaluated in the unshifted
background and hence agree with the
$\delta N$ coefficients
appearing in Eq.~\eqref{eq: zeta with backreaction Fourier}
and Eqs.~\eqref{eq: 11 type loop initial expression}--\eqref{eq: 13 base expression}.
The brackets $\langle \cdots \rangle_>$ indicate
spatial averaging only over the $>$ fields.
Similar formulas can be written for
$\langle \tilde{N}_{IJ} \rangle_>$,
$\langle \tilde{N}_{IJK} \rangle_>$
and all higher coefficients, but at 1-loop we do not need them.

We assume that $\langle \delta X^A_> \rangle_> = 0$.
However, the
short-scale variance
$\langle \delta X^A_> \delta X^B_> \rangle_>$
can depend on the long wavelength field configuration
on which it is evaluated.
Therefore, assuming a suitable soft theorem,
\begin{equation}
    \langle \delta X^A_> \delta X^B_> \rangle_>
    =
    \langle \delta X^A_> \delta X^B_> \rangle_> \big|_0
    +
    \left(
        \frac{\partial}{\partial X_0^I}
        \langle \delta X^A_> \delta X^B_> \rangle_>
    \right)\bigg|_0
    \delta X_<^I
    +
    \cdots ,
\end{equation}
where $|_0$ denotes evaluation in the unshifted background.
Then it follows from Eq.~\eqref{eq:renormalized-deltaN-preexpansion} that
\begin{multline}
    \label{eq:deltaN-average-short-noise}
    \langle N \rangle_>
    =
    \bigg(
        N_0 + \frac{1}{2} N_{AB} \langle \delta X^A_> \delta X^B_> \rangle_> + \cdots
    \bigg)
    \\
    +
    \bigg(
        N_I
        +
        \frac{1}{2!}
        N_{AB}
        \frac{\partial}{\partial X_0^I}
        \langle \delta X^A_> \delta X^B_> \rangle_>
        +
        \frac{1}{2!}
        N_{IAB}
        \langle \delta X^A_> \delta X^B_> \rangle_>
        \cdots
    \bigg)
    \delta X^I_<
    \\
    +
    \cdots .
\end{multline}
The terms quadratic in $\delta X_<$ are not needed. These would be
required for computation of the higher correlation functions,
but not the 2-point function.
We have dropped the subscript ``$0$''s to reduce notational clutter;
all coefficients in this expansion are intended to be evaluated
without any shift of zero modes.

The conclusion is that
$\zeta = \delta \langle N \rangle_>$
can be expressed in terms of $N(t)$,
averaged over a shift in the background fields.
(Notice that we shift the background field first, and then average.
This is not the same as shifting the background field by the
average $\langle \delta X_>^I \rangle$, which we have assumed to vanish.)
Therefore $\zeta$
can be expressed in
terms of a modified
first-order coefficient, which
Byrnes~{\etal} described as ``renormalized''.
We denote such coefficients with a hat, viz.
$\hat{N}_I$, $\hat{N}_{IJ}$, etc.
Up to the 1-loop order considered here, we need only write
$\zeta_< = \hat{N}_I \delta X^I_< + \cdots$,
where the
renormalized coefficient $\hat{N}_I$ satisfies
\begin{equation}
    \label{eq:renormalized-deltaN-linear}
    \hat{N}_I
    =
    N_I
    +
    \frac{1}{2!}
    N_{AB}
    \frac{\partial}{\partial X^I_0}
    \langle \delta X^A_> \delta X^B_> \rangle_>
    +
    \frac{1}{2!}
    N_{IAB}
    \langle \delta X^A_> \delta X^B_> \rangle_>
    .
\end{equation}
Eq.~\eqref{eq:renormalized-deltaN-linear}
agrees with Eq.~(25) of Ref.~\cite{Byrnes:2007tm},
except in that paper the fields were assumed to be Gaussian,
and therefore the term
involving the derivative of the two-point function was absent.
This depends on long--short mode coupling and is needed to match
the 3-point contribution to the 1-loop formula~\eqref{eq:12 13 step 3b}.
Notice that the derivative has to be correctly interpreted,
as explained below.

To reproduce the 1-loop corrected 2-point function,
Eq.~\eqref{eq:12 13 step 3b},
we compute
\begin{equation}
    \langle \zeta_{<,\vect{p}} \zeta_{<,-\vect{p}} \rangle
    =
    \hat{N}_I \hat{N}_J
    \langle \delta X^I_{<,\vect{p}} \delta X^J_{<,-\vect{p}} \rangle
    .
\end{equation}
The expected result follows immediately after
expressing the spatial average
$\langle \delta X_> \delta X_> \rangle_>$
in terms of the power spectrum,
\begin{equation}
    \langle \delta X_>^A \delta X_>^B \rangle_>
    =
    \int_{>} \frac{\d^3 q}{(2\pi)^3}
    \,
    P^{AB}(q)
    =
    \int_{\qmin}^{\qmax}
    \d \ln q \,
    \dimlessP^{AB}(q) .
\end{equation}
To compute the $\partial / \partial X_0^I$ derivative
appearing in Eq.~\eqref{eq:renormalized-deltaN-linear}
we should first compute the $q$-dependent response
of $\dimlessP^{AB}(q)$ to a change in the zero mode,
and then integrate this response
over the relevant range of $q$.
This procedure reproduces the formulae
of~\S\ref{sec: delta N 12 and 13 loops}.

Notice that this procedure does not reproduce the $(22)$-type
loop.
As explained in Ref.~\cite{Iacconi:2023ggt}, this represents
an average over short-scale noise uncorrelated with
$\delta X^I_{<,\vect{p}}$,
and hence cannot be absorbed into a
multiplicative renormalization of it.
This term could be included, if desired, by allowing
the ``constant''
first term on the right-hand side of
Eq.~\eqref{eq:deltaN-average-short-noise}
to have uncorrelated stochastic noise $\xi$.
The 1-loop $\delta N$ formula
for the long-wavelength field would then become
$\zeta = \hat{N}_I \delta X_<^I + \xi$,
with $\langle \xi \xi \rangle$
reproducing the $(22)$-type contribution.
This is very similar to renormalization
of the matter density field by stochastic terms,
as in the effective field theory of large scale structure,
or renormalized halo bias.

As presented in Ref.~\cite{Byrnes:2007tm} this argument
was largely based on
the combinatorics of Feynman-like diagrams.
However, in this context it is easy to see that it has an
interpretation in terms of back-reaction.
The moral is that, to correctly compute correlations
at wavenumber $p$,
we should first
replace $N$ by its average $\langle N \rangle_>$
obtained by integrating over structure on shorter scales
with wavenumbers $q \gg p$.

\section{Discussion}
\label{sec: discussion}

In this paper, our intention was to determine the
effect of enhanced short-scale
perturbations on a large-scale, adiabatic field
configuration produced by an (effectively) single-field inflationary
model.
The enhanced modes could be produced
by a transient non-attractor phase during inflation,
although the exact mechanism of enhancement is not important
for our analysis.
In this section, as in the rest of the paper,
we use $p$ to label the wavenumber of the large-scale
mode,
and $q$ to denote a wavenumber associated with the enhanced
short-scale band,
with $p\ll q$. 

\para{Summary}
Initially,
we work within the separate universe framework.
Employing a $\delta N$ formula for the long mode,
and accounting for back-reaction due to short
scales as in Ref.~\cite{Iacconi:2023ggt},
we compute the power spectrum of $\zeta_{\vect{p}}$ up to and
including 1-loop corrections.
We initialize the separate universe computation shortly
after the last enhanced mode
leaves the horizon.
With this choice we are able to include the entire
band of enhanced modes.
There are two contributions that are not
volume-suppressed.
The first type are due to 
non-linearities in the $\delta N$ formula,
even if the perturbations $\delta X^I$ that appear in 
it are themselves free of backreaction.
These are the $(12)$- and $(13)$-type loop terms.
The second type
comes from back-reaction to the $\delta X^I$
already at the initial time.
At 1-loop level, the only relevant such term of this type is
the $(11)$-loop;
see Eq.~\eqref{eq: total 1-loop correction}.

The $(11)$-type contribution cannot be estimated using
the separate universe framework, because it necessarily
includes contributions from times when the
back-reacting modes
are close to the horizon scale.
However, as explained
in~\S\ref{sec: 1-loop from quantum initial conditions is a boundary term},
it is possible to give an
expression for it in terms of
``multi-point propagators''
or $\Gamma$ coefficients.
This technique has already been implemented
(at tree level)
in the
\texttt{PyTransport}
code for numerical evaluation of correlation
functions during inflation~\cite{Motohashi:2025qgd}.
It represents a possible approach to numerical
evaluation of loop corrections.
We expect to return to this issue in a future
publication~\cite{in_preparation}.

Using adiabaticity of the long-wavelength field configuration,
and soft theorems for squeezed correlators and $\Gamma$
coefficients,
we show that the 1-loop correction organizes itself
into a total derivative,
as in
Eqs.~\eqref{eq: summary 11 field value}--%
\eqref{eq: summary 12+13 field value}
and
Eqs.~\eqref{eq: summary 11 momentum}--%
\eqref{eq: summary 12+13 momentum}.
Specifically, the integral
reduces to a boundary term
involving a contraction of the short-scale (cross-)
power spectrum
with a suitable coefficient tensor.

In isolation, we cannot draw quantitative conclusions
from this property,
because these contributions
must be accompanied by integrals
over infrared (wavenumbers $< \qmin$ in the language
of~\S\ref{sec: 1-loop from non-linear superhorizon evolution is a boundary term})
and ultraviolet regions (wavenumbers $> \qmax$).
These integrals must compensate for any
$\qmin$, $\qmax$
dependence of the result,
because these are arbitrary scales and cannot appear in
a physical prediction.
However, if $\qmin$, $\qmax$ bracket the whole of the enhanced band, we
can expect the IR and UV integrals to depend only
weakly on its presence,
provided the renormalization scale
is at least a little larger than $\qmax$.
Notice that, if $p$ is a typical CMB scale,
we are implicitly choosing a renormalization scheme
that is not tied to $p$.

For example, it is sometimes suggested that we
can renormalize the 2-point function by matching to
its observed amplitude $A_S^2 \sim 10^{-9}$
at the CMB pivot scale.
This is certainly possible, but
the renormalization scale is then far to the IR of the enhanced
band.
It follows that any effects
associated with this band
must be absorbed by the counterterms needed to define
the 2-point function.
Our calculation would be valid for a renormalization scheme
that is independent of the enhanced band,
for example by measuring masses associated with the
inflationary sector
on terrestrial scales $\sim q_{\text{ren}}$
where all wavenumbers are much larger
than $\qmax$.
For recent discussion of renormalization
in the context of cosmological correlation functions,
see Refs.~\cite{Ballesteros:2024zdp, Sheikhahmadi:2024peu, Inomata:2025bqw,
Inomata:2025pqa, Kristiano:2025ajj}
for ultra-slow-roll,
and Refs.~\cite{Ballesteros:2024cef, Braglia:2025cee, Braglia:2025qrb, Ballesteros:2025nhz} for the
case of slow-roll.

Eqs.~\eqref{eq: summary 11 field value}--%
\eqref{eq: summary 12+13 field value}
and
\eqref{eq: summary 11 momentum}--%
\eqref{eq: summary 12+13 momentum}
can be interpreted within the back-reaction model
proposed in \S1 of Iacconi~{\etal}~\cite{Iacconi:2023ggt}.
However, in that paper,
we considered the back-reaction of a \emph{single}
enhanced mode onto the long-wavelength mode $\vect{p}$.
In this paper we include (at least to the extent possible)
the entire enhanced band.
It is the aggregate effect of all these modes
that decouples from their detailed properties,
although the physical mechanism underlying this decoupling
is not yet entirely clear.
The conclusion is that a long-wavelength, adiabatic field
configuration decouples from enhanced structure on short scales.
This is very similar to the question posed
in early work on the effective theory of large-scale
structure~\cite{Baumann:2010tm}.%
    \footnote{Indeed, there are clear
    parallels between our back-reaction
    calculation and the effective field theory of large-scale
    structure. These are perhaps most clearly
    expressed in terms of the renormalized long-wavelength
    formula $\zeta = \hat{N}_I \delta X^I_< + \xi$
    from~\S\ref{sec:renormalized-deltaN}.
    The main conclusions of~\S\ref{sec:renormalized-deltaN}
    are that integrating out the short-wavelength modes introduces
    both additive and multiplicative renormalization of
    the tree-level operator $N_I \delta X^I_<$, with the multiplicative
    renormalization $N_I \rightarrow \hat{N}_I$
    being introduced due to long--short correlations,
    and the stochastic term $\xi$ being volume suppressed.
    These parallel equivalent conclusions in EFTofLSS~\cite{Baumann:2010tm}.}
In that case, it was argued that the short-scale structure
decoupled completely if it satisfied a virial condition.
Here, the conclusion runs in the opposite direction, because we take
the short-scale structure to be arbitrary.
However,
the condition of adiabaticity makes the long-wavelength
field configuration special.
In any case, as has been noted,
the loop correction considered here is scale invariant, and therefore
does not clearly have observable consequences.
Similar conclusions have been reached by
Tada~{\etal}~\cite{Tada:2023rgp},
Fumagalli~\cite{Fumagalli:2024jzz}
and Kawaguchi~{\etal}~\cite{Kawaguchi:2024rsv} 
in the context of the full in--in framework.

We have also emphasized that decoupling of the short-scale
structure seems to be necessary in an adiabatic scenario
to maintain compatibility with
calculations showing that the correlation functions of
$\zeta$ are constant to all orders in the loop expansion
in ``single-clock'' models,
due to Senatore \& Zaldarriaga~\cite{Senatore:2012ya}
and Assassi~{\etal}~\cite{Assassi:2012et}.
These papers both argue that,
in such models and under rather general conditions,
$\d \hat{\zeta}_{\vect{k}} / \d t \rightarrow 0$ on superhorizon
scales as an operator statement.
(For earlier work, see Refs.~\cite{Senatore:2009cf,Senatore:2012nq,Pimentel:2012tw}.)
In this paper we assume the single-clock property for the
long wavelength modes,
but not explicitly for the short ones,
and we rely on a soft theorem to understand the
response of the short-scale
modes to the long-wavelength disturbance.
Our analysis is therefore compatible with the conclusions
of these papers, but generalizes
the result to cases where the long wavelength field
configuration is adiabatic, but the short modes are not.

\para{(Non-)applicability of our results}
These results apply whenever our
fundamental assumptions are satisfied.
As explained in~\S\ref{sec: properties of long and short modes},
these are:
(1) adiabaticity of the long-wavelength field configuration,
and
(2) an appreciable hierarchy of scales, allowing application of
soft theorems.
A primary motivation for this scenario
comes from the analysis of
Kristiano \& Yokoyama~\cite{Kristiano:2022maq},
who suggested that 1-loop back-reaction might invalidate
conventional perturbation theory in models where enhanced small-scale
structure produces a cosmologically-interesting abundance of primordial
black holes.
We interpret our result to show that,
if the long-wavelength field configuration is adiabatic,
then its statistical properties do not receive large renormalizations
from the enhanced band.

Let us now discuss scenarios to which our results do \emph{not} apply.
First, our result should not be used to estimate
the back-reaction of enhanced modes onto themselves.
These modes are certainly not in an adiabatic configuration,
because they are already evolving at tree-level.
Further, there is no separation of scales.
To study this kind of back-reaction, the separate universe framework
loses all its advantages,
and one would have to return to a formulation of the full
non-equilibrium field theory.
Firmly establishing the size of such ``self''-corrections
to the enhanced modes would be very important,
because the phenomenology of models with non-attractor behaviour
(such as production of primordial black holes or scalar-induced gravitational waves)
is highly sensitive to the exact amplitude and scale dependence of the enhanced
band.
See Refs.~\cite{Inomata:2022yte, Iacconi:2023slv, Caravano:2024tlp, Caravano:2024moy, Caravano:2025diq} for studies on 1-loop effects of enhanced small-scale modes at scales around the peak, and Ref.~\cite{Fumagalli:2023loc} for near-infrared scales.

Second, in the context of multiple-field inflationary models
with active isocurvature perturbations,
$\zeta$ may already evolve at tree-level on large scales after horizon exit.
If this is the case, our results do not apply.
Indeed,
in the effective theory of the long-wavelength
modes,
one can presumably regard the averaged short-scale structure
as exciting an effective isocurvature mode that drives
evolution of $\zeta$.
We defer a full treatment of multiple-field models to future work.
 
\para{Subtraction of squeezed initial conditions}
Before closing,
we note one remaining technicality.
Because soft theorems play a critical role in our
analysis, our final formulae
in~\S\ref{sec: 1-loop from non-linear superhorizon evolution is a boundary term}
and~\S\ref{sec: 1-loop from quantum initial conditions is a boundary term}
depend on the squeezed limit of correlation functions.
Such squeezed configurations must be handled carefully.
In particular, their physical content is not clear until
an appropriate gauge transformation to physical coordinates is
performed; see, e.g., Pajer~{\etal}~\cite{Pajer:2013ana}
and Tanaka~{\etal}~\cite{Tanaka:2011aj}.
For the case of the tree-level bispectrum, this transformation
amounts to subtraction of the leading term
in Eq.~\eqref{eq: 3-point function squeezed}.

In our analysis, we do not make such transformations, but work
consistently on spatial hypersurfaces where the squeezed limit
is not suppressed.
In principle,
we should perhaps make such a transformation \emph{after}
calculation of our loop, but this would not be expected to change
the result for the 2-point function of $p$ on observable scales.
However,
there is presumably an equivalent (perhaps simpler) description
in which one works on physical hypersurfaces from the outset,
and squeezed correlations are absent when the long-wavelength
field configuration is adiabatic.
This approach will
presumably lead to the same conclusions
obtained above.
Techniques to perform the necessary subtractions were developed
in the context of the separate universe framework by
Tada \& Vennin~\cite{Tada:2016pmk}.
However, they have not yet been applied to loop calculations.
We believe this is an important issue, to which we hope to return
in future work.
 
\section*{Acknowledgments}
LI is grateful to
Guillermo Ballesteros,
Matteo Braglia,
Sebasti\'an C\'espedes,
Jes\'us Gamb\'in Egea,
Arthur Poisson,
S\'ebastien Renaux-Petel,
as well as the participants to the workshops
\href{https://indico.cern.ch/event/1433472/}{Looping in the Primordial Universe}
and
\href{https://indico.ijclab.in2p3.fr/event/11373/}{CoBALt}
for interesting and useful discussions related to this work.
DJM and LI are
supported by the Science and Technology Facilities Council
(grant number ST/X000931/1).
DS is
supported by the Science and Technology Facilities Council
(grant number ST/X001040/1).
For the purpose of open access, the authors have applied a
Creative Commons Attribution (CC-BY) licence to any
Author Accepted Manuscript version arising from this work.

\appendix
\section{Tree-level application of separate universe to a ultra-slow-roll model}
\label{app: separate universe computation of tree level power spectrum}
In this Appendix we explicitly show that separate universe was correctly implemented in Ref.~\cite{Iacconi:2023ggt} for a toy model featuring ultra-slow-roll (USR). 
To do so, we compute the linear scalar power spectrum, $\dimlessP_\zeta(k; t)_\text{tree}$, by starting a separate universe computation at time $t_i$ after the transition into USR has taken place.
We consider two choices of $t_i$, (i) the horizon-crossing time of the peak scale, $t_i \mid \kpeak=a(t_i) H $, and (ii) the end of USR, $t_i \mid k_e=a(t_i) H $. 
When expressed in terms of conformal time, these choices correspond to (i) $\eta_i = -1/\kpeak$ and (ii) $\eta_i = -1/k_e$.

\begin{figure}
\centering
\includegraphics[width=0.55\textwidth]{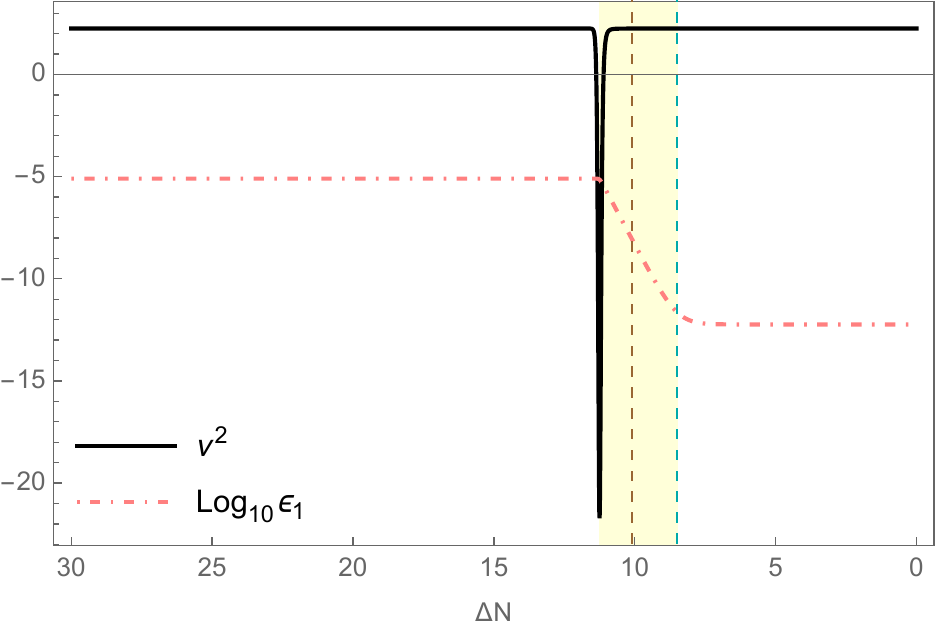}
\caption{Evolution of the Mukhanov--Sasaki mass, $\nu^2$, and slow-roll parameter $\epsilon_1$, computed for the toy USR model analysed in Ref.~\cite{Iacconi:2023ggt}. 
On the horizontal axis we display the number of e-folds to the end of inflation, $\Delta N \equiv N_\text{end}-N$. The two vertical dashed lines mark the two choices of initialization time for the separate universe computations, $\eta_i=-1/\kpeak$ (brown) and $\eta_i = -1/k_e$ (blue). The yellow region highlights USR evolution, defined by the condition $\epsilon_2<-3$.}
\label{fig:nu squared evolution} 
\end{figure}

In Fig.~\ref{fig:nu squared evolution} we display the time evolution of the first slow-roll parameter, $\epsilon_1$, and the Mukhanov--Sasaki mass, 
\begin{equation}
    \nu^ 2 = \frac{9}{4} - \epsilon_1 + \frac{3}{2} \epsilon_2 -\frac{1}{2} \epsilon_1 \epsilon_2 +\frac{1}{4} \epsilon_2^2 + \frac{1}{2} \epsilon_2 \epsilon_3 \;, 
\end{equation}
computed for the USR model studied in Appendix A of Ref.~\cite{Iacconi:2023ggt}. 
One can see that after the initial transition into USR, $\nu^2$ remains constant.

To compute $\dimlessP_\zeta(k; t)_\text{tree}$ we apply the strategy proposed in Ref.~\cite{Jackson:2023obv}, see Appendix C therein. 
In particular, at time $\eta_i$ we match the numerical solution to the Mukhanov--Sasaki equation for each mode $k<-1/\eta_i$ to its homogeneous counterpart, which provides the initial condition to describe the subsequent superhorizon evolution. 
Thanks to the semi-analytical nature of this method, at the time of the matching we discard gradient corrections and only include the homogeneous growing and decaying modes. 
\begin{figure}
\centering
\captionsetup[subfigure]{justification=centering}
    \begin{subfigure}{.48\textwidth}
        \includegraphics[width=\textwidth]{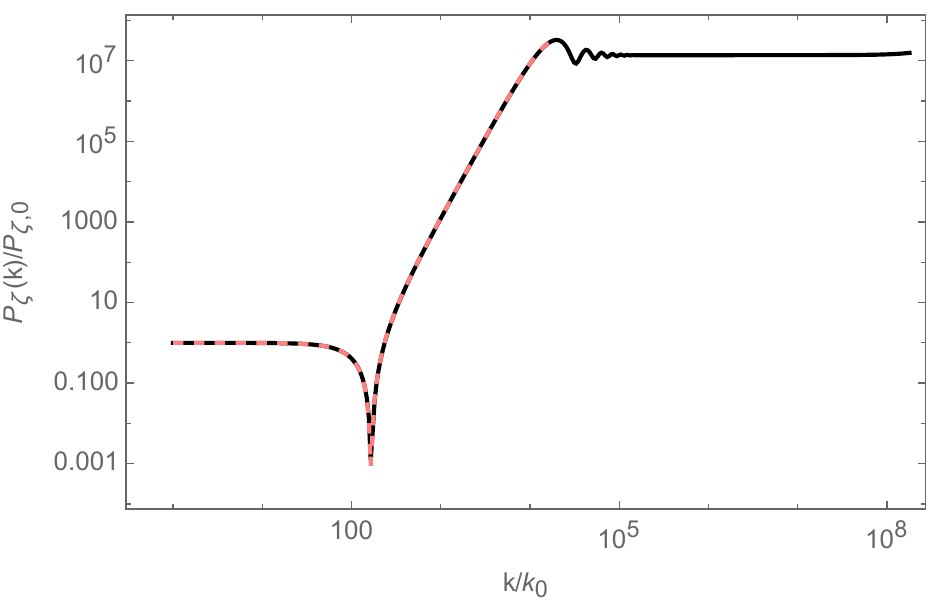}
    \end{subfigure}
    \begin{subfigure}{.48\textwidth}
        \includegraphics[width=\textwidth]{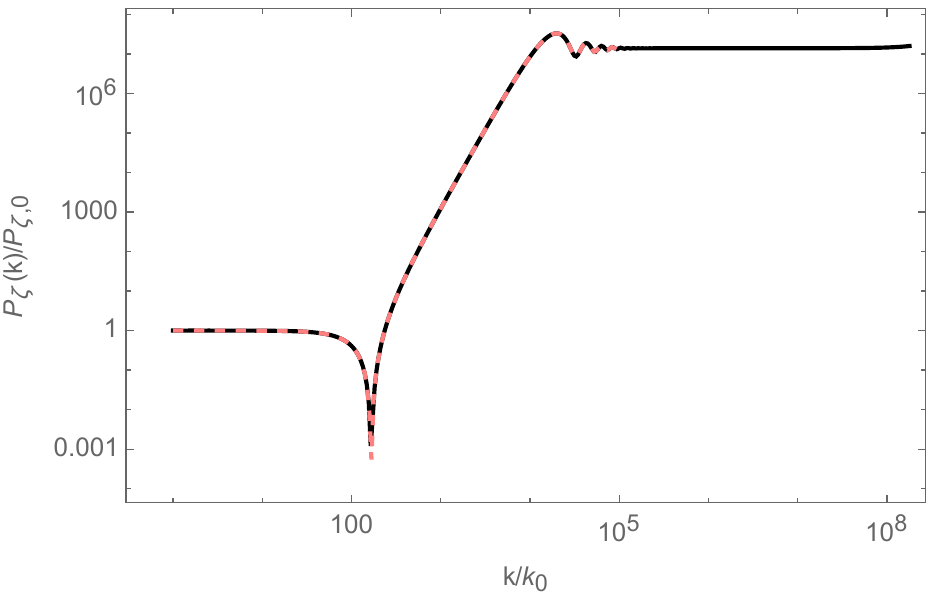}
    \end{subfigure}
\caption{Linear power spectrum, $\dimlessP_\zeta(k; t)_\text{tree}$, computed by applying the separate universe approach (pink, dashed line), initialized at $\eta_i = -1/\kpeak$ (left) and $\eta_i = -1/k_e$ (right). In both panels, the black line represents results obtained by numerically solving the Mukhanov--Sasaki equation.}
     \label{fig:linerar Pz separate universe}
\end{figure}
We represent in Fig.~\ref{fig:linerar Pz separate universe} our results for $\dimlessP_\zeta(k; t)_\text{tree}$, which show excellent agreement with the numerical values.
Note that all wavenumbers up to $k_i=-1/\eta_i$ (i.e. the inverse of the smoothing scale) are described correctly, including those which rise to the peak and those displaying scale-dependent oscillations around the peak. 

\section{Squeezed phase-space bispectrum some e-folds after horizon crossing}
\label{app: squeezed bispectrum at t_k}

In this Appendix, we compute the (tree-level)
squeezed bispectrum 
$\tensor{\alpha}{^I^J^K}(p,q,q;t_i)$, 
which appears in Eq.~\eqref{eq:12 13 step 1} as the initial condition for the (12)-type $\delta N$ loop.

Shortly after the scale $q$ has crossed the horizon,
the squeezed bispectrum can be estimated by soft-limit
arguments, see Eq.~\eqref{eq: 3-point function squeezed}. 
Nevertheless, in order to include the whole broadband of enhanced modes in our back-reaction estimate, $t_i$ might be chosen substantially later than the horizon-crossing time of some $q$ modes.
Here, we employ the multi-point propagator
expansion introduced in {\S}\ref{sec: 1-loop from quantum initial conditions is a boundary term} to evaluate
$\tensor{\alpha}{^I^J^K}(p,q,q;t_i)$,
and determine how outside-the-horizon evolution between
$t_q$ and $t_i$ contributes to the $(12)$-type loop in
Eq.~\eqref{eq:12 13 step 1}. 

Note that since all modes involved are already superhorizon
between $t_q$ and $t_i$, the multi-point coefficients
we employ are momentum independent. 
As a result, the expansion we use here is similar
to the $\delta N$ one, but formulated in phase space.
For this reason, in the following we drop momentum labels when writing
multi-point propagators.

By expanding phase-space fields at time $t_i$
in terms of those at $t_q$, and substituting these in
the 3-point correlator of interest,
we find 
\begin{equation}
\label{eq: alpha t_i step 1}
    \begin{split}
        \langle \delta X^I_{\vect{p}} \delta X^J_{\vect{q}}& \delta X^K_{-\vect{p}-\vect{q}}  \rangle_{t_i} = 
        {\tensor{\Gamma}{^I_M}}^{(t_q,t_i)}\, 
        {\tensor{\Gamma}{^J_N}}^{(t_q,t_i)} \,
        {\tensor{\Gamma}{^K_O}}^{(t_q,t_i)}\,
        \langle \delta X^M_{\vect{p}} \delta X^N_{\vect{q}} \delta X^O_{-\vect{p}-\vect{q}}  \rangle_{t_q} \\ 
        &+ \frac{1}{2} \,
        {\tensor{\Gamma}{^I_M_N}}^{(t_q,t_i)} \,
        {\tensor{\Gamma}{^J_O}}^{(t_q,t_i)} \,
        {\tensor{\Gamma}{^K_P}}^{(t_q,t_i)} \,
        \int\d^3l \; \langle \delta X^M_{\vect{l}} \delta X^N_{\vect{p}-\vect{l}} \delta X^O_{\vect{q}} \delta X^P_{-\vect{p}-\vect{q}} \rangle_{t_q} \\
        & + \frac{1}{2}\, 
        {\tensor{\Gamma}{^I_M}}^{(t_q,t_i)} \,
        {\tensor{\Gamma}{^J_N_O}}^{(t_q,t_i)} \,
        {\tensor{\Gamma}{^K_P}}^{(t_q,t_i)} \,
        \int\d^3l \; \langle \delta X^M_{\vect{p}} \delta X^N_{\vect{l}} \delta X^O_{\vect{q}-\vect{l}} \delta X^P_{-\vect{p}-\vect{q}} \rangle_{t_q} \\
        & + \frac{1}{2} \, 
        {\tensor{\Gamma}{^I_M}}^{(t_q,t_i)} \,
        {\tensor{\Gamma}{^J_N}}^{(t_q,t_i)} \, 
        {\tensor{\Gamma}{^K_O_P}}^{(t_q,t_i)} \, 
        \int\d^3l \; \langle \delta X^M_{\vect{p}} \delta X^N_{\vect{q}} \delta X^O_{\vect{l}} \delta X^P_{-\vect{p}-\vect{q}-\vect{l}} \rangle_{t_q} \;.
    \end{split}
\end{equation}
Here, we have not yet taken the limit $p\ll q$. 
The first term encodes the linear evolution of the
intrinsic phase-space bispectrum from $t_q$ to $t_i$.
The other three
describe non-Gaussianity produced by non-linear
super-horizon evolution between $t_q$ and $t_i$.
We note that higher-order contributions,
e.g., those proportional to $\Gamma^I_{JKL}$,
contribute beyond tree-level,
and therefore have been neglected here. 

After Wick contraction and taking the limit $p\ll q$, we obtain 
\begin{equation}
\label{eq: alpha t_i step 2}
\begin{split}
    \tensor{\alpha}{^I^J^K}(p,q,q; t_i) & =  
    {\tensor{\Gamma}{^I_M}}^{(t_q,t_i)}\,
    {\tensor{\Gamma}{^J_N}}^{(t_q,t_i)} \,
    {\tensor{\Gamma}{^K_O}}^{(t_q,t_i)} \,
    \tensor{\alpha}{^M^N^O} (p,q,q;t_q) \\
    & +\, 
    {\tensor{\Gamma}{^I_M_N}}^{(t_q,t_i)} \, 
    {\tensor{\Gamma}{^J_O}}^{(t_q,t_i)} \,
    {\tensor{\Gamma}{^K_P}}^{(t_q,t_i)} \, 
    P^{MO}(q;t_q) \,
    P^{NP}(q; t_q)  \\
    & + \, 
    {\tensor{\Gamma}{^I_M}}^{(t_q,t_i)}\, 
    {\tensor{\Gamma}{^J_N_O}}^{(t_q,t_i)}\,
    {\tensor{\Gamma}{^K_P}}^{(t_q,t_i)} \, 
    P^{MO}(p;t_q)\,  
    P^{NP}(q; t_q) \\
    & + \, 
    {\tensor{\Gamma}{^I_M}}^{(t_q,t_i)} \,
    {\tensor{\Gamma}{^J_N}}^{(t_q,t_i)} \,
    {\tensor{\Gamma}{^K_O_P}}^{(t_q,t_i)} \, 
    P^{MP}(p;t_q) \, 
    P^{NO}(q;t_q) \;. 
\end{split}
\end{equation}
Note we have used symmetry of the multi-point propagators
under exchange of two lower indices.
For later convenience, let us label each of the four terms as
${\#}1$, ${\#}2$, etc.
We substitute Eq.~\eqref{eq: 3-point function squeezed} into ${\#}1$, obtaining 
\begin{equation}
\label{eq: term 1 step 1}
    {\#}1 = 
    {\tensor{\Gamma}{^I_M}}^{(t_q,t_i)} \, 
    {\tensor{\Gamma}{^J_N}}^{(t_q,t_i)}\, 
    {\tensor{\Gamma}{^K_O}}^{(t_q,t_i)}\,
    P^{MR}(p;t_q) \, 
    \frac{\partial P^{NO}(q;t_q)}{\partial X^R(t_q)} \;. 
\end{equation}    
By applying the Leibniz rule, Eq.~\eqref{eq: term 1 step 1} can be rewritten as 
\begin{equation}
\label{eq: term 1 step 2}
\begin{split}
    {\#}1 & =
    {\tensor{\Gamma}{^I_M}}^{(t_q,t_i)} \, 
    P^{MR}(p;t_q) \,  
    \frac{\partial}{\partial X^R(t_q)} 
    \left[ 
    {\tensor{\Gamma}{^J_N}}^{(t_q,t_i)} \,
    {\tensor{\Gamma}{^K_O}}^{(t_q,t_i)} \, 
    P^{NO}(q;t_q) 
    \right] \\
    & \quad 
    - 
    {\tensor{\Gamma}{^I_M}}^{(t_q,t_i)} \,
    P^{MR}(p;t_q) \, 
    P^{NO}(q;t_q) 
    \frac{\partial}{\partial X^R(t_q)} \left[
    {\tensor{\Gamma}{^J_N}}^{(t_q,t_i)} \,
    {\tensor{\Gamma}{^K_O}}^{(t_q,t_i)} 
    \right] 
    \;. 
\end{split}
\end{equation}  
In the first term one can
switch the time of evaluation of the background field $X^R$ in the derivative
from $t_q$ to $t_i$.
This yields 
\begin{equation}
\label{eq: term 1 step 3}
\begin{split}
    {\#}1 & =
    {\tensor{\Gamma}{^I_M}}^{(t_q,t_i)} \, 
    P^{MR}(p;t_q) \,  
    {\tensor{\Gamma}{^S_R}}^{(t_q,t_i)} \, 
    \frac{\partial}{\partial X^S(t_i)} 
    \left[ 
        {\tensor{\Gamma}{^J_N}}^{(t_q,t_i)} \,
        {\tensor{\Gamma}{^K_O}}^{(t_q,t_i)} \, 
    P^{NO}(q;t_q) 
    \right] \\
    & \quad 
    - 
    {\tensor{\Gamma}{^I_M}}^{(t_q,t_i)} \,
    P^{MR}(p;t_q) \, 
    P^{NO}(q;t_q) 
    \frac{\partial}{\partial X^R(t_q)} \left[
    {\tensor{\Gamma}{^J_N}}^{(t_q,t_i)} \,
    {\tensor{\Gamma}{^K_O}}^{(t_q,t_i)} 
    \right]  \;.
\end{split}
\end{equation}  
In the first line,
the contractions of $\Gamma$ coefficients with $P^{MR}(p, t_q)$
and $P^{NO}(q, t_q)$ each produce a linearly
evolved power spectrum at $t_i$.
In the second line, the 2-index $\Gamma$ symbols
differentiate to a 3-index $\Gamma$ symbol.
As a result, we obtain (after relabelling summation indices)
\begin{multline}
\label{eq: term 1 step 4}
    {\#}1  = 
    P^{IS}(p;t_i)  
    \frac{\partial P^{JK}(q;t_i)}
    {\partial X^S(t_i)} 
    - 
    {\tensor{\Gamma}{^I_M}}^{(t_q,t_i)} \,
    {\tensor{\Gamma}{^J_N_O}}^{(t_q,t_i)} \,
    {\tensor{\Gamma}{^K_P}}^{(t_q,t_i)} \, 
    P^{MN}(p;t_q) \, 
    P^{OP}(q;t_q)  
    \\ 
    -
    {\tensor{\Gamma}{^I_M}}^{(t_q,t_i)} \,
    {\tensor{\Gamma}{^J_N}}^{(t_q,t_i)} \, 
    {\tensor{\Gamma}{^K_O_P}}^{(t_q,t_i)}  \, 
    P^{MO}(p;t_q) \, 
    P^{NP}(q;t_q)
    .
\end{multline}
Once ${\#}1$ is combined with
the other terms in Eq.~\eqref{eq: alpha t_i step 2},
it can be seen that
the second and third terms in Eq.~\eqref{eq: term 1 step 4}
cancel with ${\#}3$ and ${\#}4$.
Therefore, Eq.~\eqref{eq: alpha t_i step 2} reduces to  
\begin{multline}
    \label{eq: alpha t_i step 3}
    \tensor{\alpha}{^I^J^K}(p,q, q;t_i) = 
    P^{IS}(p;t_i) \, 
    \frac{\partial P^{JK}(q;t_i)}{\partial X^S(t_i)} \\
    + 
    {\tensor{\Gamma}{^I_M_N}}^{(t_q, t_i)} \, 
    {\tensor{\Gamma}{^J_O}}^{(t_q, t_i)} \,
    {\tensor{\Gamma}{^K_P}}^{(t_q, t_i)} \, 
    P^{MO}(q;t_q) \,  
    P^{NP}(q; t_q)  
    \;.
\end{multline}
The first term has the same structure as
Eq.~\eqref{eq: 3-point function squeezed}, but is evaluated at $t_i$.
The second term introduces a correction due to non-linear evolution
between $t_q$ and $t_i$,
in which two short scales at $t_q$ combine
to produce a long mode at $t_i$. 

The 1-loop correction to $\dimlessP_\zeta(p;t)$
induced by the second term in Eq.~\eqref{eq: alpha t_i step 3}
is volume-suppressed in the limit of $p\to 0$.
This can be seen by substitution in Eq.~\eqref{eq:12 13 step 1},
which yields 
\begin{equation}
    \label{eq:12 second contribution 1}
    \dimlessP_\zeta(p;t)_\text{12} 
    \supset  
    \frac{p^3}{2\pi^2} 
    \tensor{N}{_I}^{(t_i,t)} 
    \tensor{N}{_J_K}^{(t_i,t)}
    {\tensor{\Gamma}{^I_M_N}}^{(t_q, t_i)} \,
    {\tensor{\Gamma}{^J_O}}^{(t_q, t_i)} \,
    {\tensor{\Gamma}{^K_P}}^{(t_q, t_i)} 
    \int \d^3 q \; 
    P^{MO}(q;t_q) \,
    P^{NP}(q; t_q) \;,
\end{equation}
To conclude, we have shown that the squeezed bispectrum at
$t_i$ is given by two contributions, see Eq.~\eqref{eq: alpha t_i step 3}.
Of these, the second leads to volume-suppressed back-reaction,
which justifies our use of Eq.~\eqref{eq:soft limit alpha at t_i}
in {\S}\ref{sec: 1-loop from non-linear superhorizon evolution is a boundary term}.

\bibliographystyle{JHEP}
\bibliography{refs} 

\end{document}